\documentclass[aps,prb,amsmath,twocolumn,amssymb,floatfix,superscriptaddress,showpacs]{revtex4}

\newcommand{\degree}{\ensuremath{^\circ} } 
\newcommand{\pd}[2]{\frac {\partial #1} {\partial #2}}
\newcommand{\linecite}[1]{Ref.~\onlinecite{#1}}
\newcommand{\linecitedouble}[2]{Refs.~\onlinecite{#1} and \onlinecite{#2}}

\usepackage{color}

\usepackage{ amssymb }
\usepackage{graphicx}
\usepackage{enumerate}
\usepackage{subfigure}
\usepackage{bbold}
\usepackage{verbatim}
\usepackage{float}
\usepackage{dsfont}
\usepackage{multirow} 
\begin{document}
\title{
%Novel magnetic supersolid 
Competition between supersolid phases and magnetization plateaux 
in the frustrated easy-axis antiferromagnet on a triangular lattice}
\author{Luis Seabra}
\affiliation{H.\ H.\ Wills Physics Laboratory, University of Bristol,  Tyndall Av, BS8--1TL, UK.}
\author{Nic Shannon}
\affiliation{H.\ H.\ Wills Physics Laboratory, University of Bristol,  Tyndall Av, BS8--1TL, UK.}

\date{\today}
\begin{abstract}   
The majority of magnetic materials possess some degree of magnetic anisotropy, either at the level of a 
single ion, or in the exchange interactions between different magnetic ions.    
Where these exchange interactions are also frustrated, the competition between them and anisotropy can 
stabilize a wide variety of new phases in applied magnetic field.
Motivated by the hexagonal delafossite  2H-AgNiO$_2$, we study the Heisenberg antiferromagnet on a 
layered triangular lattice with competing first- and second-neighbour interactions and single-ion easy-axis 
anisotropy.
Using a combination of classical Monte Carlo simulation, mean-field analysis, and Landau theory, 
we establish the magnetic phase diagram of this model as a function of temperature and magnetic field
for a fixed ratio of exchange interactions, but with values of easy-axis anisotropy $D$ extending from the 
Heisenberg ($D$=0) to the Ising ($D$=$\infty$) limits.
We uncover a rich variety of different magnetic phases.
These include several phases which are magnetic supersolids (in the sense of Matsuda and Tstuneto 
or Liu and Fisher), one of which may already have been observed in AgNiO$_2$.
We explore how this particular supersolid arises through the closing of a gap in the spin-wave 
spectrum, and how it competes with rival collinear phases as the easy-axis anisotropy is increased.  
The finite temperature properties of this phase are found to be different from those of any previously 
studied magnetic supersolid.
\end{abstract}


\maketitle

%%%%%%%%%%%%%%%%%%%%%%%%%%%%%%%%%%%%%%%%%%%%%%%%%%%%%%

\section{Introduction}
\label{sec:introduction}

%%%%%%%%%%%%%%%%%%%%%%%%%

The very real possibility of finding new quantum phases and excitations has made frustrated magnetism one of the most exciting and 
dynamic research areas in contemporary condensed matter physics\cite{diepbook,balents10}.   
Often, frustrated magnets have the potential to support many different ground states, and can be tuned at will between these 
states by applying a magnetic field.
Relative to other approaches, such as chemical doping, this field control of phase transitions has the great advantage 
that it does not introduce new complications --- such as disorder --- into the system.   
The advent of powerful new magnets, in some cases capable of reaching many hundreds of Tesla, has made it possible to 
study new magnetic phases and the (quantum) phase transitions between them in a wide range of materials for the first time.  
These high-field experiments have already yielded new insights into phenomena as diverse as 
Bose-Einstein condensation~\cite{giamarchi08}, and the delicate interplay between magnetism and 
structural transitions in Cr spinels~\cite{penc04}.

%%%%%%%%%%%%%%%%%%%%%%%%%

Triangular lattice antiferromagnets have been central to our understanding of frustration since the pioneering work of Wannier\cite{wannier50}.  
A wide range of novel phases have first been proposed in the context of the triangular lattice, including the celebrated
``resonating valence bond'' (RVB) spin-liquid~\cite{anderson73,fazekas74}, two-dimensional chiral phases~\cite{kawamura84}, 
magnetisation plateaux~\cite{kawamura85,chubukov91}, and a collection of different multipolar states~\cite{momoi99,tsunetsugu06,
momoi06,lauchli06}.
Recently, spin-liquid behaviour has been established experimentally in a variety of triangular-lattice systems, 
including the quasi-2D organics \mbox{$\kappa$-(ET)$_2$Cu$_2$(CN)$_3$}~\cite{shimizu03} 
and EtMe$_3$Sb[Pd(dmit)$_2$]$_2$ \cite{yamashita10}, 
thin films of $^3$He~[\linecite{ishida97}], 
and the layered~pnictide~NiGa$_2$As$_4$~\cite{nakatsuji05}.

%%%%%%%%%%%%%%%%%%%%%%%%%%

While the majority of this work assumes $SU(2)$ invariance --- i.e. the absence of magnetic 
anisotropy --- this is rarely true of real materials.   
Generically, these exhibit some degree of magnetic anisotropy, either at the level of a single ion, 
or in their exchange interactions~\cite{collins97}.  
It has been known for a long time that the interplay between magnetic field, frustration and magnetic 
anisotropy can lead to novel magnetic phases on the triangular lattice, even in simple classical models.
The classical Heisenberg antiferromagnet with easy-axis anisotropy, first studied by Miyashita and 
Kawamura\cite{miyashita85,miyashita86}, exhibits a competition between the coplanarity favoured 
by magnetic interactions \cite{kawamura84}, and collinear order driven by anisotropy and magnetic 
field\cite{sheng92}.  
More recent studies have highlighted successive Berezinskii-Kosterlitz-Thouless phase transitions as 
a function of temperature \cite{melchy09,stephan00} and exotic quantum effects at large single-ion 
anisotropy \cite{sen09}.

%%%%%%%%%%%%%%%%%%%%%%%%%
 
One possible outcome of the competition between exchange interactions, magnetic anisotropy and 
magnetic field is a state which breaks both spin and lattice symmetries, exhibiting a finite (staggered) 
magnetisation perpendicular to the field, and broken translational symmetry.  
Matsuda and Tstuneto~\cite{matsuda70} and Liu and Fisher~\cite{liu73} argued, via a 
quantum lattice gas mapping, that such a mixed-symmetry phase could be an exact analogue of the 
supersolid phase proposed in $^4$He~[\linecite{andreev71,chester70,leggett70}].   
In the model considered by Liu and Fisher, a magnetic supersolid was found to interpolate between a 
collinear antiferromagnet and its high-field, canted, ``spin-flop" state.   
This magnetic supersolid was argued to terminate in a finite-temperature tetracritical point, where 
collinear, supersolid, spin-flop and paramagnetic phases meet\cite{liu73}.    

%%%%%%%%%%%%%%%%%%%%%%%%%

Recent, controversial, experiments on $^4$He [\linecite{chan04}] have lead to a renaissance of 
interest in supersolid phases~\cite{balibar10} and their magnetic analogues. 
Quantum Monte Carlo 
simulations now provide clear evidence for magnetic supersolids in $S$=$1/2$, easy-axis models 
on the triangular lattice\cite{wessel05,heidarian05,melko05, boninsegni05}.   
In this case, following the original logic of Andreev~\cite{andreev71}, the transition into a magnetic 
supersolid can be interpreted as a field-driven Bose-Einstein condensation of magnons within a 
magnetic solid --- a collinear, magnetically ordered phase.  
Magnetic supersolids have also been associated with biconical phases\cite{holtschneider07,holtschneider08}. 

%%%%%%%%%%%%%%%%%%%%%%%%%

Among triangular lattice magnets, AgNiO$_2$ is of special interest.  
Combining metallicity and localised magnetism, its ground state is a collinear antiferromagnetic 
``stripe" phase \cite{wawrzynska07}.  
When applying magnetic field a complicated and  as of yet unexplained cascade of phase transitions is observed~\cite{coldea09}.  
In a previous work we have argued that  the low-field transition observed in AgNiO$_2$ converts the collinear ground state into a novel state 
which is a magnetic supersolid in the sense of Matsuda and Tstuneto or Liu and Fisher~\cite{seabra10}.  
In this case, the expected first-order spin-flop transition is replaced by a continuous transition into a magnetic supersolid.    
This occurs through the Bose-Einstein condensation of magnons with a {\it finite} momentum, leading to a novel magnetic supersolid 
distinct from any of the cases considered above.

%%%%%%%%%%%%%%%%%%%%%%%%%

In this paper we extend the analysis began in \linecite{seabra10} to explore systematically how competing exchange interactions 
and easy-axis anisotropy provide a route to novel magnetic phases under applied field in, e.g., hexagonal delafossites.   
We consider explicitly the model used to fit low-energy magnetic excitations in AgNiO$_2$, with competing first-neighbour ($J_1$) and second-neighbour ($J_2$) exchange interactions on a triangular lattice, small but finite inter-layer coupling ($J_\perp$), and a varying degree of easy-axis anisotropy $D$ [\linecite{wheeler09}].   
While there have been some theoretical studies on the interplay of further-neighbour interactions and anisotropy on the 
triangular lattice \cite{takagi95,haraldsen09}, the possibility of finding new phases in magnetic field remains largely unexplored.  

%%%%%%%%%%%%%%%%%%%%%%%%%

In this paper we constrain the ratio between exchange parameters to values relevant to AgNiO$_2$ and explore how 
the magnetic phase diagram evolves as a function of easy-axis anisotropy.   
The resulting behaviour is extremely rich, exhibiting a wide variety of magnetic supersolid and collinear plateau phases, 
(almost) all of which share a common four-site unit cell.  
Since our main interest is in finite-temperature phases and phase transitions, we consider the classical limit 
$S \rightarrow \infty$, which opens the problem to large-scale Monte Carlo simulation.  
This is complemented by zero-temperature mean-field theory, low-temperature spin-wave expansions, and Landau theory.   
Particular attention is paid to the novel magnetic supersolid phase introduced in \linecite{seabra10}. 
This is shown to have qualitatively different properties at finite temperature from any previously studied supersolid, and to be 
remarkably robust against changes in parameters.

%%%%%%%%%%%%%%%%%%%%%%%%%

\begin{figure*}[hbt!]
\includegraphics[width= 17cm]{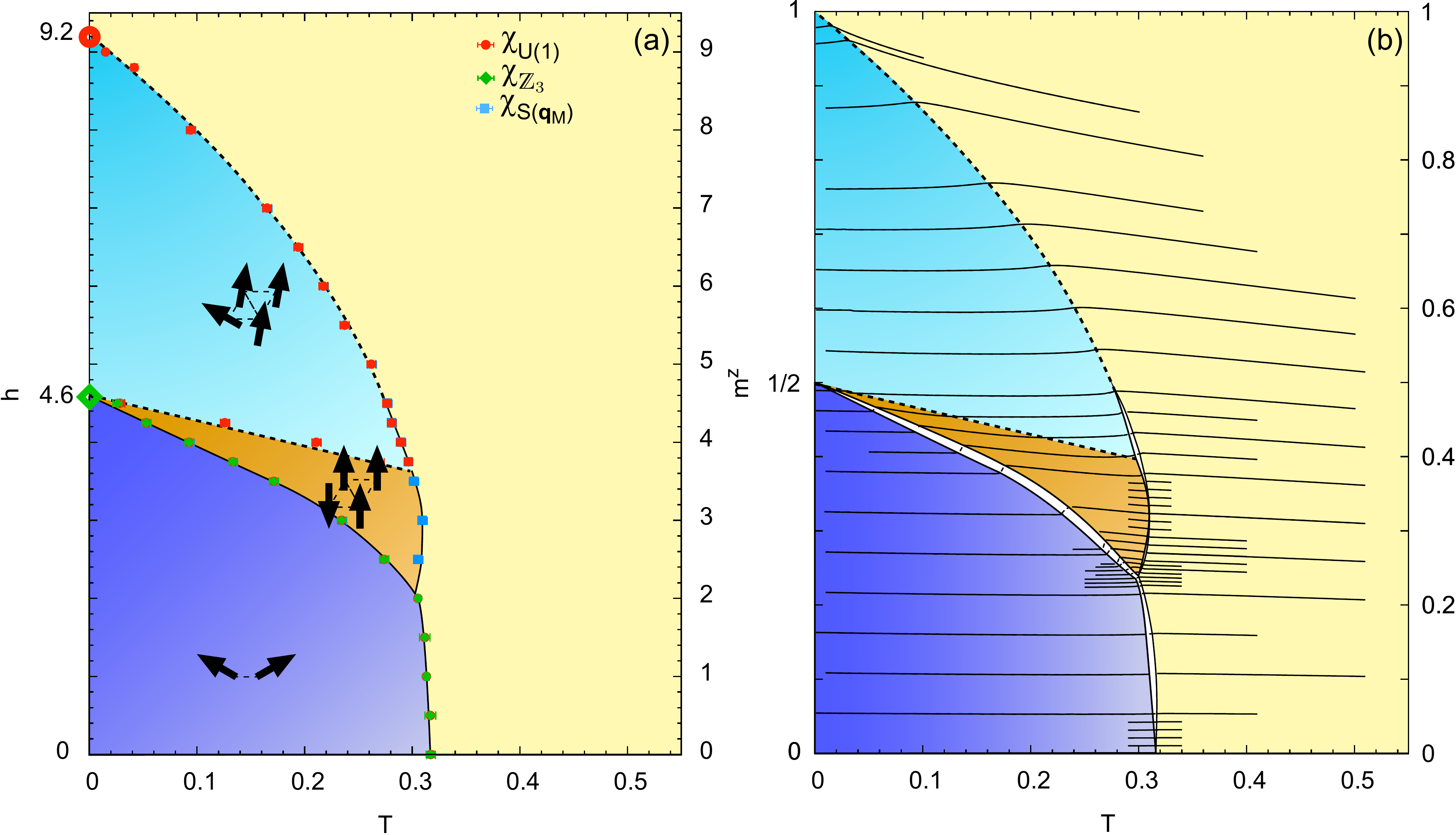}
\caption{\footnotesize{(color online) Magnetic phases of a layered triangular-lattice antiferromagnet with exchange interactions $J_1$=$1$, $J_2$=$0.15$, $J_\perp$=$-0.15$ and \mbox{easy-axis anisotropy $D$=0}.
(a)  Phase diagram as a function of temperature and magnetic field.     
Open symbols on the h-axis show transitions obtained in mean-field theory.   
Phase boundaries at finite temperature are obtained from Monte Carlo simulation for a cluster of $24$$ \times$$ 24$$ \times 8$ 
spins, and determined by peaks in the relevant order parameter susceptibilities.  
All phase transitions are first-order, except where shown with a dashed line.  
All phases share a 4-sublattice unit-cell.  
(b) Phase diagram as a function of temperature and magnetization.   
Solid lines running left-right show cuts at constant magnetic field $h$ taken from simulations.    
The coexistence regions associated with first order phase transitions are coloured white. }}
\label{fig:D0-panel}
\end{figure*}

\begin{figure*}[htp!]
\includegraphics[width= 17cm]{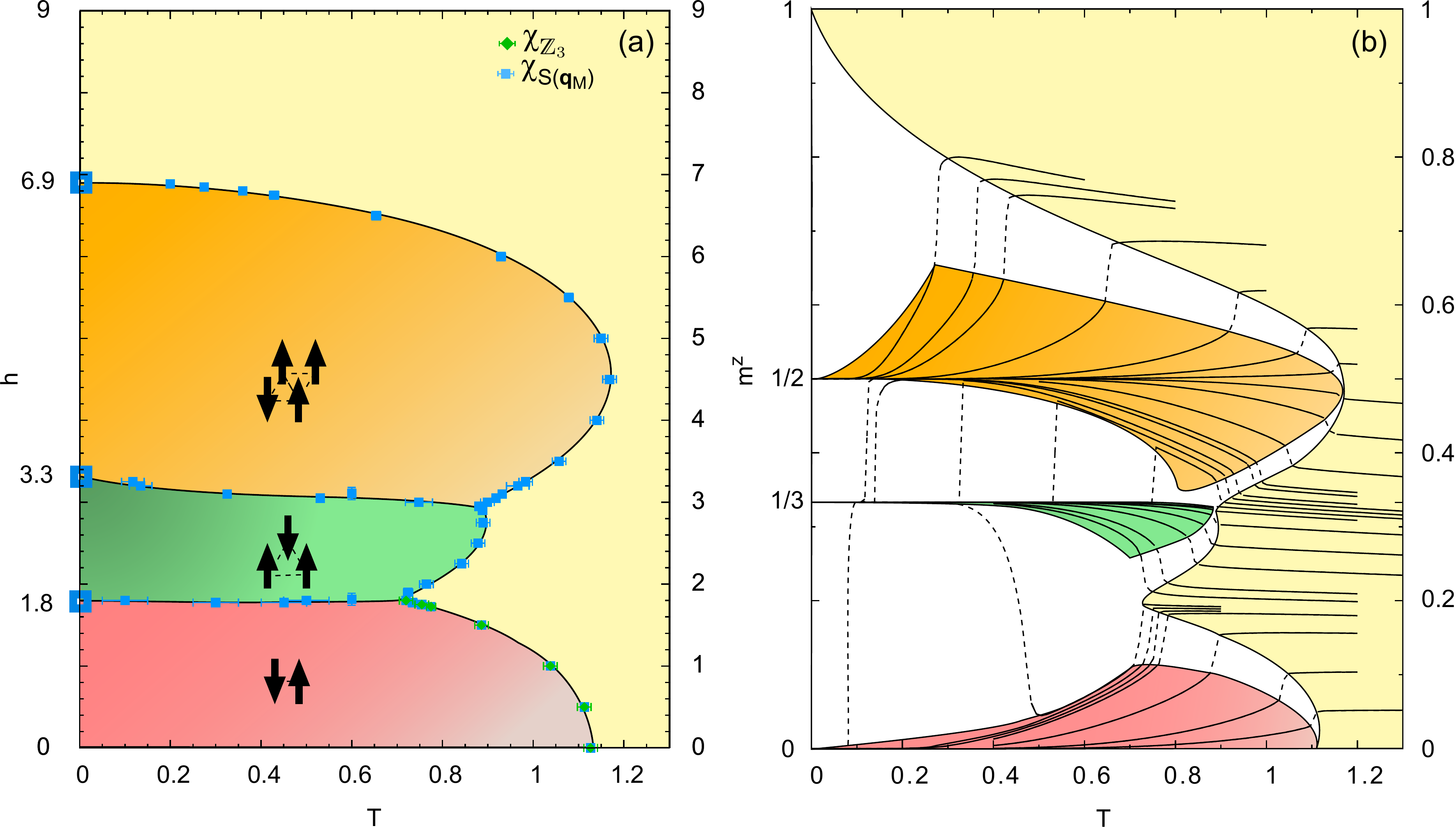}
\caption{\footnotesize{ (color online) (a) Magnetic phases of a layered triangular-lattice Ising antiferromagnet with  $J_1$=$1$, $J_2$=$0.15$, $J_\perp$=$-0.15$.
(a)  Phase diagram as a function of temperature and magnetic field.  
Open symbols on the h-axis show transitions obtained in mean-field theory.   
Phase boundaries at finite temperature are obtained from Monte Carlo simulation for a cluster of $24$$ \times$$ 24$$ \times 8$ spins, and determined by peaks in the relevant order parameter susceptibilities.  
All phase transitions are first-order.  
(b) Phase diagram as a function of temperature and magnetization.   
Solid lines running left-right show cuts at constant magnetic field $h$ taken from simulations.    
The coexistence regions associated with first order phase transitions are coloured white.}}
\label{fig:Ising-panel}
\end{figure*}

The plan of the paper is as follows: in Section~\ref{sec:model+mft} we  introduce the model, discuss the choice of parameters 
and present mean-field arguments.
We also briefly outline the numerical and analytical methods employed. 
In Sections.~\mbox{\ref{D=0}--\ref{Ising}} we explore the finite-temperature and finite-field behaviour of the model for 
representative values of easy-axis anisotropy, interpolating from the Heisenberg model [$D=0$] to the Ising limit  [$D=\infty$].  
The main results of the paper are summarised in a corresponding series of phase diagrams :
Fig.~\ref{fig:D0-panel} [$D=0$],
Fig.~\ref{fig:D002-panel} [$D=0.02$], 
Fig.~\ref{fig:D025-panel} [$D=0.25$], 
Fig.~\ref{fig:D05-panel} [$D=0.5$], 
Fig.~\ref{fig:D0675-panel} [$D=0.65$], 
Fig.~\ref{fig:D15-panel} [$D=1.5$] and 
Fig.~\ref{fig:Ising-panel} [$D=\infty$].
We summarise the key points of the results and discuss the remaining open questions in Section~\ref{results}.
In Section~\ref{experiment} we consider the relevance of these results to experiments on AgNiO$_2$.  
We conclude in Section~\ref{conclusions}.
Technical details of calculations of spin stiffness and low-temperature spin wave 
expansions are discussed in Appendix~\ref{spinstiffness} and Appendix~\ref{appendix}.

%%%%%%%%%%%%%%%%%%%%%%%%%%%

\section{Model and methods}
\label{sec:model+mft}

%%%%%%%%%%%%%%%%%%%%%%%%%%%

\begin{figure}[t!]
\includegraphics[height= 4cm]{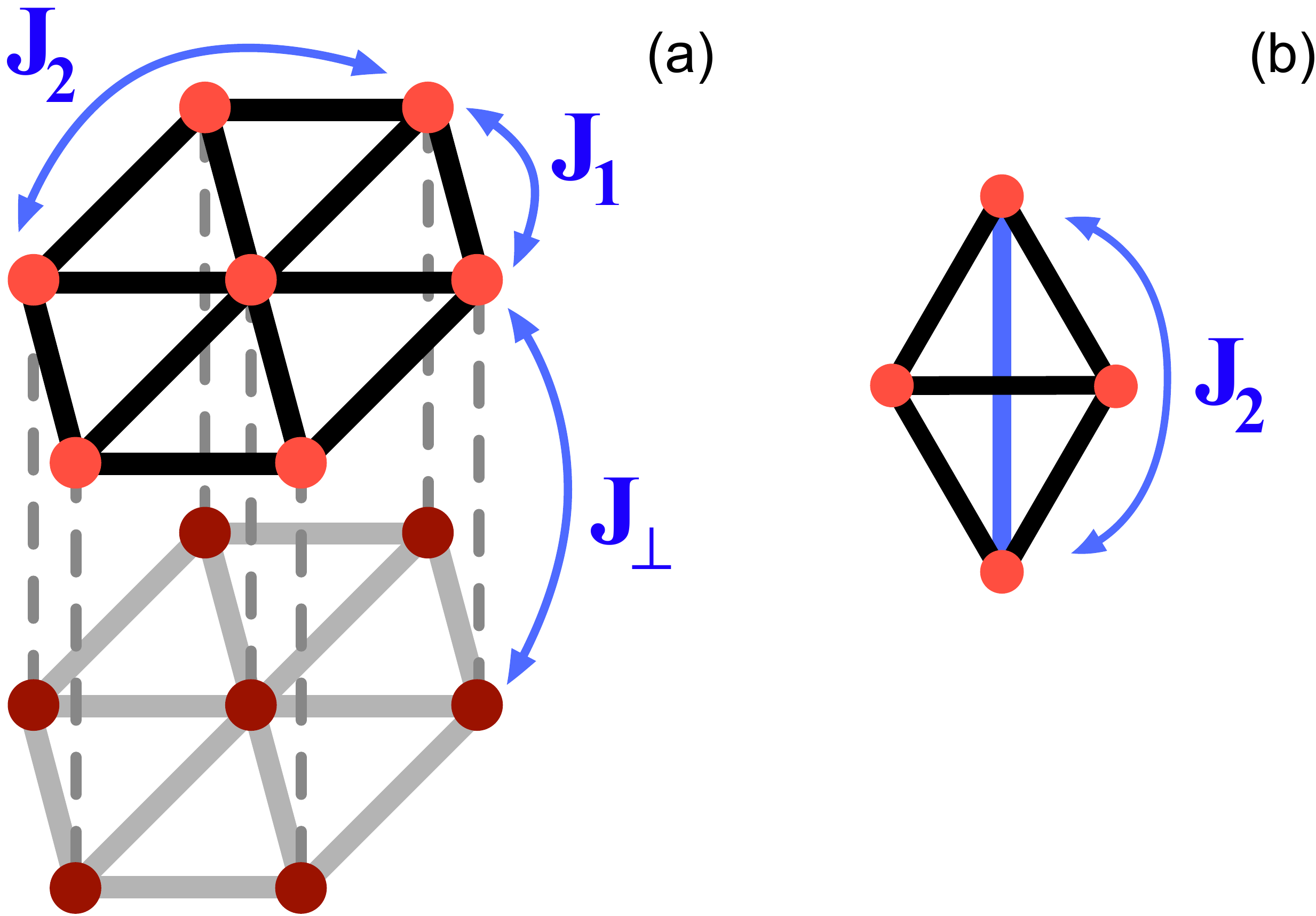}
\caption{\footnotesize{(color online) (a) Layered triangular lattice considered in this paper, with first- and 
second- neighbour interactions $J_1$ and $J_2$ in each plane, coupled by interactions between  
perpendicular \mbox{neighbours $J_\perp$}. 
(b) Second-neighbour interactions  transform the natural unit cell within the plane into a diamond plaquette. }}
\label{fig:lattice}
\end{figure}

We consider the classical Heisenberg model on a layered triangular lattice
\begin{align} 
\label{eq:H}
\mathcal{H} =& J_1\sum_{\langle ij \rangle_1} {{\bf S}_i} . {\bf S}_j + J_2\sum_{\langle ij \rangle_2} {\bf S}_i . {\bf S}_j
+ J_\perp\sum_{\langle ij \rangle_\perp} {\bf S}_i . {\bf S}_j \nonumber\\
 &-D \sum _i ({S_i^z})^2   - h\sum_{i} S^z_i,
\end{align}
with competing antiferromagnetic interactions $J_1>0$ and $J_2 >0$ on the first and second neighbour bonds of  
a triangular lattice plane [Fig.~\ref{fig:lattice}]. 
Neighbouring planes are coupled by an interlayer exchange $J_\perp$. 
Easy-axis anisotropy is characterised by a single-ion term $D$ and the magnetic field $h$ is assumed to lie along the $z$ axis. 
We set $|{\bf S}|$$ = $$S $=$1$ throughout, and all energy scales including field $h$ and temperature $T$ are measured in units of $J_1$ hereafter. 

%%%%%%%%%%%%fig-lattice %%%%%%%%%%%%

%%%%%%%%%%%%%%%%%%%%%%

We fix first-neighbour interactions to be $J_1 \equiv 1$ and set interlayer exchange $J_\perp$=$-0.15$.   
This leaves 
three distinct free parameters: second-neighbour coupling $J_2$, easy-axis anisotropy $D$ and external field $h$.  
We note that the $h=0$,  $J_\perp=0$,  $D \to \infty$ limit of Eq.~(\ref{eq:H}) corresponds to the frustrated Ising 
model previously studied in e.g. \linecitedouble{metcalf74}{takasaki86}. 

Second-neighbour interactions change the ``natural" description of the lattice from edge-sharing triangles to edge-sharing
 four-site ``diamond'' plaquettes, which can be thought of as a 2D projection of an irregular tetrahedron.   
Thus, at a mean-field level,  the exchange terms of Eq.~(\ref{eq:H}) can be rewritten in terms of the total 
magnetisation of each plaquette
\begin{align}
\mathcal{H}_{ex}=   \frac{J_1+J_2}{4} \sum^{N/4}_{\lozenge} | \mathbf{S}_A+\mathbf{S}_B+\mathbf{S}_C+\mathbf{S}_D |^2 - 4.
\label{eq:four-sl-energy}
\end{align}
This allows an interesting parallel with the physics of tetrahedra embedded in a pyrochlore lattice in a bilinear-biquadratic 
model \cite{penc04,motome06,shannon10}, explored further below.

\begin{figure}[ht]
\includegraphics[width= 7.5cm]{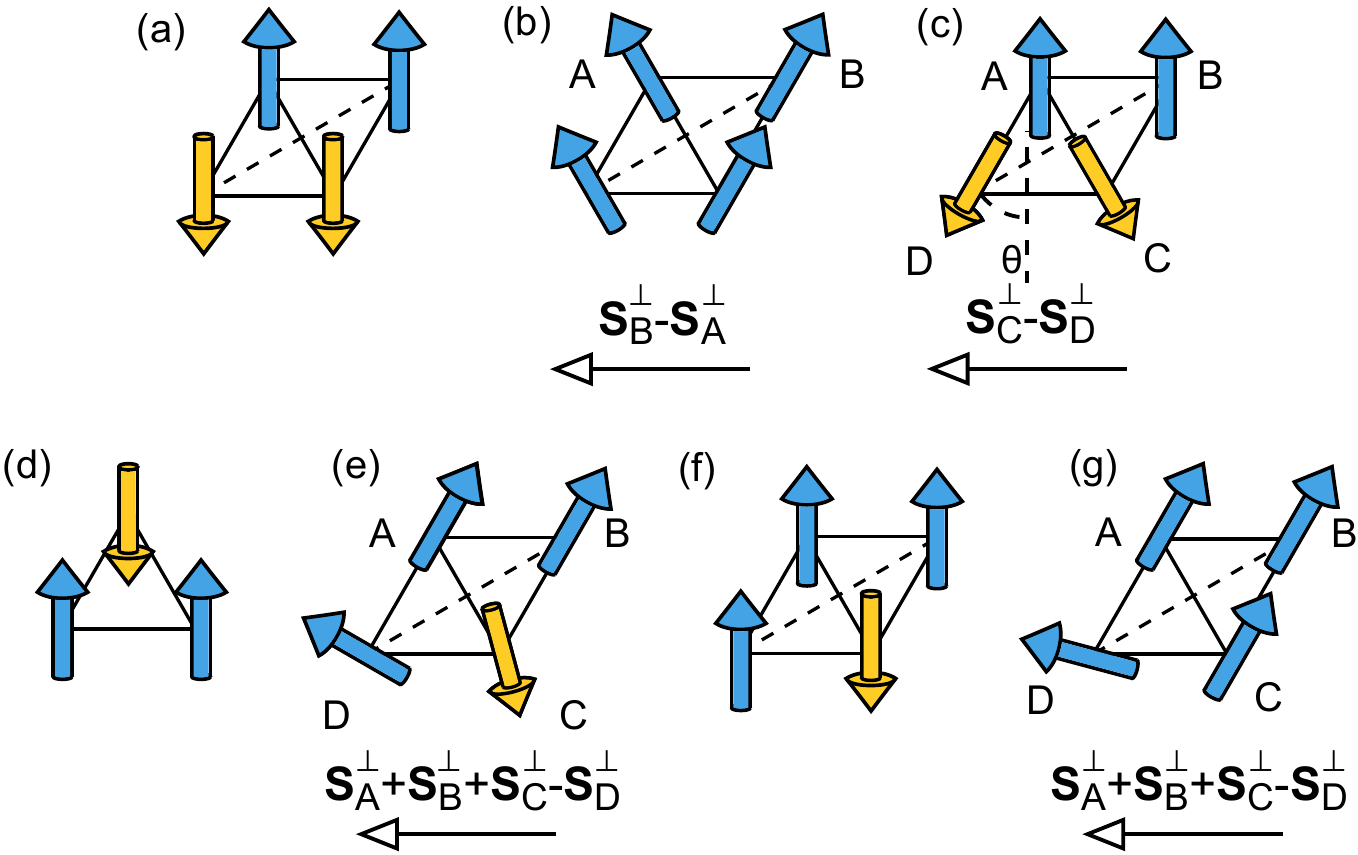}
\caption{\footnotesize{(color online) Cartoons of different three- and four-sublattice ground states found at mean-field 
level as a function of field. 
(a) Collinear stripe phase, 
(b) ``Spin-flopped" version of the stripe phase.
(c) novel magnetic supersolid state, the main focus of this paper. 
(d) $m$=1/3 magnetisation plateau. 
(e) 2:1:1 canted phase. 
(f) $m$=$1/2$ magnetisation plateau. 
(g)  3:1 canted phase. 
The staggered moment in the $S^x$$-$$S^y$ plane is shown as a vector 
for states (b), (c), (e) and (g).
}}
\label{fig:configs}
\end{figure}

The application of a magnetic field reveals a cascade of new states, summarised in Fig.~\ref{fig:configs}.
%and Table~\ref{table:orderparameters}.
%
These consist of :
\begin{enumerate}[(a)]
%\begin{list}{(a)}{}
%
\item  A two-sublattice, collinear antiferromagnetic ``stripe'' state [Fig.~\ref{fig:configs}(a)], which is the 
ground state in the absence of magnetic field, and which breaks both the translational and rotation 
symmetries of the lattice.
\item A two-sublattice canted ``spin-flop'' state [Fig.~\ref{fig:configs}(d)] which breaks both the rotational
symmetries of the lattice, {\it and} spin-rotation symmetry in the $S^x$-$S^y$ plane.
\item A novel magnetic supersolid state [Fig.~\ref{fig:configs}(e)] with a four-site unit cell 
which breaks both the translational and rotational symmetries of the lattice, 
{\it and} spin-rotational symmetry in the $S^x$-$S^y$ plane
\item A three-sublattice, $m$=1/3 magnetisation plateau [Fig.~\ref{fig:configs}(b)] which breaks the 
translation symmetries of the lattice.
\item A four-sublattice, 2:1:1 canted magnetic supersolid state [Fig.~\ref{fig:configs}(f)], which breaks 
both the rotational symmetries of the lattice {\it and} spin-rotational symmetry in the $S^x$-$S^y$ plane
\item A four-sublattice, $m$=1/2 magnetisation plateau [Fig.~\ref{fig:configs}(c)] which breaks both the 
translational and rotational symmetries of the lattice.
\item A four-sublattice, 3:1 canted magnetic supersolid state [Fig.~\ref{fig:configs}(g)],  which breaks 
both the rotational symmetries of the lattice {\it and} spin-rotational symmetry in the $S^x$-$S^y$ plane
\end{enumerate}
Each of these phase will be described in more detail below.  
In what follows we shall focus in particular on the novel magnetic supersolid state [Fig.~\ref{fig:configs}(c)], 
previously introduced in \linecite{seabra10} as an explanation for the low-field transition observed in 
AgNiO$_2$~\cite{coldea09}.

%%%%%%%%%%%%%%%
%%%%%%%%%%%%%%%%%%

\begin{figure}[ht!]
\includegraphics[width= 6cm]{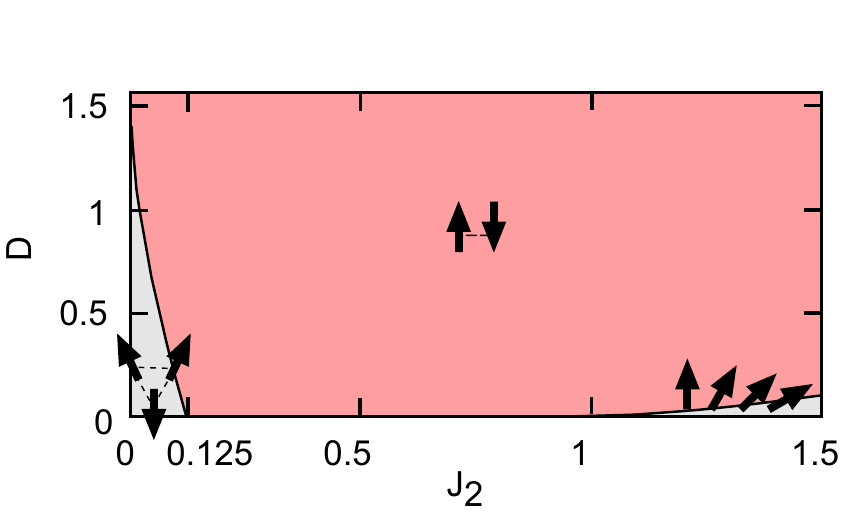}
\caption{\footnotesize{(color online) Zero field $J_2$$-$$D$ phase diagram  for $J_1$=1 and $J_\perp$=-0.15  from zero temperature mean-field theory. The model supports three different ground states : a three-sublattice coplanar state, a two-sublattice collinear ``stripe'' 
antiferromagnet and an incommensurate helicoidal phase.  }}
\label{fig:mft}
\end{figure}

\begin{figure}[h!]
\includegraphics[width= 6.5cm]{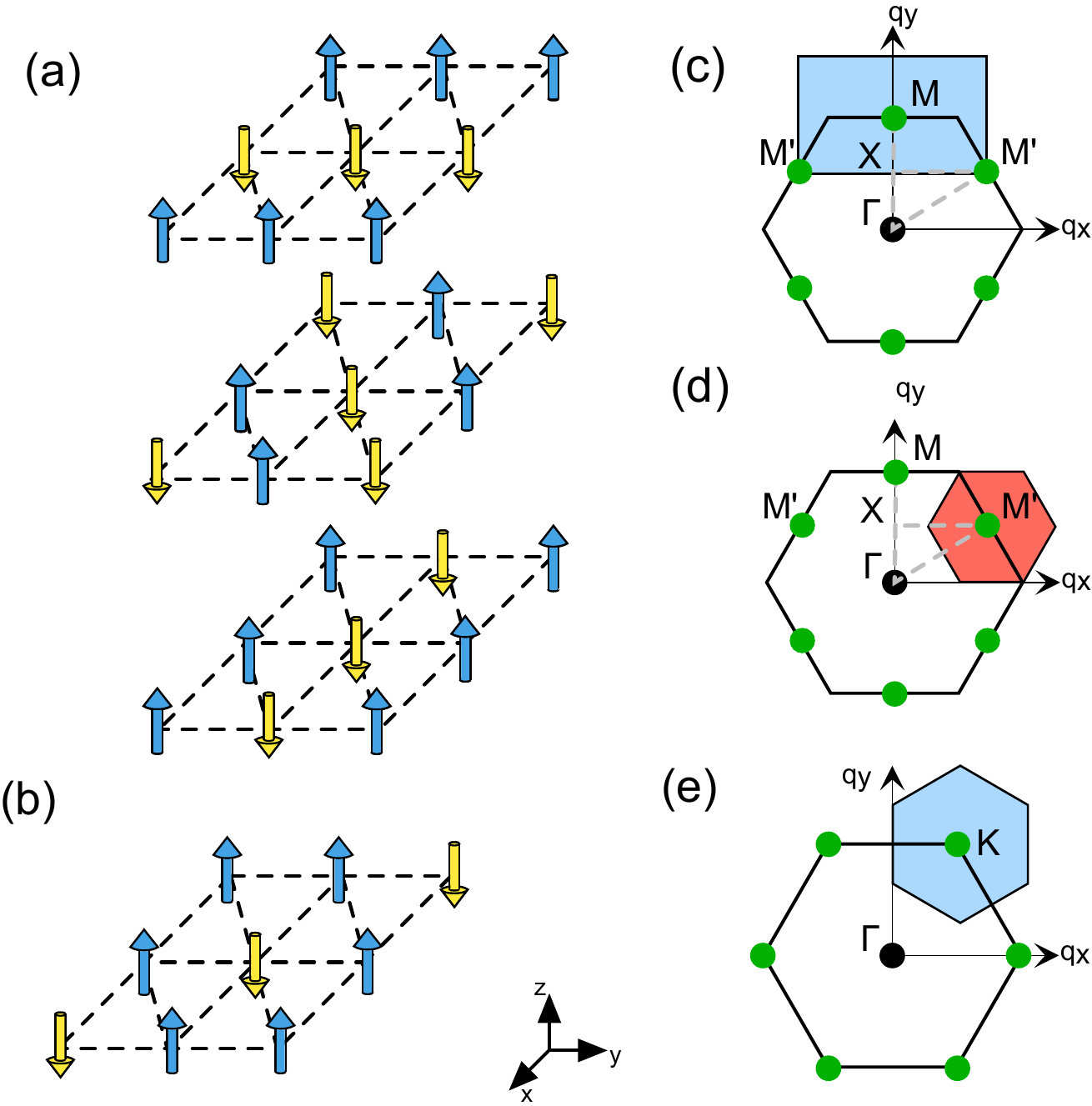}
\caption{\footnotesize{(color online) 
(a) Collinear stripe phase favoured by second-neighbour interactions, showing the three possible stripe orientations.   
(b) Three-sublattice  $m$=1/3 plateau state with two spins up and one down.   
(c) First Brillouin zone in $q_x$-$q_y$ plane showing momentum vectors \emph{M} and \emph{M'} associated with the stripe phases. 
The blue rectangle shows the first magnetic Brillouin zone for the stripe phase with ordering vector~$M$ (top cartoon).    
(d)  The red hexagon shows the first magnetic Brillouin zone for a generic four-sublattice state centred on \emph{M'}.  
(e) Ordering vectors \emph{K} of the $m$=1/3 plateau state and related first magnetic Brillouin zone, shaded blue. }}
\label{fig:BZ-stripes}
\end{figure}

To get a first idea of how these phases fit together, let us first consider the phase diagram in the absence of field, 
for a range of  $J_2$ and $D$, within zero-temperature mean field theory  [Fig. \ref{fig:mft}].  
For vanishing magnetic anisotropy ($D=0$) and $J_2 < 1/8$, we find the well-known coplanar three-sublattice 
120\degree ground state\cite{kawamura84}.   
For $1/8 < J_2 < 1$ this gives way to a set of classically-degenerate two-sublattice antiferromagnetic configurations.   
These are locked together, by thermal \cite{loison94} 
or quantum fluctuations, to form collinear ``stripe'' order [Fig.\ref{fig:BZ-stripes}(a)] \cite{jolicoeur90,chubukov92}.  
For $J_2$$>$1 the ground state is a one-dimensional coplanar spiral with wave number 
\mbox{$q= (2/\sqrt{3}) \cos^{-1} [-(J_1+J_2)/(2J_2)]$} [\linecite{jolicoeur90,loison94}].  
The inclusion of easy-axis anisotropy further stabilizes the collinear stripe state, and converts the 120\degree state 
and coplanar spirals into a three-sublattice coplanar fan\cite{miyashita86}, and a one-dimensional helicoidal phase, 
respectively [Fig. \ref{fig:mft}].

%%%%%%%%%%%%%%%%%%%%%%%%%%%
%%%%%%%%%%%%%%%%%%%%%%%%%%%%%%%

%%%%%%%%%%%%%%%

%%%%%%%%%%%%%%%%%t

\begin{figure}[ht!]
\includegraphics[width= 6cm]{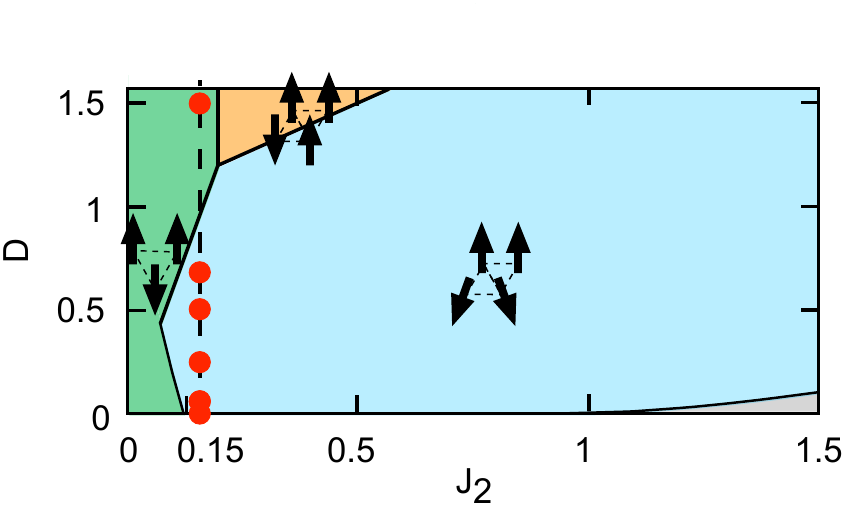}
\caption{\footnotesize{(color online) Finite-field $J_2$$-$$D$ phase diagram  for  $J_1$=1 and $J_\perp$=-0.15, showing the first mean-field instability with applied field. Three different states are found :  $m$=1/3 and $m$=$1/2$ magnetisation plateaux and magnetic supersolid state. Red dots along dashed vertical line show studied anisotropy values. }}
\label{fig:mft-field}
\end{figure}

Now let us consider how this zero-temperature phase diagram evolves with field.  
A reasonable expectation would be that, for sufficiently large field, the collinear stripe state [Fig.~\ref{fig:configs}(a)] 
would undergo a first-order ``spin-flop'' transition into a two-sublattice canted state [Fig.~\ref{fig:configs}(b)].
However this is not what happens.  
The first new phase encountered as magnetic field is increased, within mean field theory, is shown as a function 
of $J_2$ and $D$ in Fig.~\ref{fig:mft-field}.
At small $J_2$, we find the familiar collinear $m=1/3$ magnetisation plateau [Fig.~\ref{fig:BZ-stripes}(c)]
and for large $D$, moderate $J_2$,  a collinear $m=1/2$ magnetisation plateau [Fig.~\ref{fig:configs}(f)]; for large $J_2$ and small 
$D$ the helicoidal phase interpolates all the way to saturation.
However, away from the margins of the phase diagram, a novel four-sublattice, partially canted phase [Fig.~\ref{fig:configs}(c)] 
completely dominates, displacing the more conventional spin-flop phase.   
This partially canted state is a magnetic supersolid phase, according to the definition of 
Matsuda and Tsuneto~\cite{matsuda70} or Liu and Fisher~\cite{liu73}.
While the stripe state breaks only discrete lattice symmetries, and the canted spin-flop state only continuous 
spin-rotational ones, the partially canted supersolid breaks \emph{both}.
The reasons for the emergence of this unusual --- and very robust --- new state form the main focus of this work.   

%%%%%%%%%%%%%%%%%%%%%%%%%%%%%%%

 %%%%%%%%%%%%%%%%%%%%%%%%%%%
%%%%%%%%%%%%%%%%%%%%%%%%%%%%%%%

%%%%%%%%%%%%%%%%%%%%%%%%%%%%%%%

For the purposes of this paper we choose to fix the nearest-neighbour interaction $J_2$=$0.15$ and vary the 
anisotropy strength, selecting values along the vertical dashed line in Fig.~\ref{fig:mft}.
This line includes the parameters $J_2$=0.15$J_1$ and $D$=$2/3J_1$ relevant to the magnet AgNiO$_2$ 
[\linecite{wheeler09}] discussed above  
\footnote{Quantum corrections of order 1/S to linear spin-wave theory 
renormalise the anisotropy value to $D$=$3/2J_1$.}. 
Moreover, since the $J_2=0.15$ line lies close to several phase boundaries in $J_2$-$D$ space, it opens the 
possibility of field-driven competition between different phases.

%%%%%%%%%%%%%%%%%%%%%%%%%%

The main results of this paper are a set of finite-temperature, finite-field phase diagrams for Eq.~(\ref{eq:H})
obtained using classical Monte Carlo (MC) simulation. 
Because of the strong magnetic frustration and magnetic anisotropy, and 
a large number of first-order phase transitions, MC simulations of this model are very challenging. 
To overcome  severe slowing down in simulation dynamics we employ a parallel tempering MC scheme \cite{hukushima96}, 
which combines local Metropolis updates with parallel simulation of multiple replicas of the system at different temperatures.  
The exchange of replicas at different temperatures allows the system to avoid local free energy minima traps. 
To further reduce the autocorrelation between spin configurations, especially in the equilibration phase, every local-update sweep 
is followed by successive over-relaxation sweeps \cite{kanki05}.  
These are entirely deterministic, and comprise the reflection of each spin in turn in the plane formed 
by the $S^z$ easy axis and the local field from its neighbouring spins.
This over-relaxation scheme is a micro-canonical update, and reversible, and therefore the global Markov chain
for parallel tempering and over-relaxation will also obey detailed balance.

%%%%%%%%%%%%%%%%%%%%%%%%%%

%%%%%%%%%%          D=0 phase diagram              =====================]
%%%%%%%%%%%%%%%%%%

%%%%%%%%%%%%%%%%%%%

Simulations  of from 48 to 160 replicas were performed for  rhombohedral clusters of \mbox{$3L\times 3L\times L=9L^3$} 
spins with periodic boundary conditions, where $L$=4,6,8,10 is the  cluster thickness, i.e. counts the number of triangular 
lattice planes. 
This cluster geometry was chosen to reflect  the higher correlation length in the $\mathbf{\hat{x}}$$-$$\mathbf{\hat{y}}$ 
plane relative to the $\mathbf{\hat{z}}$ axis for $J_\perp/J_1$$\approx$0.1. 
Typical simulations involved 4$\times$10$^6$ steps, half of which were discarded for thermalization.   
Each step consisted of one local-update sweep of the lattice followed by 
two over-relaxation sweeps, with replicas at different temperatures exchanged every 10 steps. 
Typically, simulations were started from initial random configurations. 
Nevertheless, the competition between three and four-sublattice states leads to strongly first order transitions involving large 
internal energy discontinuities at low temperature, difficult to overcome even with parallel tempering.
In order to overcome this, the results in this region were obtained through careful comparison between several runs with 
different initial configurations. Some runs were initialised with fully ordered states and others with mixed phase configurations 
between both types of order.

%%%%%%%%%%%%%%%%%%%%%%%%%%

%%%%%%%%%%%%%%%%%%%%%%%%%%%%
These simulations are complemented by analytic calculations at low temperature. 
The $T$=0 ground state of the model can be calculated within mean-field theory for 2, 3 and 4 sublattice states. 
This provides a direct check on the $T$=0 limit of the phase transitions found in simulations.
The Gaussian fluctuations about these mean-field states are calculated using the low-temperature spin-wave technique outlined in Appendix A. 
We use this approach to calculate magnetisation curves at finite field, and to parametrize a Landau theory for 
continuous phase transitions. 
These results are compared with phase boundaries obtained from simulation in the phase diagrams below.

%%%%%%%%%%%%%%%%%%%%%%%%%%
%%%%%%%%%%%%%%%%%%%%%%%%%%
\section{Heisenberg limit, $D$=0}
\label{D=0}
%%%%%%%%%%%%%%%%%%%%%%%%%%
%%%%%%%%%%%%%%%%%%%%%%%%%%

The two natural limits of the Hamiltonian Eq.~(\ref{eq:H}) are the Heisenberg ($D$=0) and Ising~($D$=$\infty$) models. 
We consider a range of $D$ interpolating between these, beginning with a pure Heisenberg model.
For $D$$=$$0$, the phase diagram is dictated by the order-from-disorder selection of a small number of
magnetically ordered phases from a disordered manifold of competing ground states.  
In the absence of magnetic field, like previous authors~\cite{jolicoeur90,loison94}, we find that fluctuations favour 
a collinear antiferromagnetic ``stripe'' ground state.
Once magnetic field is applied, these stripes cant.
At a field of $h$=$4(J_1+J_2)=4.6$ $(T$$=$$0^+)$, there is a first order transition from the 2-sublattice canted state into a 
4-sublattice collinear half-magnetization plateau state.
As field is increased further, this plateau undergoes a continuous phase transition into a 3:1 canted state, 
which interpolates to saturation.
Finally the system saturates at a field of $h$$=$$9.2$ $(T$$=$$0)$.  
These results are summarised in Fig.~\ref{fig:D0-panel}, and described in more detail below.

%%%%%%%%%%%%%%%%%%%%%%%%%%%%%%%

 \begin{figure}[hbt]
\includegraphics[width=5.2cm]{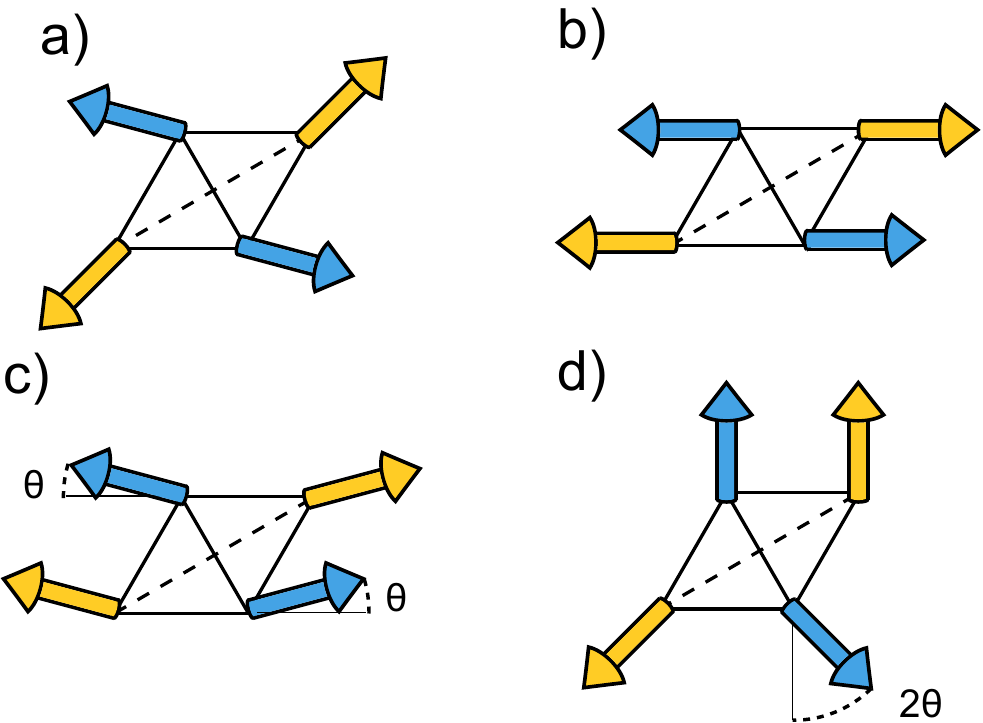}
\caption{\footnotesize{(color online)  
Structure of the ground state of the $J_1$--$J_2$ Heisenberg model on a triangular lattice
for $J_2/J_1 > 1/8$.      
(a) Degenerate manifold of states for $T$=0, $h$=0, built of two independent N\'eel sublattices.  
(b) Collinear antiferromagnetic stripe phase selected by thermal fluctuations ($h$=0).   
(c) ``Spin flop'' phase formed by direct canting of antiferromagnetic stripes in applied magnetic field.  
(d) Supersolid phase formed from the same two canted sublattices.
}}
\label{fig:twosublattice}
\end{figure}

%%%%%%%%%%%%%%%%%%%%%%%%%%%%%%%

\begin{figure}[hbt]
\includegraphics[width= 7.1087cm]{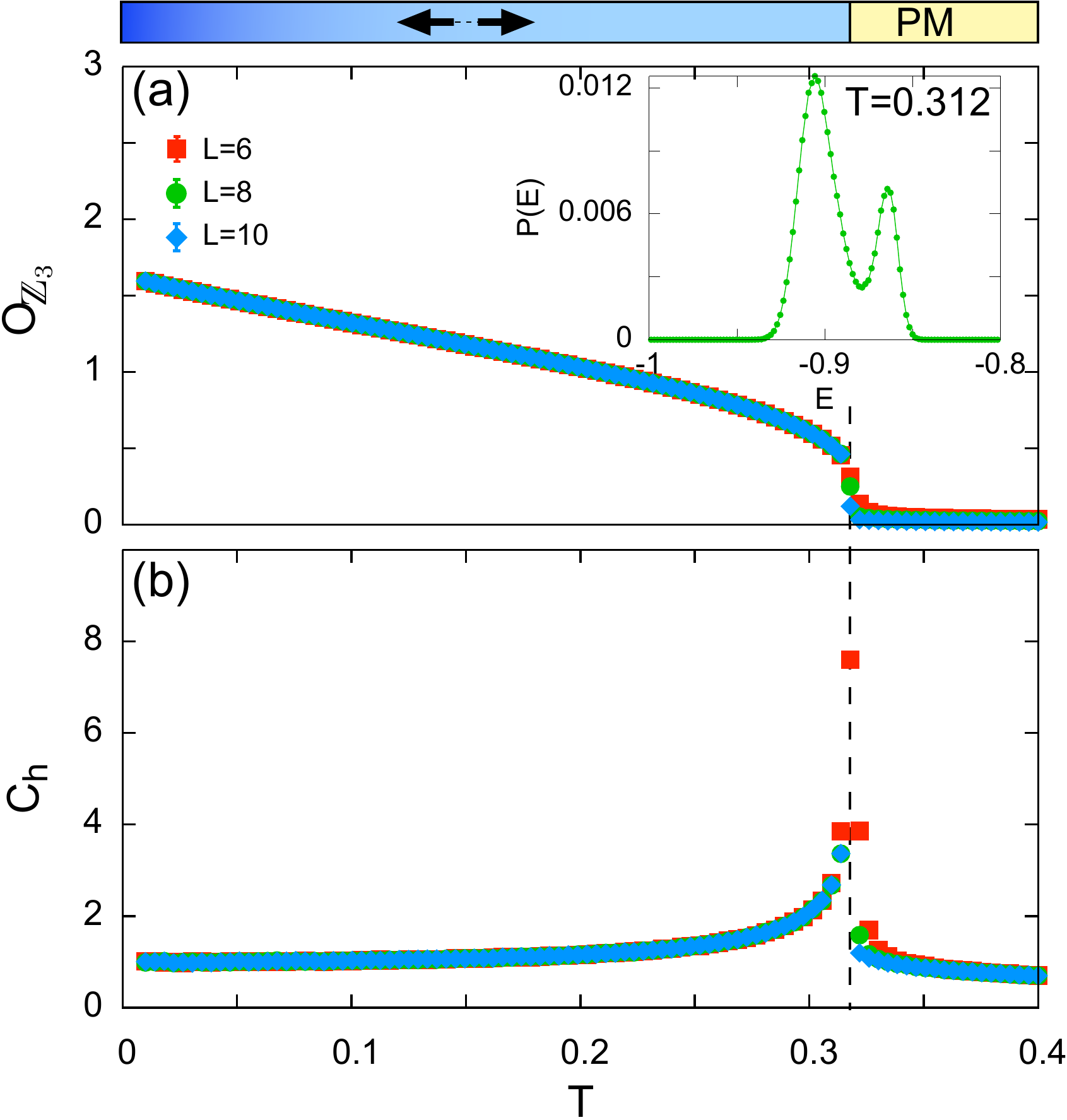}
\caption{\footnotesize{(color online)  Phase transition at \mbox{$D=0,h=0$} from the paramagnet  into the isotropic 
collinear antiferromagnetic stripe state. 
(a) Stripe order parameter from broken $\mathds{Z}_3$  lattice symmetry,  Eq.~(\ref{eq:Z3}), and (b) heat capacity. 
The first-order behaviour is clear in the bimodal distribution of the energy histograms close to the critical temperature, inset to (a). }}
\label{fig:D0-h0}
\end{figure}

In the absence of magnetic field ($h$$=$$0$) the zero-temperature mean-field ground state of Eq.~(\ref{eq:H}) is a manifold of states built of two decoupled N\'eel sublattices [Fig.~\ref{fig:twosublattice}(a)].  
Thermal fluctuations select the collinear stripe phase [Fig.~\ref{fig:twosublattice}(b)] which maximizes the entropy 
accessible to spin wave excitations.  
This stripe state breaks both $O(3)$ spin-rotation symmetry {\it and} the (discrete) 
rotation symmetry of the lattice --- the stripes can be orientated in three different ways [Fig.~\ref{fig:BZ-stripes}(a)].
To study the finite temperature transition into this state, we therefore introduce a complex order parameter 
\mbox{$\psi=\psi_{1}+ i \psi_{2}$} based on a two-dimensional irreducible representation of the $C_3  \cong \mathds{Z}_3$ lattice 
rotation group
\begin{align}
\psi_{1} &=\frac{1}{\sqrt{6}N}
\sum_i 2\mathbf{\mathbf{S}}_i.\mathbf{S}_{i+\delta_1}-\mathbf{S}_i.\mathbf{S}_{i+\delta_2}-\mathbf{S}_i.\mathbf{S}_{i-\delta_1-\delta_2}, \nonumber\\
\psi_{2} &=-\frac{1}{\sqrt{2}N}
\sum_i \mathbf{S}_i.\mathbf{S}_{i+\delta_2}-\mathbf{S}_i.\mathbf{S}_{i-\delta_1-\delta_2}, \nonumber 
\end{align}
Here the primitive vectors of the triangular lattice are  ${\bf \delta}_1 = (1,0)$, ${\bf \delta}_2 = (1/2,\sqrt{3}/2)$.
Since parallel tempering effectively restores this lattice symmetry, we measure the magnitude of the order parameter
\begin{align}
O_{\mathds{Z}_3}& =\Big<   \sqrt{ | \psi_1 |^2 +  | \psi_2 |^2 }\hspace{0.1cm} \Big>,
\label{eq:Z3}
\end{align}
which is normalised to $4/\sqrt{6}$ for perfect stripe order.
In  Fig.~\ref{fig:D0-h0} we present MC simulation results for the finite temperature transition from the paramagnetic 
into the collinear antiferromagnetic state.    
The transition is found to be first order, as expected for a 3-state Potts model in 3D (see e.g. \linecite{loison94}). 

%%%%%%%%%%%%%%%%%%

%%%%%%%%%%%%%%%%%%

Applying a magnetic field causes the spins in each N\'eel sublattice to cant.   
This can happen in two distinct ways.   
Both stripes can cant equally, to give the spin flop state shown in Fig.~\ref{fig:twosublattice}(c).
Alternatively, the sublattices can rotate so that half of the spins are aligned with the magnetic field, while
the others are canted away from it --- the supersolid state shown in Fig.~\ref{fig:twosublattice}(d).
These two states are degenerate, but thermal fluctuations favour the spin-flop state over the supersolid, 
since it has a higher density of spin wave excitations at low energies.

%%%%%%%%%%%%%%%%%%

In the spin-flop phase all spins have equal $S^z$ components, and the lattice symmetries are broken only 
by spin components in the $S^x$-$S^y$ plane.   
We therefore modify Eq.~(\ref{eq:Z3}) to read
\begin{align}
\psi_{1}^\perp &=\frac{1}{\sqrt{6}N}
\sum_i 2\mathbf{\mathbf{S}}^\perp_i.\mathbf{S}^\perp_{i+\delta_1}-\mathbf{S}^\perp_i.\mathbf{S}^\perp_{i+\delta_2}-\mathbf{S}^\perp_i.\mathbf{S}^\perp_{i-\delta_1-\delta_2}, \nonumber\\
\psi_{2}^\perp &=-\frac{1}{\sqrt{2}N}
\sum_i \mathbf{S}^\perp_i.\mathbf{S}^\perp_{i+\delta_2}-\mathbf{S}^\perp_i.\mathbf{S}^\perp_{i-\delta_1-\delta_2}, \nonumber 
\end{align}
where $\mathbf{S}_i^\perp=(S_i^x,S_i^y)$, and consider the order parameter
\begin{align}
O_{\mathds{Z}_3}^\perp& =\Big<   \sqrt{ | \psi_1^\perp  |^2 +  | \psi_2^\perp  |^2  } \hspace{1mm} \Big>.
\label{eq:Z3-Sperp}
\end{align}
The spin-flop phase also separately breaks spin-rotation symmetry in the $S^x$-$S^y$ plane.   
This can be determined using an order parameter for the staggered in-plane magnetization
\begin{align}
O_{U(1)}=\frac{1}{N}\sum^{N/4}_{\lozenge} | \mathbf{S}_B^\perp - \mathbf{S}_A^\perp |,
\label{eq:U1-op}
\end{align}
where $A$ and $B$ label the two sublattices of the spin-flop state, as illustrated in Fig.~\ref{fig:configs} (b).
The breaking of spin-rotation symmetry implies the existence of a finite spin stiffness  $\rho_S$, 
as defined in Appendix~\ref{spinstiffness}.

%%%%%%%%%%%%%%%%%%%%%%%%%%%%%%%%%%%%%%%%%%%%%%%%%%

\begin{figure}[h!]
\includegraphics[width= 8cm]{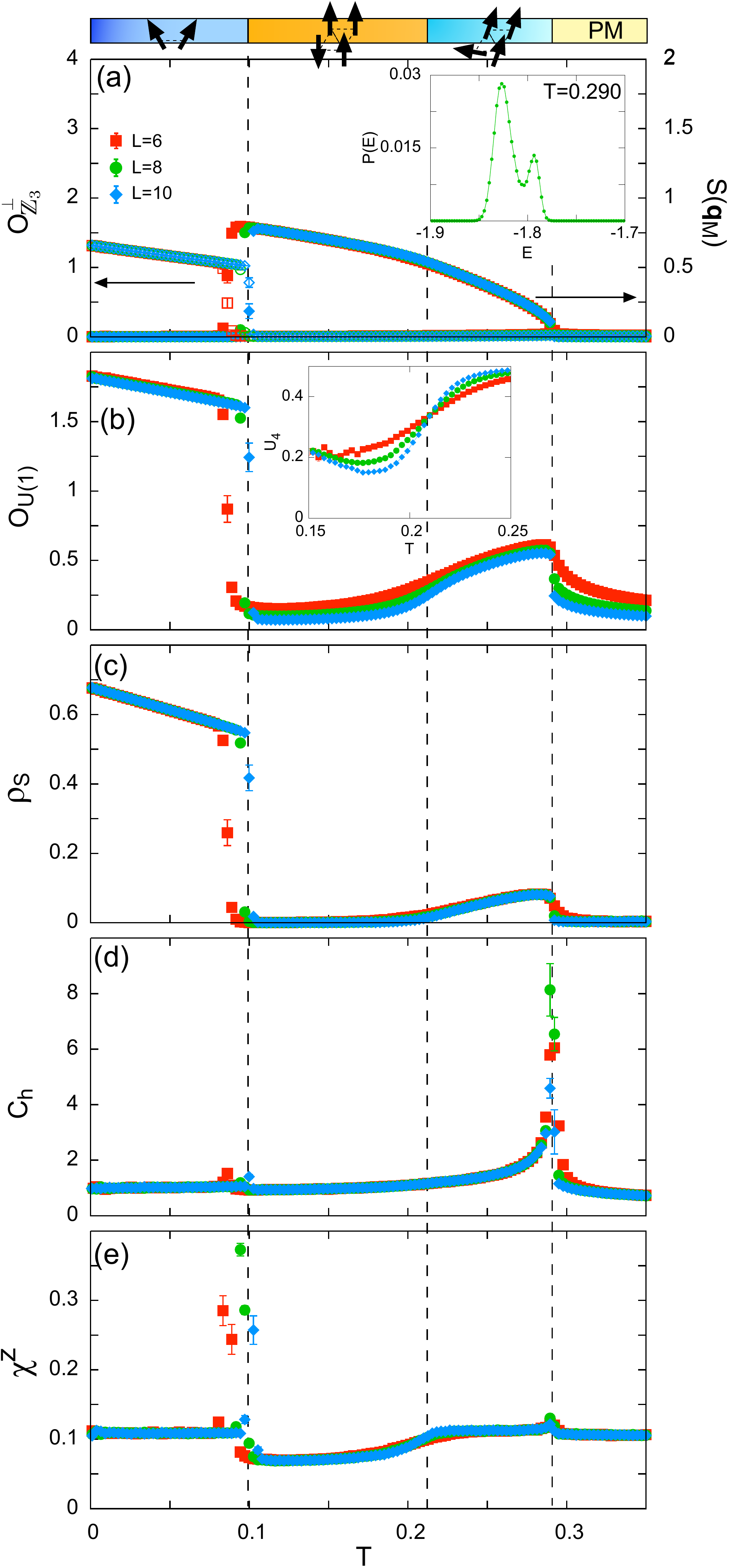}
\caption{\footnotesize{(color online) 
Selection of simulation results for $D=0$, $h=4.0$,  spanning the canted ``spin-flop'' phase, collinear $m$=1/2 
plateau, 3:1 canted state, and high temperature paramagnet.
(a) $\mathds{Z}_3$ order parameter $O_{\mathds{Z}_3}$ associated with broken lattice rotation symmetry
in the spin flop state [Eq.~(\ref{eq:Z3})], and spin structure factor ${\mathcal S}^{zz}(\mathbf{q}_{\sf M})$ associated with 
broken translational symmetry in the 3:1 canted and collinear $m$=1/2 
plateau states [Eq.~(\ref{eq:SqM})];  
(b) in-plane staggered magnetization $O_{U(1)}$ [Eq.~(\ref{eq:U1-op})];  
(c) spin stiffness $\rho_S$;  
(d) heat capacity $C_h$;  
(e) magnetic susceptibility $\chi$.  
The inset to (a) shows the energy distribution at the transition from paramagnet to 3:1 canted phase.
The inset to (b) shows the crossing of the Binder cumulants associated with $O_{U(1)}$ at the 
transition from the 3:1 canted phase to the collinear $m$=1/2 plateau state.
}}
\label{fig:D0-h4}
\end{figure}

%%%%%%%%%%%%%%%%%%%%%%%%%%%  

At higher values of magnetic field, thermal fluctuations select another new ordered state from a large set of competing alternatives.
This is a collinear state with a 4-site unit cell in which three spins point ``up'' and one ``down'' --- commonly referred to as 
a half-magnetization ($m$=1/2) plateau [Fig.~\ref{fig:configs}(f)].  
In the limit of vanishing temperature, this first occurs at a magnetic field value of $h$=$4(J_1+J_2)$=4.6.
This collinear phase breaks translational symmetries, with associated ordering vectors 
\mbox{$\{ \mathbf{q}_{\sf M} \}$}, cf. Fig.~\ref{fig:BZ-stripes}.
The $S^z$--$S^z$ component of the static spin structure factor at these wave vectors 
\begin{align}
{\mathcal S}^{zz} (\mathbf{q}_{\sf M} )
= \Big \langle\sum_{\{ q_M\} } \Big|   \frac{1}{N} \sum_i S_i^z
\textrm{e}^{-i \textbf{q}_{\sf M} \cdotp \textbf{r}_i} \Big| ^2  \Big\rangle,
\label{eq:SqM}
\end{align}
acts as (the square of) an order parameter for this state.
Since the state is collinear, and aligned with the magnetic field, its spin stiffness $\rho_S \equiv 0$.   
We note that, since the ``down'' spin is unique within the 4-site unit cell, this collinear state 
also breaks the permutation symmetries of the bonds within the unit cell.
It is therefore possible to define a complementary $\mathds{Z}_4$ order parameter 
for this state, based on the irreps of the relevant permutation group.
For present purposes, however, the combination of a vanishing $\rho_S$ and a finite 
${\mathcal S}^{zz}(\mathbf{q}_{\sf M})$ are sufficient to identify the state.

%%%%%%%%%%%%%%%%%%%%%%%%%%%%

The catalogue of ordered phases for $D$$=$$0$ is completed by a 3:1 canted phase [Fig.~\ref{fig:configs}(g)], 
which is connected to collinear $m$=1/2 plateau by a continuous phase transition as a function of magnetic field.
All spin wave modes are gapped in the collinear plateau state.
With increasing magnetic field, the gap associated with one of the zone-centre spin waves closes.
Since this spin wave is a transverse excitation, it converts the collinear state into one with a staggered
magnetization in the $S^x$--$S^y$ plane, and a finite spin stiffness.
And, since, it is a zone-centre mode, the 3:1 canted phase inherits the broken translational symmetry 
of the $m$=1/2 plateau.  
The resulting state is therefore a supersolid, in the sense of Matsuda and Tstuneto or Liu and Fisher, breaking 
both translational and spin-rotational symmetries.

%%%%%%%%%%%%%%%%%%%%%%%%%%

In Fig.~\ref{fig:D0-h4} we show a selection of finite temperature simulation results for $h$=4.0, a cut 
through the $T$--$h$ plane which spans all three ordered states.   
The transition from the high temperature paramagnet to the 3:1 canted phase at $T$=0.290(3) 
can be seen in the temperature dependence of ${\mathcal S}^{zz}(\mathbf{q}_{\sf M})$ [Fig.~\ref{fig:D0-h4}(a)], 
staggered magnetization $O_{U(1)}$ [Fig.~\ref{fig:D0-h4}(b)], spin stiffness $\rho_S$  [Fig.~\ref{fig:D0-h4}(c)], 
and a sharp feature in $C_h$  [Fig.~\ref{fig:D0-h4}(d)].  
While symmetry permits a second-order phase transition, the phase transition observed in simulation
for this value of $h$ is first-order, as shown by the energy histograms in the inset to  Fig.~\ref{fig:D0-h4}(a).  
The transition from 3:1 canted phase to $m$=1/2 plateau is distinguished by a collapse of the spin 
stiffness  [Fig.~\ref{fig:D0-h4}(c)], $O_{U(1)}$  [Fig.~\ref{fig:D0-h4}(b)]
and a suppression of the magnetic susceptibility $\chi$  [Fig.~\ref{fig:D0-h4}(e)].
The transition is continuous, with the Binder cumulant for the two-component $U(1)$ order parameter 
$U_4=1-\frac{\langle O_{U(1)}^4 \rangle }{ 2\langle O_{U(1)}^2 \rangle^2 }$ crossing at a single temperature 
$T$=0.210(3) for a range of system sizes [inset to Fig.~\ref{fig:D0-h4}(b)]. 
A strong finite size dependence of this Binder cumulant is observed for all observed continuous phase transitions. 
These finite-size corrections are especially large in ordered phases (e.g. plateau) where the relevant $U(1)$ order parameter vanishes in the thermodynamic limit.  

%%%%%%%%%%%%%%%%%%%%%%%%%%%
\begin{figure*}[ht!]
\includegraphics[width= 17cm]{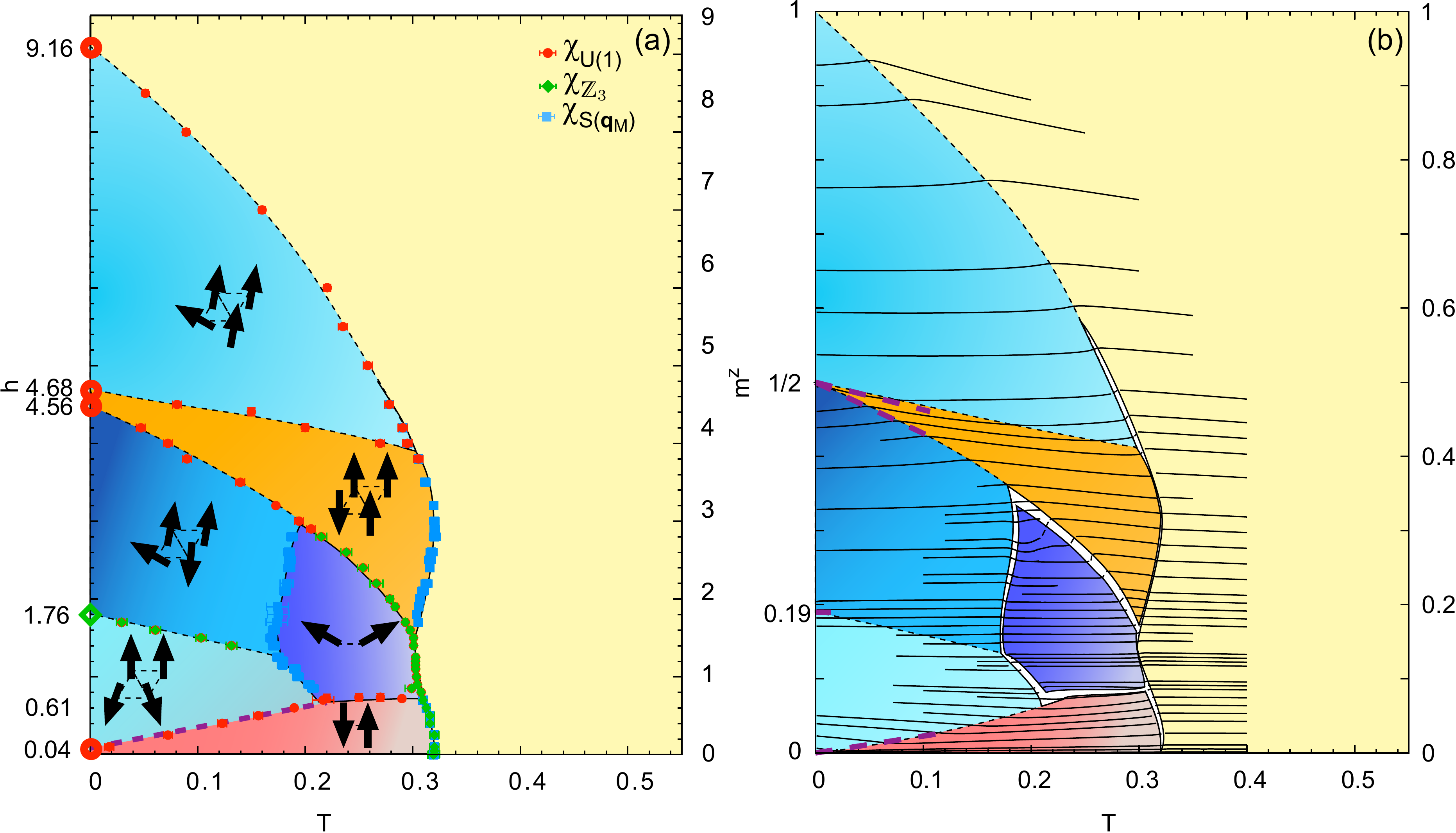}
\caption{\footnotesize{(color online) 
Magnetic phases of a layered triangular-lattice antiferromagnet with exchange interactions $J_1$=$1$, $J_2$=$0.15$, 
$J_\perp$=$-0.15$ and \mbox{easy-axis anisotropy $D$=0.02}.
(a)  Phase diagram as a function of temperature and magnetic field.
Open symbols on the h-axis show transitions obtained in mean-field theory.   
Phase boundaries at finite temperature are obtained from Monte Carlo simulation for a cluster of $24$$ \times$$ 24$$ \times 8$ spins, and determined by peaks in the relevant order parameter susceptibilities.  
All phase transitions are first-order, except where shown with a dashed line.    
Thick purple dashed line is obtained through a Landau expansion for the supersolid transition. 
(b)  Phase diagram as a function of temperature and magnetization.
Solid lines running left-right show cuts at constant magnetic field $h$ taken from simulations. 
Thick purple dashed lines show phase boundaries  obtained through a  low-T expansion.  
Finite anisotropy lifts $T$=0 degeneracy in favour of the stripe and supersolid phases, relegating the spin-flop to finite-temperature only. }}
\label{fig:D002-panel}
\end{figure*}

%%%%%%%%%%%%%%%%%%%%%%%%%%

Finally, the abrupt, first-order transition from the $m$=1/2 plateau to the spin flop phase at $T$=0.100(3) is visible 
in the sudden onset of $O_{\mathds{Z}_3}$  [Fig.~\ref{fig:D0-h4}(a)], $O_{U(1)}$  [Fig.~\ref{fig:D0-h4}(b)] and 
spin stiffness  [Fig.~\ref{fig:D0-h4}(c)].
Here a continuous phase transition can be ruled out on symmetry grounds.

%%%%%%%%%%%%%%%%%%%%%%%%%%

The magnetic phase diagram can also be plotted as a function of temperature and magnetisation [Fig.~\ref{fig:D0-panel}(b)]. 
In this case a discontinuity in the magnetisation as a function of temperature or magnetic field 
implies a first-order phase transition, and leads to regions of phase coexistence in the $T$--$m$ plane, 
coloured white in Fig.~\ref{fig:D0-panel}(b).
The transition from the paramagnet to the $m$=1/2 plateau is thus seen to be first order, although with a very 
small magnetisation jump. 
The phase transition from the spin-flop phase to the paramagnet is also first order for $h$$>$0.

The transition between the $m$=1/2 plateau and the 3:1 canted state is continuous, as is the finite-temperature
transition from the 3:1 canted state to the paramagnet, for $h\gtrsim$5 ($m\gtrsim$0.5).
The 3:1 canted phase finally saturates at zero temperature, at a field $h_{\sf SAT}=9.2$.

%%%%%%%%%%%%%%%%%%%%%%%%%%%%%

It is interesting to note that the same succession of phases in magnetic field occurs in a model 
of spin-lattice coupling in Cr spinels~\cite{penc04,motome06,shannon10}. 
In that case the four-site cell is  the tetrahedron from the which the pyrochlore lattice
is built, and the role of thermal fluctuations is played by a biquadratic spin interaction. 

%%%%%%%%%%%%%%%%%%%%%%%%%%%%%%
%%%%%%%%%%%%%%%%%%%%%%%%%%%%%
\section{Small anisotropy, $D$=0.02}
\label{D=0.02}
%%%%%%%%%%%%%%%%%%%%%%%%%%%%%%%
%%%%%%%%%%%%%%%%%%%%%%%%%%%%%%%%

The phase diagram for $D$=0.02, presented in Fig.~\ref{fig:D002-panel}, is extremely rich.   
It exhibits all but one of the phases catalogued in Fig.~\ref{fig:configs} --- a collinear stripe phase 
at low values of field, giving way to either a supersolid or a spin-flop phase as magnetic
field is increased; a novel 2:1:1 canted state, a collinear $m$=1/2 plateau state and 
finally, approaching saturation, a 3:1 canted phase.
The key to understanding the richness of this magnetic phase diagram is to recognise that 
none of these phases found for $D=0$ are selected by energetic considerations alone 
--- all of them owe their stability to thermal fluctuations.
The difference in entropy between these phases and the other degenerate alternatives is, however, very small.
For this reason the introduction of even a vanishingly small easy-axis anisotropy has 
profound consequences for the phase diagram, especially at low temperatures.  

%%%%%%%%%%%%%%%%%%%%%%%%%%

To explore how this works, we first consider the $T$=0, mean-field energies per spin of the two competing 
states at low magnetic field, the canted spin flop state $({\sf SPF})$ and the partially canted magnetic supersolid $({\sf SSD})$  
\begin{align}
 \label{eq:SF-energy}
E_{\sf SPF}&=-J_1 -J_2 + J_{\perp} -\frac{h^2}{16(J_1+J_2)-4D},\\
\label{eq:SS-energy} 
E_{\sf SSD}&=\hspace{-0.5mm}\frac{\hspace{-0.5mm}-\hspace{-0.5mm}16(J_1\hspace{-0.5mm}+\hspace{-0.5mm}J_2) (J_1+J_2\hspace{-0.5mm}\hspace{-0.5mm}-\hspace{-0.5mm}J_\perp)\hspace{-0.5mm}- \hspace{-0.5mm}8D(J_1\hspace{-0.5mm}+\hspace{-0.5mm}J_2 \hspace{-0.5mm}+\hspace{-0.5mm} J_\perp)\hspace{-0.5mm}+\hspace{-0.5mm}4D^2}{16(J_1+J_2)-8D},\nonumber\\
&-\frac{h^2+4Dh}{16(J_1+J_2)-8D}. 
\end{align}
Both of these state are built of the same, canted N\'eel sublattices, but with different relative orientations, 
as illustrated  in Fig.~\ref{fig:twosublattice} (c) and (d).  
Their energies should be compared with the (field-independent) energy of the collinear stripe state $({\sf STR})$
\begin{align}
E_{\sf STR}&=-J_1 -J_2 + J_{\perp} -D \label{eq:UD-energy}.
\end{align}

%%%%%%%%%%%%%%%%%%%%%%%%%%

In the absence of anisotropy (i.e. for $D$=0), the spin flop and supersolid states are degenerate.   
It is thermal fluctuations which select both the collinear stripe state in zero field, and the canted 
spin flop state for $h$$>$0.
Introducing a finite anisotropy $D$ singles out the stripe phase as the ground state, and lifts the degeneracy 
between the canted spin flop and supersolid states.
However, while thermal fluctuations favour the spin flop state, anisotropy favours the supersolid, since
the spins lie closer to the easy axis.
This sets up an interesting tension between energy and entropy which, for very small values of $D$, 
compete on an equal footing.

%%%%%%%%%%%%%%%%%%%%%%%%%%%
 
\begin{figure}[h!]
\includegraphics[width= 7.5cm]{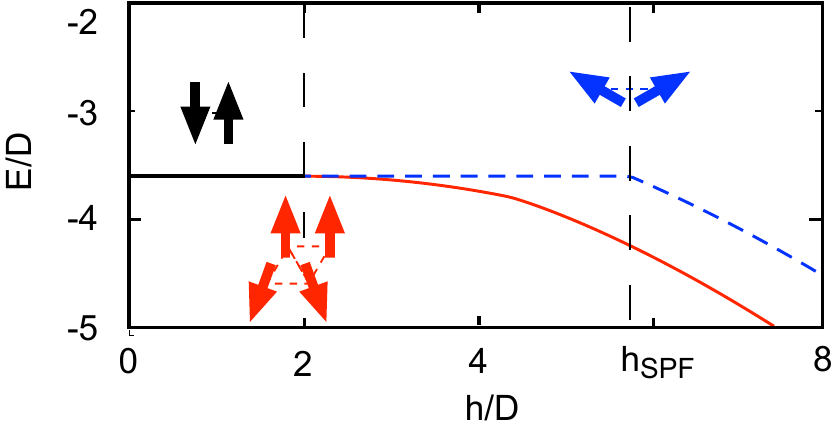}
\caption{\footnotesize{(color online)  Comparison of zero-temperature mean field energies for the
canted spin flop (blue, dashed line) and magnetic supersolid (red solid line) states for parameters 
$J_2$=0.15  and \mbox{$J_\perp$=-0.15} as a function of magnetic field $h$.
A continuous phase transition into the supersolid state occurs at $h_{\sf SSD}$=2D, anticipating
the first order spin flop transition which would otherwise have occurred at $h_{\sf SPF}$, Eq.~(\ref{eq:transition-fields-sf}).}}
\label{fig:ss-U-mft}
\end{figure}

%%%%%%%%%%%%%%%%%%%%%%%%%%%%%%%

%%%%%%%%%%%%%%%%%%%%%%%%%%

At zero temperature phase transitions are controlled by energy alone.   The first transition as a function
of magnetic field is into the supersolid state, at a critical field of 
\begin{align}
\label{eq:transition-fields-ss}
h_{\sf SSD}&=2D,
\end{align}
In contrast, the field at which the spin flop state first becomes energetically favourable is
\begin{align}
\label{eq:transition-fields-sf}
h_{\sf SPF}&=4\sqrt{D(J_1+J_2-D/4)}.
\end{align}
Moreover, the transition into the supersolid is continuous, whereas the (avoided) spin flop 
transition is first order.
This hierarchy of transitions is illustrated in Fig.~\ref{fig:ss-U-mft}.

\begin{figure}[ht]
\includegraphics[width= 7.5cm]{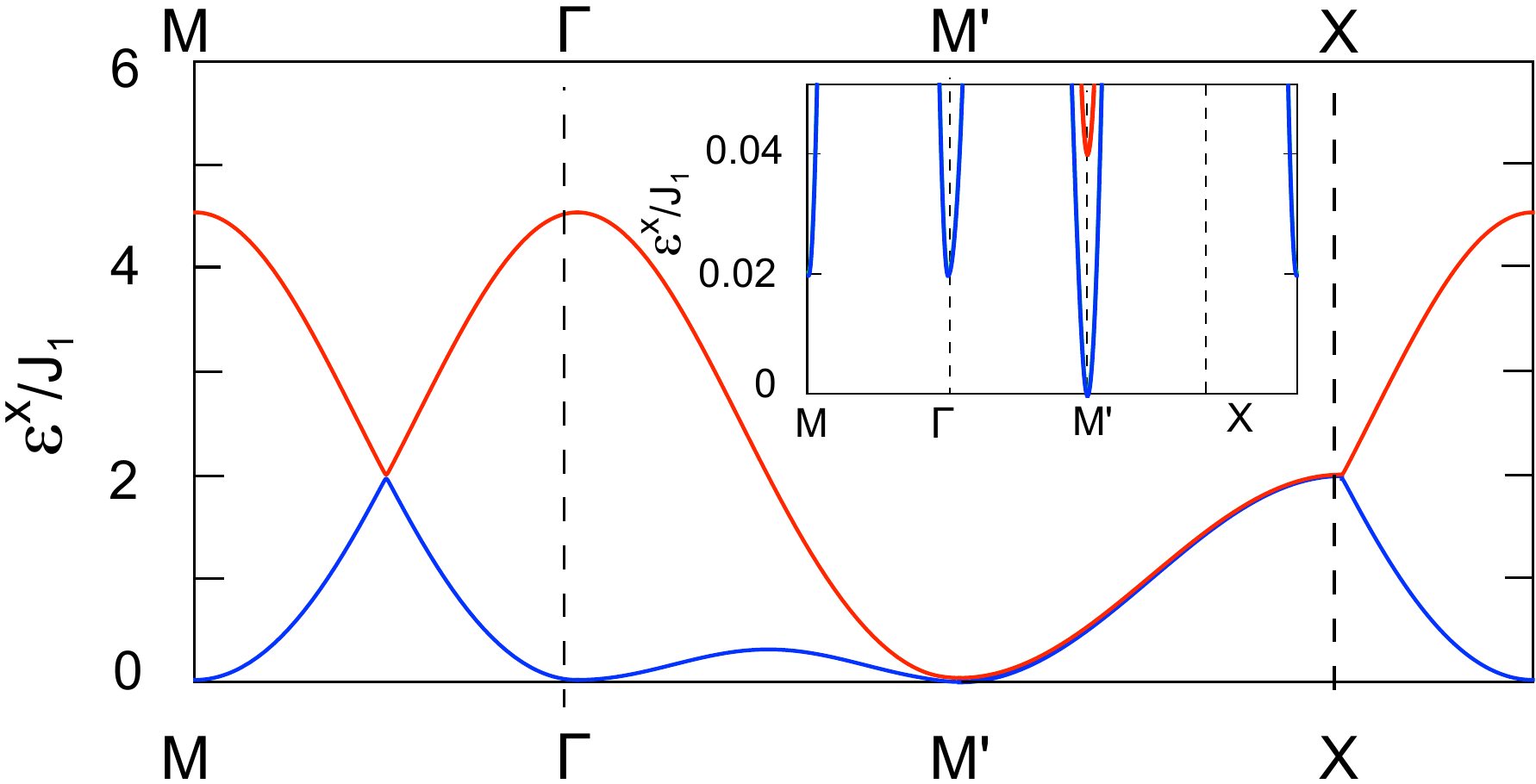}
\caption{\footnotesize{(color online) Classical spin-wave dispersion ($S^x$ modes only) for $J_2$=0.15, $J_\perp$=-0.15 and $D=0.02$, at a magnetic field $h$=2$D$ associated with continuous transition from collinear stripe phase into supersolid through closing of the spin gap at $M'$. Inset shows low-energy detail.}}
\label{fig:D002-CSW}
\end{figure}

The seeds of this supersolid state can also be found in a spin-wave expansion around the collinear state for $D$$>$0. 
At zero field this yields a gapped dispersion with minima at \emph{both} the ordering momentum vector $M$ and 
at the symmetry related momentum $M'$ [Fig.\ref{fig:BZ-stripes}(b)]. 
This degeneracy is lifted by an applied field in favour of $M'$, with the spin gaps evolving as
\begin{align}
\Delta_M&=D+2(J_1+J_2)-\frac{1}{2}\sqrt{16(J_1+J_2)^2+ h^2}, \label{eq:gap-M}\\
\Delta_{M'}&=D-\frac{h}{2}.
\label{eq:gaps}
\end{align}
The gap at $M'$ closes at a field $h_{\sf SSD}$=$2D$, while the gap at $M$ would, hypothetically, close at $h_{\sf SPF}$ [Eq.~(\ref{eq:transition-fields-sf})].  
It is therefore the closing of the gap at $M'$ which mediates the phase transition.
And, since this spin wave is a transverse mode with finite momentum, it converts the collinear stripe phase into a supersolid state 
with a finite (staggered) magnetization in the $S^x$--$S^y$ plane.
The spin-wave dispersion at $h=h_{\sf SSD}$, when the gap first closes, is illustrated in Fig.~\ref{fig:D002-CSW}.
An exactly analogous phase transition occurs in the quantum model studied with (linear) spin-wave analysis, where it is 
interpreted as Bose-Einstein condensation of magnons \cite{seabra10}, to give a condensate which  
breaks the transverse $U(1)$ symmetry. 
Similar mode-softening transitions have been observed in other models of frustrated triangular lattices \cite{haraldsen09}. 

%%%%%%%%%%%%%%%%%%%%%%%%%%%%%%%

%%%%%%%%%%%%%%%%%%%%%%%%%%%%%%%%

At finite temperature entropy again enters into the argument.   
Entropy favours both the collinear stripe state and the canted spin flop state over the supersolid.
This has two consequences.   
At low temperatures, it leads to a critical field $h_{\sf SSD}(T)$ which increases with temperature.
And, at higher temperatures, for sufficiently small $D$, entropy can drive a first order transition from
the supersolid into the spin flop state.
We can describe the first of these two effects quantitatively within a Landau theory.

%%%%%%%%%%%%%%%%%%%%%%%%%%
\begin{figure*}[t!]
\includegraphics[width= 17cm]{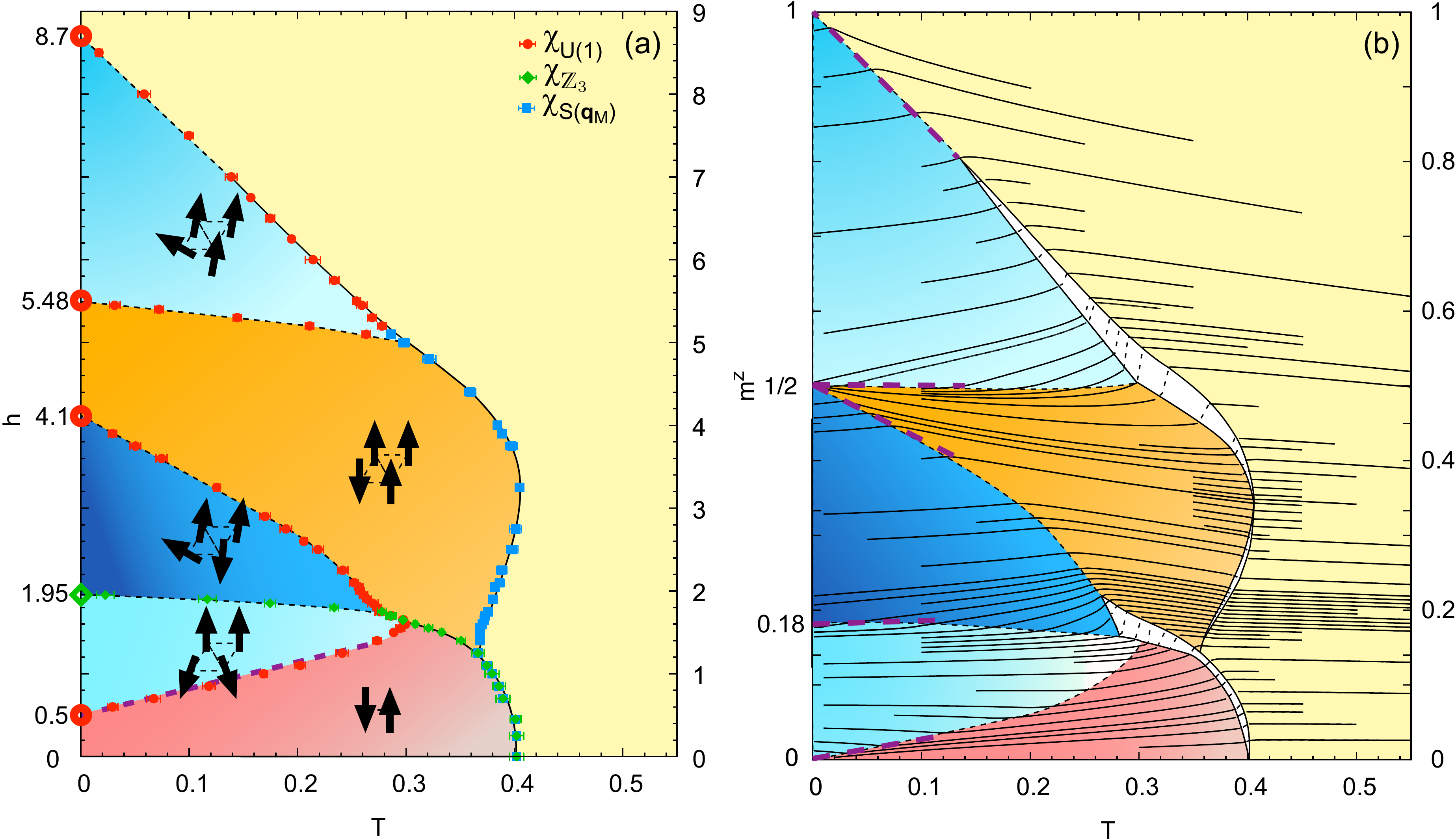}
\caption{\footnotesize{(color online) 
Magnetic phases of a layered triangular-lattice antiferromagnet with exchange interactions  
$J_1$=$1$, $J_2$=$0.15$, $J_\perp$=$-0.15$ and \mbox{easy-axis anisotropy $D$=$0.25$}.
(a)  Phase diagram as a function of temperature and magnetic field.   
Open symbols on the h-axis show transitions obtained in mean-field theory.   
Phase boundaries at finite temperature are obtained from Monte Carlo simulation for a cluster of 
$24$$ \times$$ 24$$ \times 8$ spins, and determined by peaks in the relevant order parameter susceptibilities.  
All phase transitions are first-order, except where shown with a dashed line.  
Thick purple dashed line is obtained through a Landau expansion for the supersolid transition. 
(b) Phase diagram as a function of temperature and magnetization.   
Solid lines running left-right show cuts at constant magnetic field $h$ taken from simulations.    
The coexistence regions associated with first order phase transitions are coloured white.  
Thick purple dashed lines show phase boundaries  obtained through a  low-T expansion.  
Easy-axis anisotropy stabilizes the new phases for a wide-range of values. }}
\label{fig:D025-panel}
\end{figure*}

%%%%%%%%%%%%%%%%%%

In analogy with Eq.~(\ref{eq:U1-op}), we define an order parameter
\begin{align}
O_{U(1)}=\frac{1}{N}\sum^{N/4}_{\lozenge} | \mathbf{S}_C^\perp - \mathbf{S}_D^\perp |,
\end{align}
where $C$ and $D$ label the two canted sublattices of the supersolid state, as illustrated in Fig.~\ref{fig:configs}(c).
The Landau expansion of the free energy can then be written as
\begin{align}
\mathcal{F} =  \mathcal{F}_0 + \frac{a}{2}|O_{U(1)}|^2 + \frac{b}{4}|O_{U(1)}|^4 + ...  \hspace{0.5cm},
\end{align}
where the coefficient $a =h_{\sf SSD}(T)-h$ is calculated {\it exactly} within a spin-wave expansion about the supersolid state.
Details of this are given in Appendix \ref{appendix}. 
For $D$=0.02 we find a critical field of
\begin{align}
h_{\sf SSD}(T)=0.04+3.02(4)\times T.
\end{align}

The increase of the critical field with temperature is a direct consequence of the fact that 
the supersolid state has a lower entropy than the collinear stripe phase. 
The decrease in entropy on entering the supersolid state can be traced back to a 
shift of spectral weight to higher energies in the spin-wave spectrum.

%%%%%%%%%%%%%%%%%%%%%%%%%%%%%%%%

We can also independently estimate $h_{\sf SSD}(T)$ from the peak in the related order parameter
susceptibility 
\begin{align}
\chi_{O_{U(1)}} = N \frac{\langle  O_{U(1)}^2  \rangle - \langle O_{U(1)} \rangle^2}{T},
\end{align}
which we calculate from Monte Carlo simulation.
For $T$$\lesssim$0.2 the analytic and numerical calculations are in perfect agreement, as shown by the 
purple dashed line in Fig.~\ref{fig:D002-panel}(a). 
However for $T$$\gtrsim$0.2 the continuous phase transition into the supersolid phase
is replaced by a first-order transition into the entropically favoured, spin flop state.  

%%%%%%%%%%%%%%%%%%%%%%%%%%%%%%%%

As a direct result, there is now a finite temperature transition from the supersolid 
into the spin flop state for $T$$\approx$0.2.   
As might be expected, this phase transition is strongly first order, and exhibits marked
hysteresis.
The character of this phase transition is most clearly seen in the coexistence region for 
$T$$\approx$0.2, $m$$\approx$0.1 in Fig.~\ref{fig:D002-panel}(b).
In conclusion, although the spin flop state is entirely eliminated at low temperatures by the introduction
of anisotropy, its greater entropy enables it to survive at finite temperatures --- for sufficiently small values of $D$.  
From simulations, we estimate that the spin flop phase survives at finite temperature 
for $D\lesssim0.045$.
The great wealth of other phases shown in Fig.~\ref{fig:D002-panel} are also present at higher values 
of $D$, where simulations are easier to perform.  
We therefore defer the detailed analysis of these phases to the remaining sections of this paper.

%%%%%%%%%%%%%%%%%%%%%%%%%%%%%%
%%%%%%%%%%%%%%%%%%%%%%%%%%%%
\section{Moderate anisotropy, $D$=0.25}
\label{D=0.25}
%%%%%%%%%%%%%%%%%%%%%%%%%%%%
%%%%%%%%%%%%%%%%%%%%%%%%%%%%%%

%%%%%%%%%%%%%%%%%%%%%%%%%%%

\begin{figure}[ht!]
\includegraphics[width= 8.0cm]{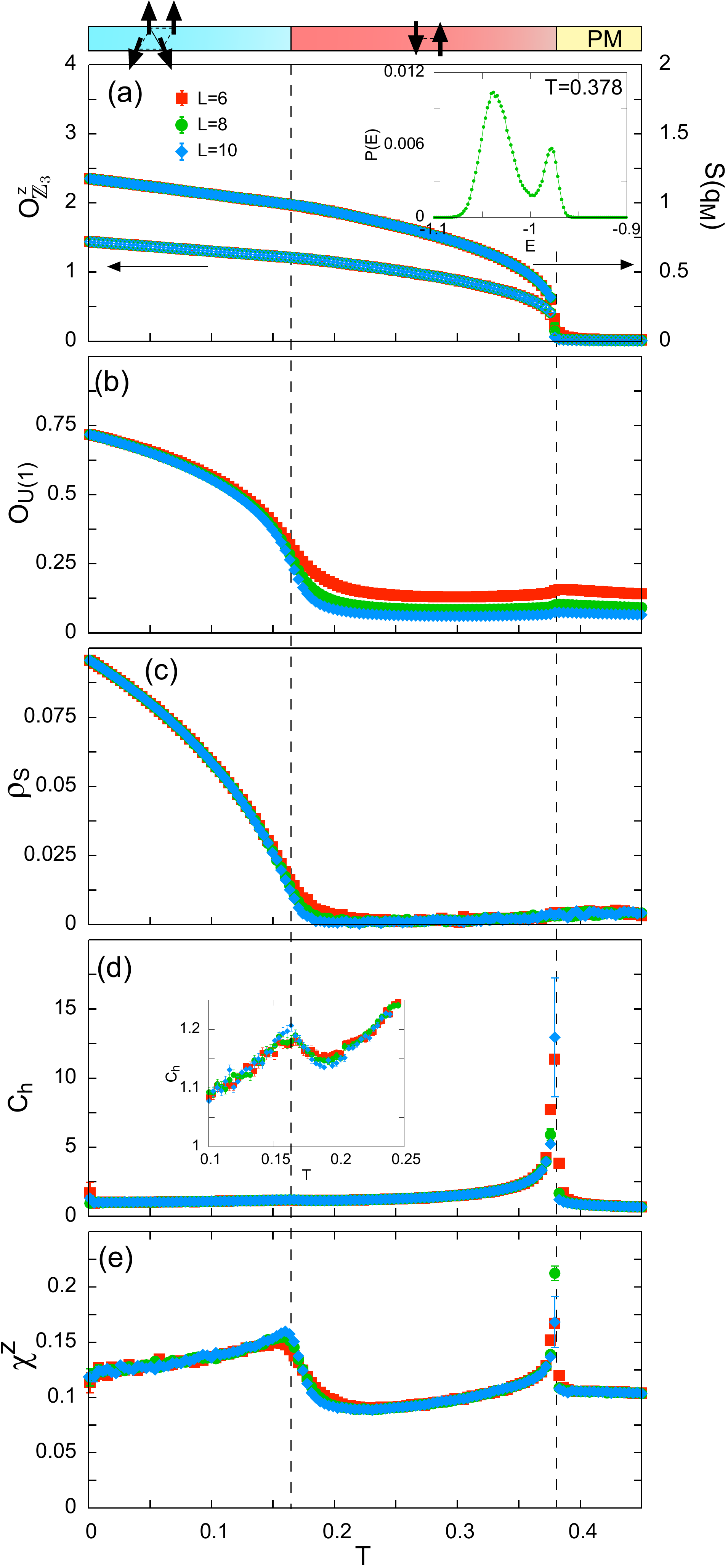}
\caption{\footnotesize{(color online) Double transition from the paramagnet into the stripe collinear antiferromagnet and then into the supersolid phase for $D=0.25,h=1.0$.  Translational $\mathcal{S}^{zz}(\mathbf{q}_{\sf M})$  and rotational $\mathds{Z}_3$ [Eq.~\ref{eq:Z3-Sparallel}] symmetries are broken at the same temperature $T$$\approx$$0.37$ (a), where system transitions into the collinear phase. The first-order character of the outer transition is confirmed by the double peak in energy histograms near the critical temperature in the inset to (a). A further continuous symmetry is broken at $T\approx0.16$ at the onset of the supersolid phase, as observed in the $U(1)$ order parameter (b)  and spin stiffness (c).  The different nature of these transitions is clear in heat capacity (d) and magnetic susceptibility (e).   }}
\label{fig:D025-h1}
\end{figure}

%%%%%%%%%%%%%%%%%%%%%%%%%%%

The  $D$=0 results showcase the canonical features of frustrated magnets in applied magnetic field --- see e.g.
\linecitedouble{kawamura85}{chubukov91} --- a collinear plateau state is stabilized at intermediate fields by thermal fluctuations, 
while coplanar canted phases are found at both lower and higher fields. 
However, in this model even a very small easy-anisotropy  leads to strikingly different results, as 
demonstrated for $D$=0.02.  
We now analyse thoroughly the phases driven by anisotropy for a representative, moderately small value of anisotropy $D$=0.25.
Our results are summarized in the phase diagrams shown in Fig.~\ref{fig:D025-panel}.   
For this value of $D$, the spin flop phase is entirely suppressed, and the low temperature 
physics is dictated by considerations of energy, rather than entropy.
As a function of increasing magnetic field, we find a collinear stripe state, the supersolid phase described above, an exotic 
2:1:1 canted state, a collinear $m$=1/2 plateau and finally a 3:1 canted state which interpolates to saturation.  
This large set of phases gives rise to a correspondingly large number of phase transitions, which we analyze below.

%%%%%%%%%%%%%%%%%%
 
We begin by considering the finite-temperature phase transition from the paramagnet into 
the collinear stripe state at low values of magnetic field, and the subsequent low 
temperature transition from the stripe phase into the supersolid. 
Both of these transition may have been observed in AgNiO$_2$ [\linecite{wawrzynska07}].   
In Fig.~\ref{fig:D025-h1} we show a selection of simulation results for $h=1.0$, and a range of temperatures 
spanning all three phases.  
The transition from paramagnet to collinear stripe phase at $T$=0.379(4) is evident in sharp peaks
in both the heat capacity [Fig.~\ref{fig:D025-h1}(d)] and the magnetic susceptibility 
[Fig.~\ref{fig:D025-h1}(e)].  
This transition is first order, as evidenced by a bimodal energy histogram 
[inset to Fig.~\ref{fig:D025-h1}(a)].  
The collinear stripe phase breaks both the translational and the rotational 
symmetry of the lattice.
Both of these symmetries are broken at the same temperature, 
as demonstrated in Fig.~\ref{fig:D025-h1}(a) by the temperature dependence of the 
spin structure factor ${\mathcal S}^{zz}({\bf q}_{\sf M})$ and a suitably modified 
order parameter for lattice rotations
\begin{align}
O_{\mathds{Z}_3}^z& =\Big< \sqrt{ | \psi^z_1 |^2 + |\psi^z_2 |^2  } \hspace{1mm} \Big>,
\label{eq:Z3-Sparallel}
\end{align}
where
\begin{align}
\psi_{1}^z &=\frac{1}{\sqrt{6}N}
\sum_i 2S^z _i.S^z _{i+\delta_1} - S^z_i .S^z_{i+\delta_2} -S^z _i .S^z_{i-\delta_1-\delta_2}, \nonumber\\
\psi_{2}^z &=-\frac{1}{\sqrt{2}N}
\sum_i S^z_i.S^z_{i+\delta_2}-S^z_i.S^z_{i-\delta_1-\delta_2}. \nonumber
\end{align}
The first order behaviour of the phase transition seen here chimes with the known first order phase transition
from the paramagnet into the collinear stripe phase of AgNiO$_2$ [\linecite{coldea09}].

\begin{figure}[t!]
\includegraphics[height= 5.0cm]{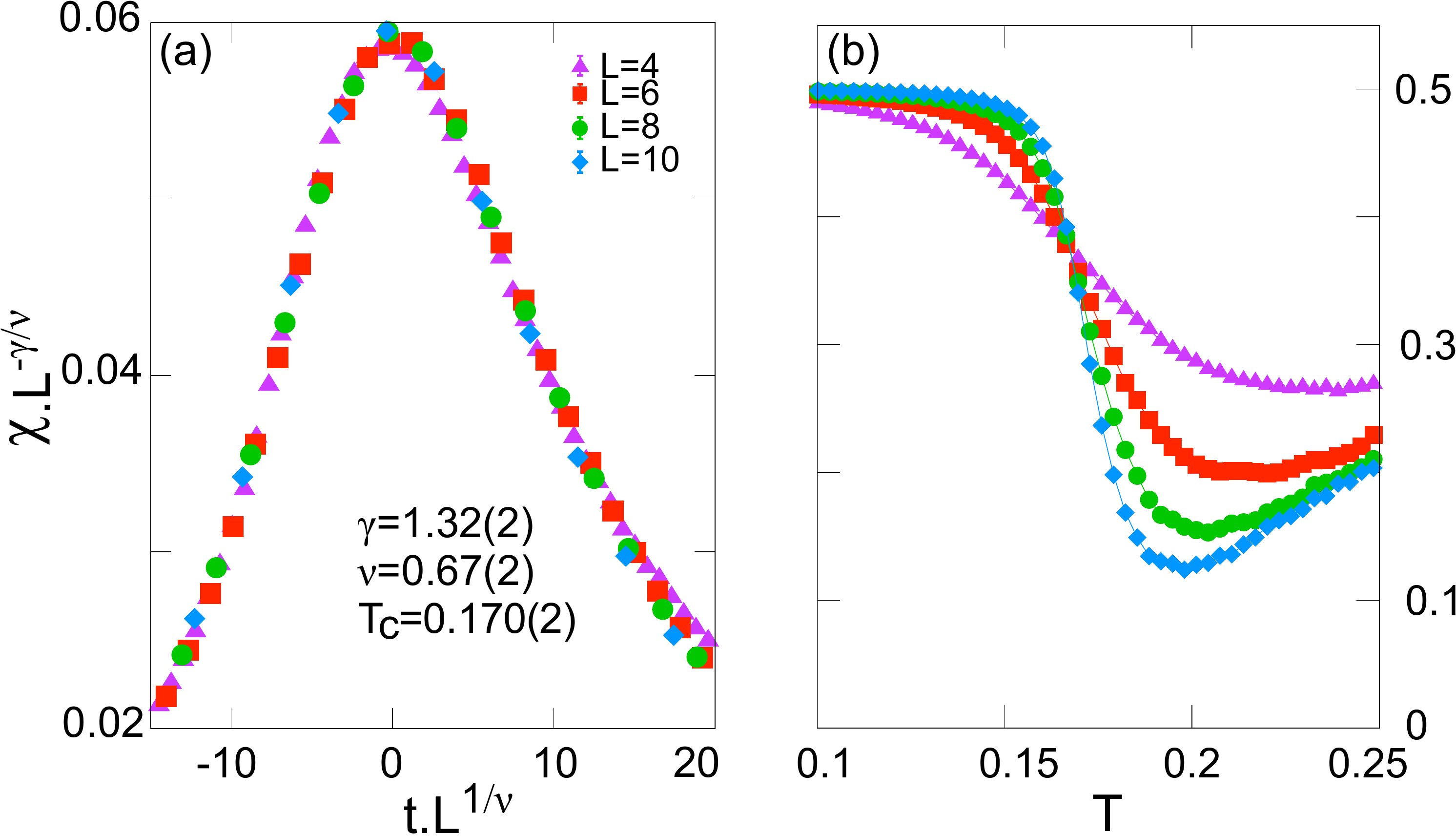}
\caption{\footnotesize{(color online) 
Continuous behaviour of the collinear-supersolid transition for  \mbox{$D=0.25,h=1.0$} at  $T$$\approx$$0.17$. 
(a) Data collapse for the supersolid $U(1)$ order parameter susceptibility using 3D XY universality class critical exponents. 
(b) The Binder cumulant for the same order parameter calculated for different lattice sizes all cross at a single temperature $T=$0.168(1).}}
\label{fig:D025-2nd-order}
\end{figure}

The second, internal, phase transition at $T$=0.171(4) from collinear stripe  into the supersolid state is continuous, exhibiting
only shows weak anomalies in specific heat [Fig.~\ref{fig:D025-h1}(d)] and magnetic susceptibility [Fig.~\ref{fig:D025-h1}(e)].
The new broken symmetry --- a staggered in-plane magnetization --- is heralded by the smooth rise of both the $U(1)$ order parameter 
$O_{U(1)}$ [Fig.~\ref{fig:D025-h1}(b)] and the spin stiffness [Fig.~\ref{fig:D025-h1}(c)].
In this context, the susceptibility associated with the $U(1)$ order parameter can be analysed with a finite-size scaling ansatz 
\begin{align}
\chi_{O_{U(1)}}=L^{\gamma/\nu} \tilde{\chi}(L^{1/\nu}t),
\label{eq:chi-u1-fss}
\end{align}
where $t$ is the reduced temperature $t=(T-T_c)/T_c$. 
The critical exponents associated with the order parameter susceptibility ($\gamma$), and correlation length ($\nu$)
are extracted by fitting this expression to simulation results for different lattice sizes [Fig.~\ref{fig:D025-2nd-order}]. 
The exponents obtained, $\nu$=1.32(2) and $\gamma$=0.67(2), are in excellent agreement with the 3D XY universality class.

%%%%%%%%%%%%%%%

%%%%%%%%%%%%%%%%%%

\begin{figure}[hb!]
\includegraphics[width= 7.5cm]{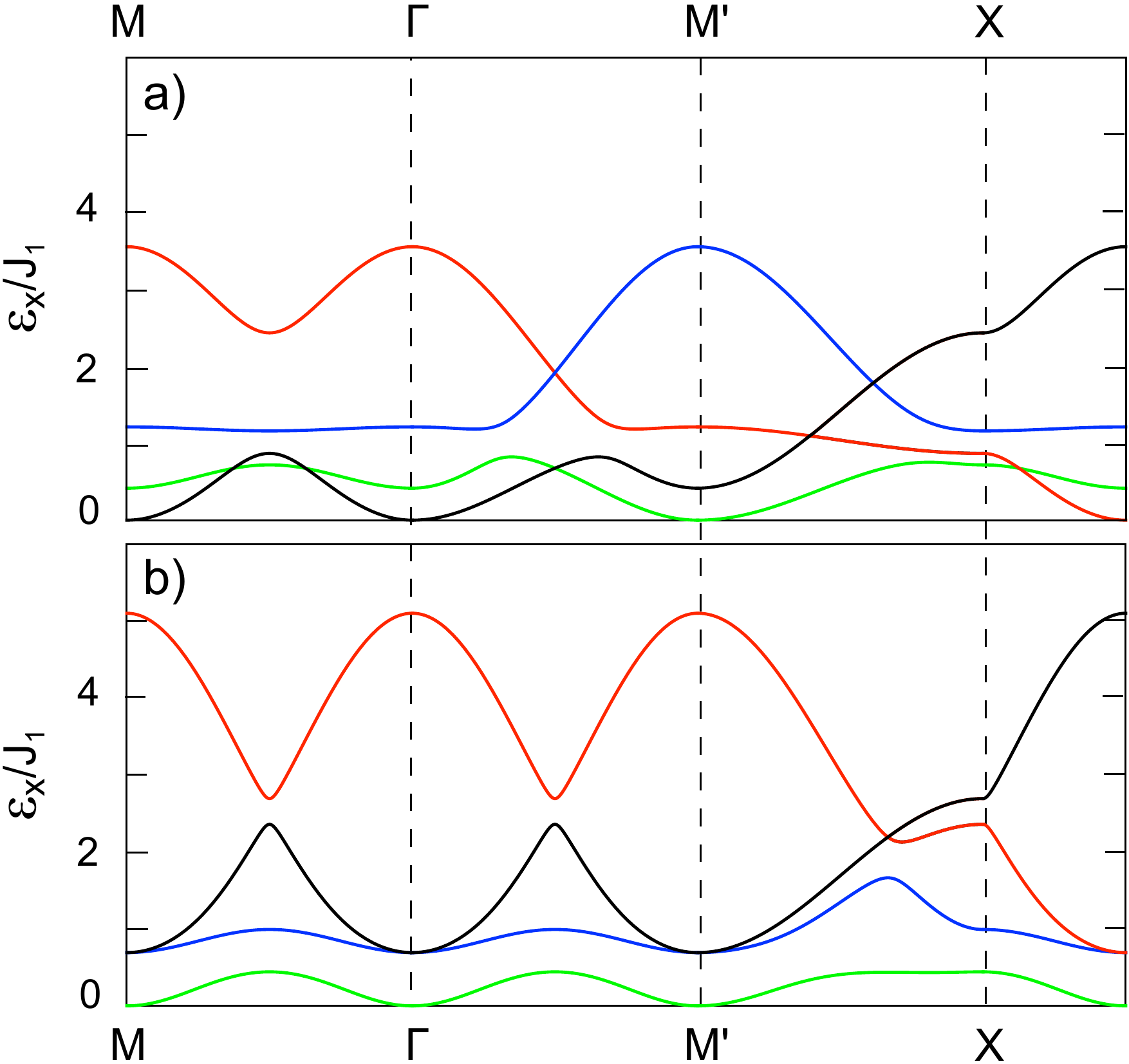}
\caption{\footnotesize{(color online) 
Classical spin-wave dispersion ($S^x$ modes only) at $D$=0.25 for values of magnetic field associated with continuous transitions 
at high field. 
(a) Closing of the spin gap at $M$ for $h$=1.95, associated with transition from the supersolid phase into the 2:1:1 phase. 
(b) Closing of the spin gap at $\Gamma$ for $h$=5.48, associated with transition from the $m$=1/2 plateau into the 3:1 canted phase. 
The points $M$, $M'$ and $\Gamma$ are defined in Fig.~\ref{fig:BZ-stripes}(b)}.}
\label{fig:D025-CSW}
\end{figure}

%%%%%%%%%%%%%%%%%%

 %%%%%%%%%%%%%%%
 
\begin{figure}[htp!]
\includegraphics[width= 7.1087cm]{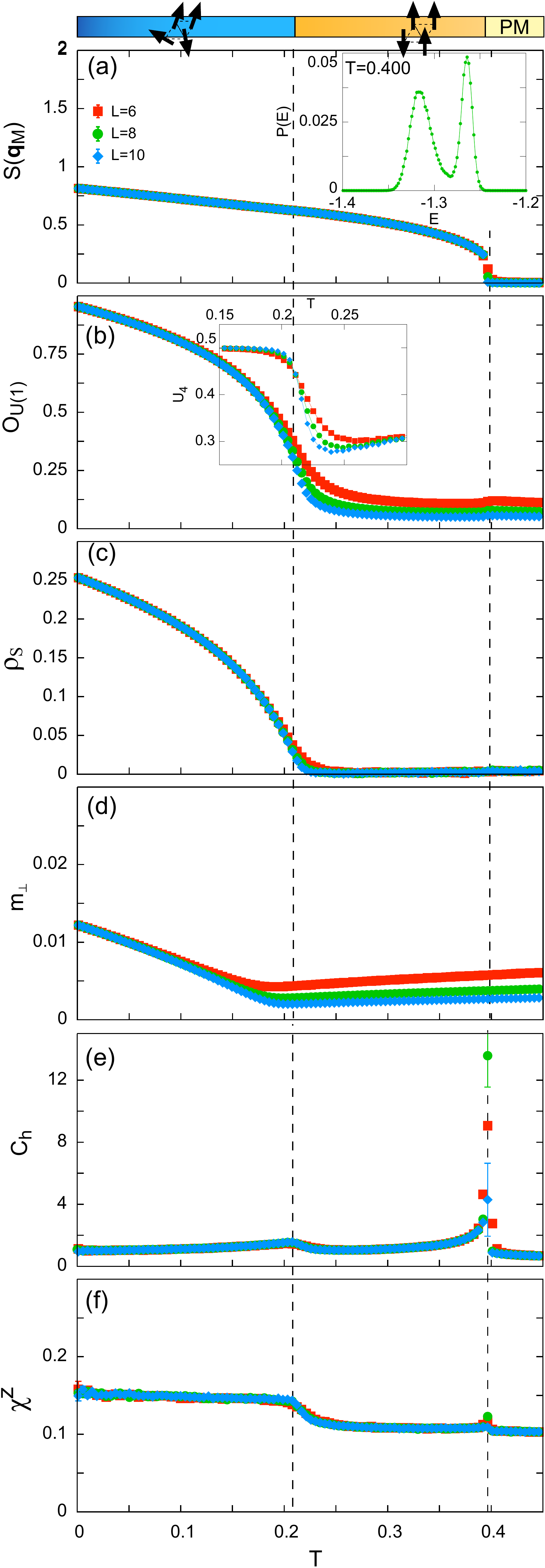}
\caption{\footnotesize{(color online) 
Double phase transition at \mbox{$D=0.25,h=2.5$}, from the paramagnet into $m$=$1/2$ plateau and then into 
2:1:1 canted state at lower temperature. 
(a) Translational symmetry measured by $\mathcal{S}^{zz}(\mathbf{q}_{\sf M})$  is broken at $T$$\approx$$0.4$ in the first-order transition -- cf. bimodal energy distribution (inset) -- into the plateau. 
(b) $U(1)$ continuous symmetry  and (c) spin stiffness are broken in the continuous transition --
cf.  crossing of $U(1)$ order parameter Binder cumulants in inset to (b)
 -- from the plateau into the canted phase 
at $T$$\approx$$0.21$. 
(d) The 2:1:1 canted state is also characterised by in-plane magnetisation $m_\perp$ (d).
The different nature of these transitions is resolved in heat capacity (e) and magnetic susceptibility (f). 
}}
\label{fig:D025-h25}
\end{figure}

%%%%%%%%%%%%%%%%%%

In previously studied models with \emph{XXZ} anisotropy, supersolid phases were found as intermediate phases 
interpolating between N\'eel antiferromagnetic and spin-flop phases \cite{liu73,holtschneider07}. 
However, in the present model, a different phase emerges above the supersolid. 
This is a 2:1:1 canted phase with a four-spin unit cell where two spins are parallel to each other and have positive $S^z$, 
while a third is orientated in the negative $S^z$ direction, and the remaining spin rotates between those positions [Fig.~\ref{fig:configs}(e)].  
At $T$=0 this transition is continuous, and is observed in both mean-field and spin-wave calculations, which reveal a soft 
mode at momentum $M$ [Fig.~\ref{fig:D025-CSW}(a)]. 
With further increase in magnetic field, the 2:1:1 canted phase evolves smoothly into to the collinear $m$=1/2 plateau.

Since the 2:1:1 canted phase breaks both translational and in-plane symmetries it can also be labelled as a supersolid. 
Furthermore, it possesses a (very small) moment
\begin{align}
m_{\perp}=\frac{1}{N} \Big\langle  \big|  \sum_i \mathbf{S}_i^\perp  \big|  \Big\rangle.
\label{eq:mperp}
\end{align}
in the $S^x$$-$$S^y$ plane.  
This behaviour distinguishes it from the lower-field supersolid state, but is also seen in the easy-axis nearest-neighbour 
triangular lattice, where it was attributed to the non-trivial degeneracy of the $T$=0 ground state \cite{miyashita86}. 
In addition to $m_\perp$, this phase is characterised by finite values of the staggered in plane magnetization, 
spin stiffness $\rho_S$ and structure factor $\mathcal{S}^{zz}(\mathbf{q}_{\sf M})$.

In Fig.~\ref{fig:D025-h25} we present simulation results for $h=2.5$, 
for a range of temperatures spanning the paramagnetic, $m$=1/2 plateau and 2:1:1 canted phases.
The phase transition from the paramagnet to the $m$=1/2 plateau is marked by an 
abrupt rise in ${\mathcal S}^{zz}({\bf q}_{\sf M})$ at $T$= 0.397(4).  
The transition is first order, as evidenced by a bimodal energy histogram [inset to Fig.~\ref{fig:D025-h25}(a)], 
and accompanied by a sharp feature in the heat capacity [Fig.~\ref{fig:D025-h25}(e)].  
The phase transition from the $m$=1/2 plateau to the 2:1:1 canted state is continuous, 
with the Binder cumulants for the associated $U(1)$ order parameter crossing at $T$=0.211(1) in
inset to Fig.~\ref{fig:D025-h25}(b). 
 
%%%%%%%%%%%%%%%%%%%%%%%%%%

This continuous phase transition is mediated by a soft spin wave mode within the collinear $m$=1/2 plateau, 
just as the transition into the supersolid is mediated by a soft spin-wave mode within the collinear stripe phase.
In this case the relevant spin-wave gap occurs at the zone centre, and closes with {\it decreasing} magnetic field, 
at a $T$=0 critical field of 
\begin{align}
h_{c}=4 (J_1 + J_2) - 2D. 
\label{eq:SS2->1/2}
\end{align}
Easy-axis anisotropy stabilizes the plateau at $T$=0 for a finite range of field, and its collinearity ensures it is entropically 
favoured against the 2:1:1 canted state at finite temperature.  
The critical field therefore slopes downward with increasing temperature --- the opposite of what is seen in the 
transition from the collinear stripe phase into the supersolid [Fig.~\ref{fig:D025-panel}].  

%%%%%%%%%%%%%%%%%%%%%%%%%%%%%%%%%%%%%

\begin{figure}[htp!]
\includegraphics[width= 7.1087cm]{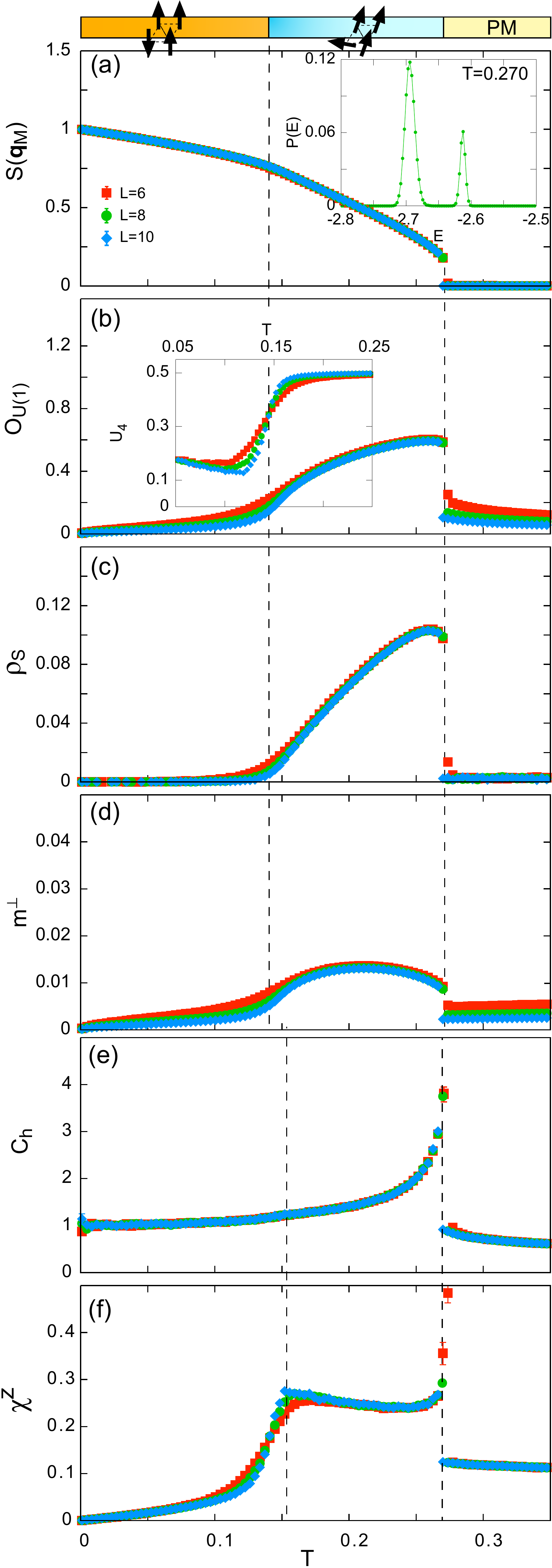}
\caption{\footnotesize{(color online) 
Double  phase transition at \mbox{$D=0.25,h=5.3$} from the paramagnet to 3:1 canted state and then into $m$=$1/2$ plateau.  
The 3:1 canted state is heralded by its discrete $\mathcal{S}^{zz}(\mathbf{q}_{\sf M})$ (a) and continuous U(1) order parameters (b), and also by finite spin stiffness (c). 
This state also has a finite in-plane magnetisation $m_\perp$ (d).  
The upper transition first-order character is clear in the double-peaked energy energy distribution close to the critical temperature 
in inset to (a), while the  continuous character of the lower transition is shown by the Binder cumulant for the $U(1)$ order parameter, in inset to (b).
The different nature of these transitions is resolved in heat capacity (e) and magnetic susceptibility (f). 
}}
\label{fig:D025-h53}
\end{figure}

%%%%%%%%%%%%%%%%%%%%%%%%%%%%%%%%%%%%%

The transition from the collinear $m$=1/2 plateau into the 3:1 canted state at high field also occurs through 
the condensation of a zone-centre spin wave mode [Fig. \ref{fig:D025-CSW}(b)], this time at a $T$=0 critical field of 
\begin{align}
h_{c}=2(J_1+J_2)+2\sqrt{D^2+4D(J_1+J_2)+(J_1+J_2)^2}.
\label{eq:1/2->SS3}
\end{align}
This critical field is only weakly dependent on temperature and, perhaps surprisingly, slopes downwards  [Fig.~\ref{fig:D025-panel}].
As noted in Section~\ref{D=0}, the 3:1 canted phase breaks both discrete translational and continuous spin rotational symmetries 
[cf. Fig.~\ref{fig:D025-h53}].   
It is therefore a third supersolid, in the sense of Matsuda and Tstuneto or Liu and Fisher.  
However, in contrast to the $D$=0 case, for finite anisotropy the 3:1 canted phase possesses a finite value of in-plane 
magnetisation $m_\perp$.

%%%%%%%%%%%%%%%%%%%%%%%%%%%

\begin{figure}[t!]
\includegraphics[width= 7.1087cm]{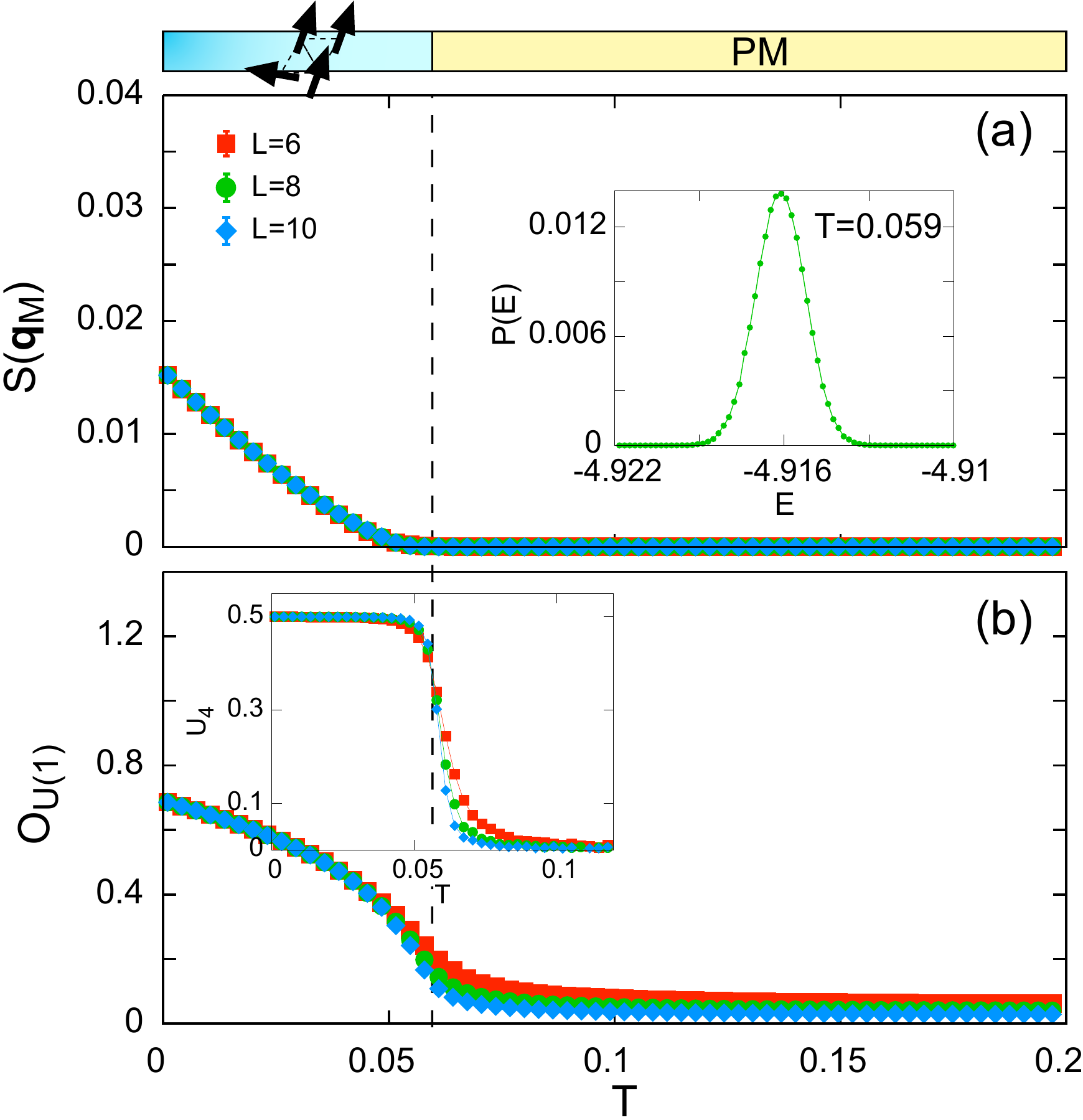}
\caption{\footnotesize{(color online) Temperature dependence of order parameters associated with 3:1 canted phase close to saturation. The continuous nature of the phase transition is confirmed by the single-valued energy histogram close to the critical temperature  in inset to (a),  and crossing of Binder cumulants associated with $U(1)$ order parameter  in inset to~(b).
 }}
\label{fig:D025-h80}
\end{figure}

%%%%%%%%%%%%%%%%%%

%%%%%%%%%%%%%%%%%%%%%%%%%%%%%%%

\begin{figure}[b!]
\includegraphics[width= 7.8cm]{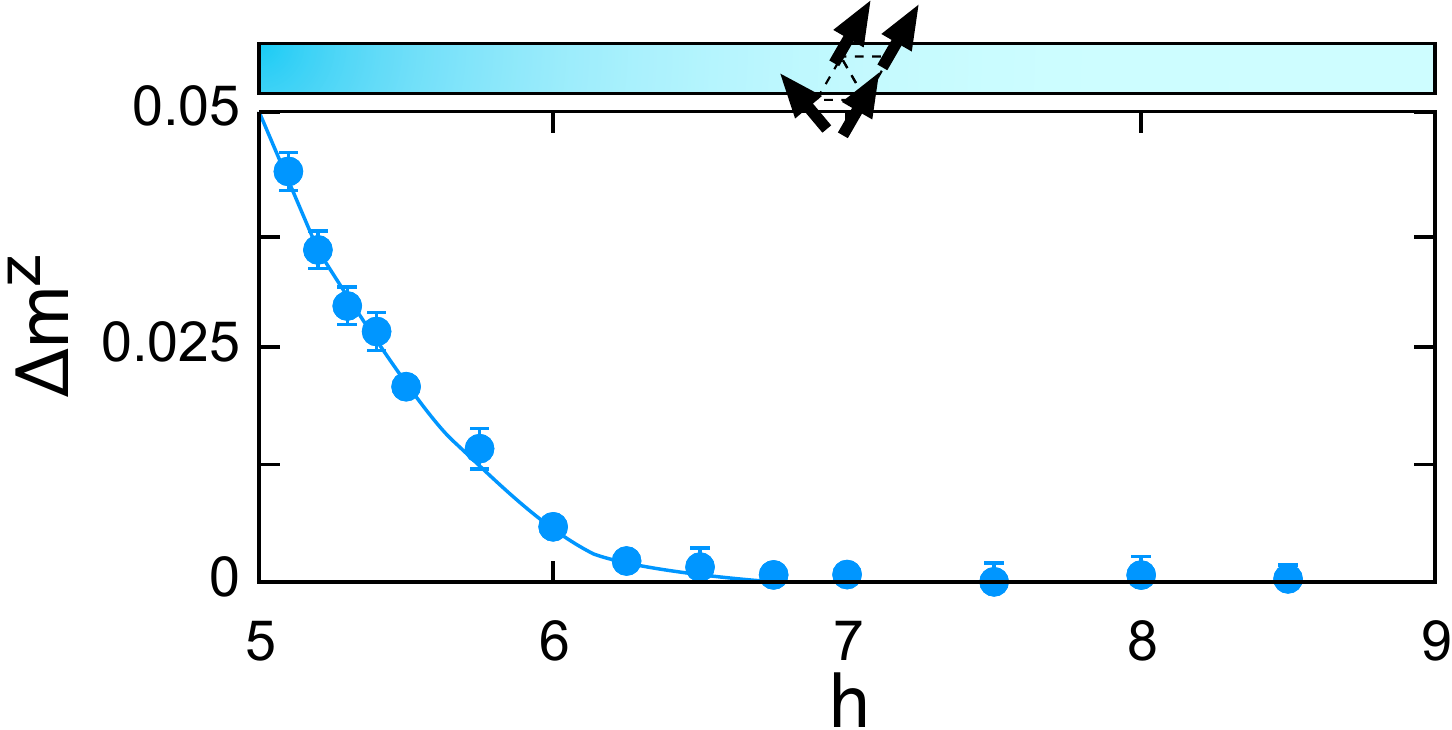}
\caption{\footnotesize{(color online) 
Vanishing discontinuities in magnetisation for $D$=$0.25$, extracted  from temperature cuts in Fig.~\ref{fig:D025-panel}, indicate a crossover from first to second-order phase transition at the paramagnet - 3:1~canted state transition, where the critical endpoint is found at $T$$\approx$$0.14$, $h\approx$$6.5$. Line is a guide to the eye. }}
\label{fig:D025-deltaM-high-field}
\end{figure}
%%%%%%%%%%%%%%%%%%%%%%%%%%%%%%

The  3:1 canted state is the only phase with supersolid character which is directly connected to the paramagnetic region.  
This phase transition is clearly first-order at low field, e.g. $h$=5.3, from simulation results [Fig.~\ref{fig:D025-h53}].   
Nevertheless a continuous phase transition is permitted by symmetry, and the transition at high fields is indeed 
continuous [Fig.~\ref{fig:D025-h80}].
At $T$=0 it is easy to see that this continuous transition corresponds to the opening 
of a spin-wave gap at the 4-sublattice ordering vector(s) \mbox{$\{ \mathbf{q}_{\sf M} \}$}, 
within the saturated state.
For $T$=0, this occurs at 
\begin{eqnarray}
h_{\sf SAT}=8(J_1+J_2)-2D.
\label{saturation}
\end{eqnarray}
and the critical field slopes sharply downwards, as required by the higher entropy 
of the paramagnetic phase.
We conclude that fluctuations drive this continuous phase transition first order for $T$$\gtrsim$0.12 ($h$$\lesssim$6.5) 
--- the point at which a finite jump in the magnetization $\Delta m^z$ is first observed when going from the 3:1 canted phase 
to paramagnet [Fig.~\ref{fig:D025-deltaM-high-field}].

%%%%%%%%%%%%%%%%%%%%%%%%%%%%%

\begin{figure}[b!]
\includegraphics[width= 8.0cm]{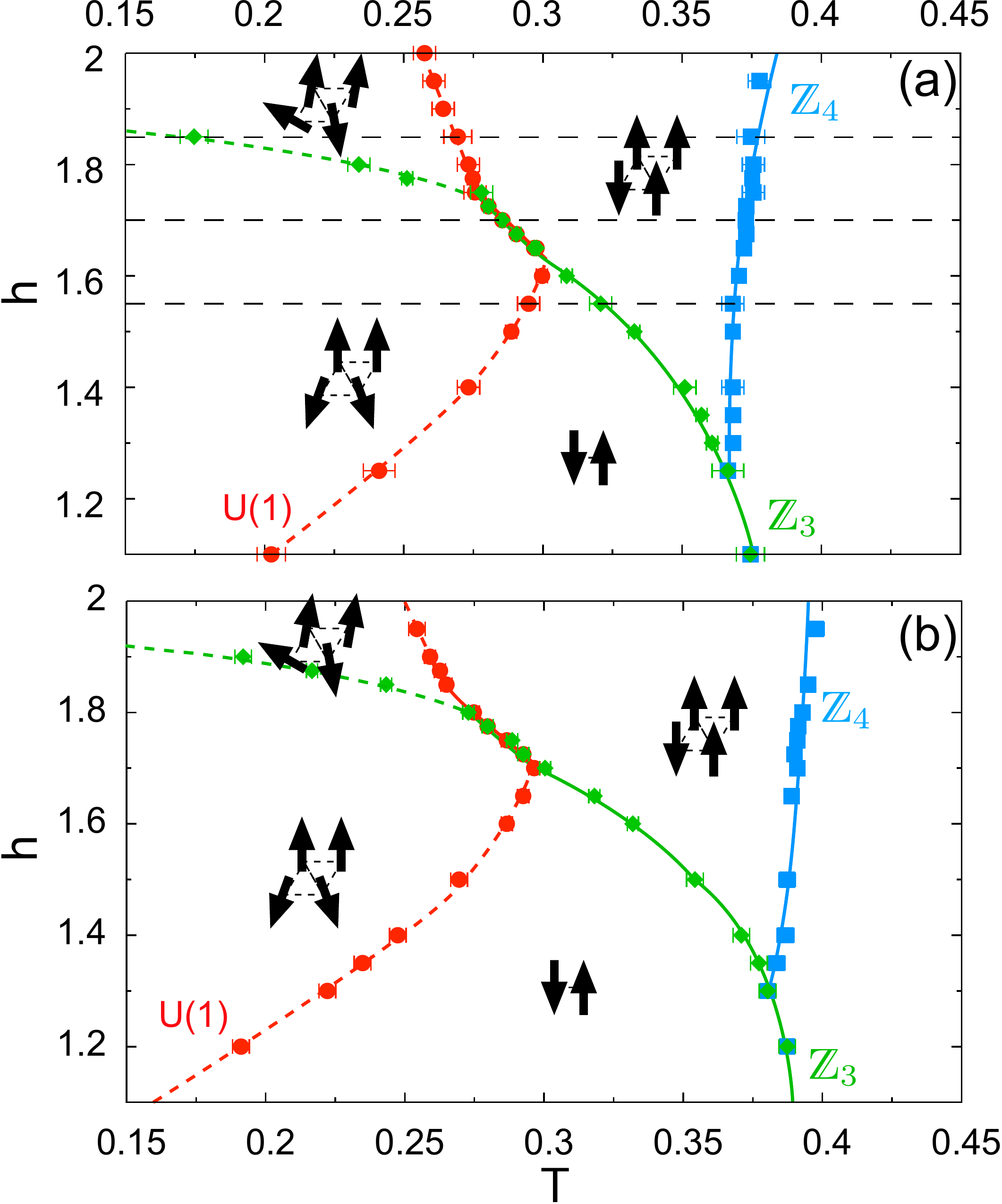}
\caption{\footnotesize{(color online) 
Region where four phases compete at intermediate field for $D$$=$$0.25$ (a)  and $D$$=$$0.325$~(b) for system size $L$=8. 
Phase transitions are labelled according to the symmetries broken.  
Three distinct critical points are identified in a narrow field range, however these never merge into a single tetracritical point. 
This structure is insensitive to changes in the easy-axis anisotropy as long as the $T$$=$$0$ mean field  analysis yields the same phases. 
All phase transitions are first-order, except where shown with a dashed line.  
Dashed horizontal lines in (a) show cuts at fixed field studied in Fig.~\ref{fig:D025-tetraregion-field-cuts}.}}
\label{fig:D025-tetraregion}
\end{figure}

%%%%%%%%%%%%%%%%%%%%%%%%%%

Up to this point we have chiefly concentrated on symmetry breaking at low temperatures, where simulation 
results can be reliably compared with mean-field theory and spin-wave calculations. 
We now turn to the more delicate question of how the full symmetry of the system is recovered with  
increasing temperature, starting from the low-field supersolid phase.
In the original finite-temperature scenario of Liu and Fisher, the supersolid phase extends all the way to the 
paramagnet, where it terminates in a tetracritical point\cite{liu73}. 
Exactly at this tetracritical point four phases (collinear antiferromagnet, supersolid, spin-flop and paramagnet) meet, and 
all phase transitions between them are continuous.  
The behaviour of the present model at high temperatures is markedly different.  
From the phase diagrams shown in Fig.~\ref{fig:D025-panel}, it is clear that the collinear phases completely 
engulf the non-collinear ones at high temperature due to their higher entropy, so the supersolid phase
never touches the paramagnet.
However, for intermediate values of field four phases --- the collinear stripe phase, supersolid, 2:1:1 canted phase, 
and the collinear $m$=1/2 plateau --- are found in very close proximity to one another.
It is therefore worth asking whether any other multicritical points arise in this model.

%%%%%%%%%%%%%%%%%%%%%%%%%%%

\begin{figure}[t!]
\includegraphics[width= 7.8cm]{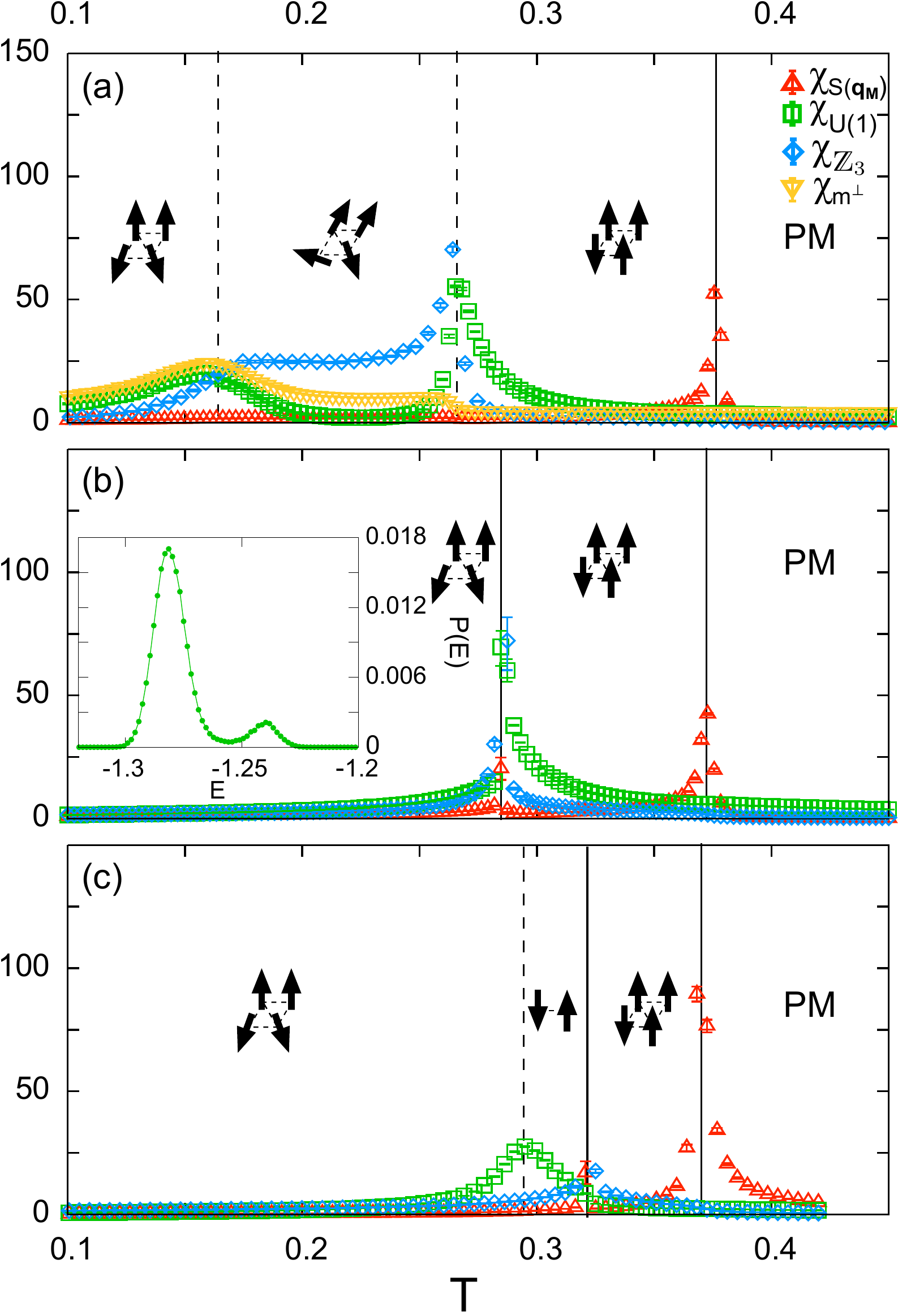}
\caption{\footnotesize{(color online) 
Successive phase transitions at $D$=0.25 for system size $L$=8 as a function of field as resolved in the susceptibilities associated with breakdown of translational symmetry $\mathcal{S}^{zz}(\mathbf{q}_{\sf M})$, spin-rotation symmetry $U(1)$, lattice rotation symmetry $\mathds{Z}_3$ and in-plane magnetisation $m_\perp$.
(a) At high field, $h$=1.85, the $m$=1/2 plateau transforms into the 2:1:1 canted state, followed by the supersolid phase.
(b)  For intermediate field, $h$=1.70, there is a direct first-order phase transition from the $m$=1/2 plateau into the supersolid state, as evidenced by the coincidence  the susceptibility peaks at $T\approx0.28$.
(c) At lower field, $h$=1.55, the system transitions from paramagnet into $m$=1/2 plateau, then into the stripe phase and finally into into supersolid state. 
Phase transitions are first order except where indicated with dashed line.
}}
\label{fig:D025-tetraregion-field-cuts}
\end{figure}

%%%%%%%%%%%%%%%%%%%%%%%%%%%%

\begin{figure}[b!]
\includegraphics[width= 7.8cm]{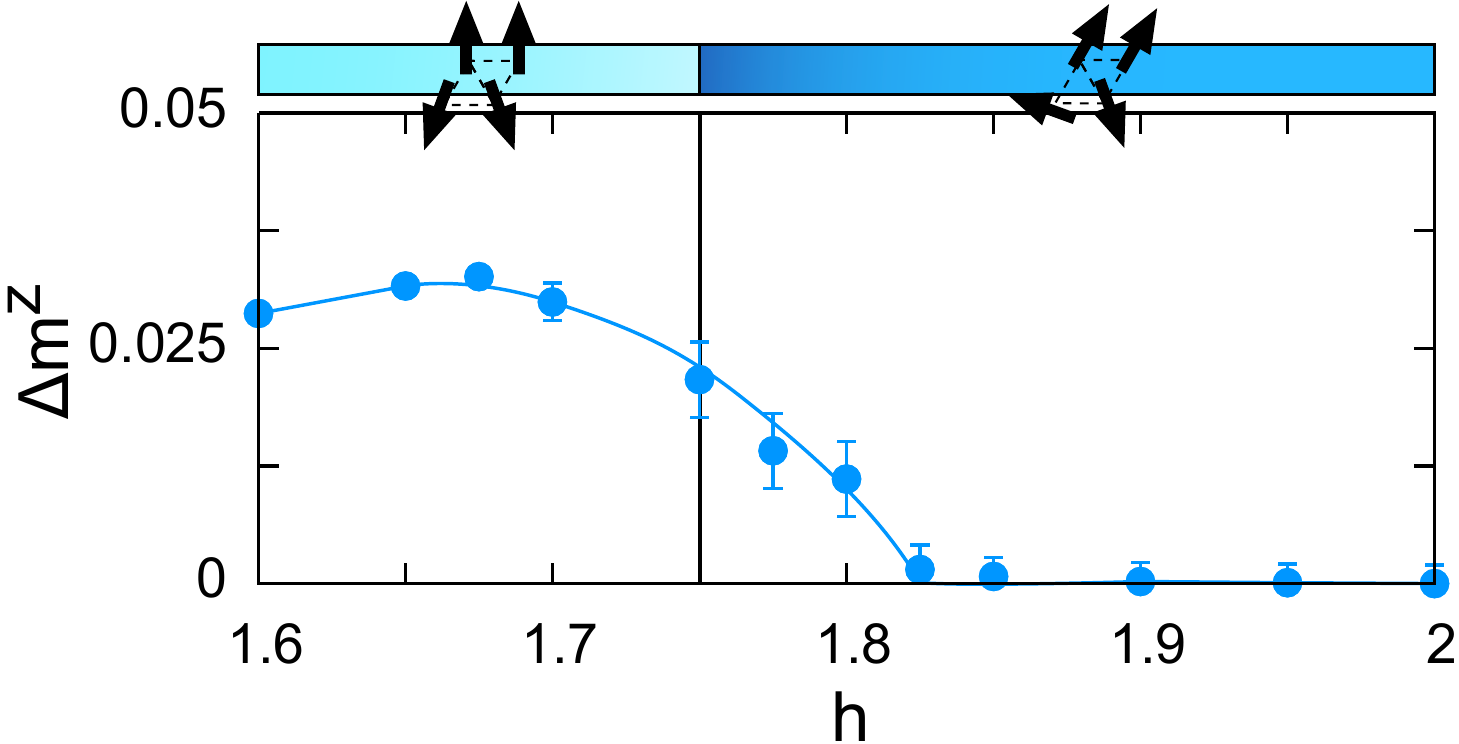}
\caption{\footnotesize{(color online) 
Discontinuities in magnetisation $\Delta m^z$ on entering the $m$=1/2 plateau from the supersolid and 2:1:1 canted phases for 
$D$=$0.25$ and intermediate field, extracted from temperature cuts in Fig.~\ref{fig:D025-panel}.   
The collapse in $\Delta m^z$ for $h$=1.81(2) indicates that the transition from the 2:1:1 canted phase to the 
$m$=1/2 plateau becomes first order {\it before} merging with the transition from the supersolid
to the $m$=1/2 plateau. 
This precludes a multicritical point.
The line is a guide to the eye.
}}
\label{fig:D025-deltaM}
\end{figure}

%%%%%%%%%%%%%%%%%%%%%%%%%%%

In Fig.~\ref{fig:D025-tetraregion}(a), we present a detailed study of the intermediate field region, $1.2$$<$$h$$<$$2$. 
Associated field cuts for $D$=0.25 are shown in Fig.~\ref{fig:D025-tetraregion-field-cuts}.
The $\mathds{Z}_3$ transition between the two collinear states is clearly first order, while the transition between 
the supersolid and 2:1:1 canted phases is continuous.  
The phase transition between the collinear stripe state and the supersolid terminates on the $\mathds{Z}_3$~line.  
The transition between the 2:1:1 canted state and the collinear $m$=1/2 plateau is continuous at high fields, but becomes 
first order shortly before terminating on the $\mathds{Z}_3$~line for $h$=1.76(1) [Fig.~\ref{fig:D025-deltaM}].  
For $h$=1.70 [Fig.~\ref{fig:D025-tetraregion-field-cuts}(b)], there is a clear first order transition between 
the $m$=1/2 plateau and supersolid, where all observed susceptibility peaks merge.  
Given the difficulty in simulating this parameter region with many neighbouring phases, the possibility of a vanishingly 
narrow ``strip" of 2:1:1 canted phase extending between the supersolid and plateau phases  is hard to rule out definitely.  
However, we find no evidence of finite $m_\perp$, characteristic of the 2:1:1 state, in this range of temperature and field.

%%%%%%%%%%%%%%%%%%%%%%%%%%%

From this analysis we conclude that the majority of these phase transitions remain first order where the four phases
converge, and that no more than two phases meet at a single point via a continuous phase transition. 
Even if the 2:1:1 canted phase were to stretch to lower fields, the first-order nature of the phase transitions
between the $m$=1/2 plateau and 2:1:1 canted phases, and $m$=1/2 plateau and collinear stripe phases 
would preclude a tetracritical point --- at most we would have a ``bicritical endpoint", where two continuous 
phase transitions meet two first-order ones. 
This behaviour persists throughout the anisotropy range where these four phases are present, as illustrated in 
Fig.~\ref{fig:D025-tetraregion}(b) for $D$=0.325.

%%%%%%%%%%%%%%%%%%%%%%%%%%%

%%%%%%%%%%%%%%%%%%%%%%%%%%%%%%

\begin{figure*}[ht!]
\includegraphics[width= 17cm]{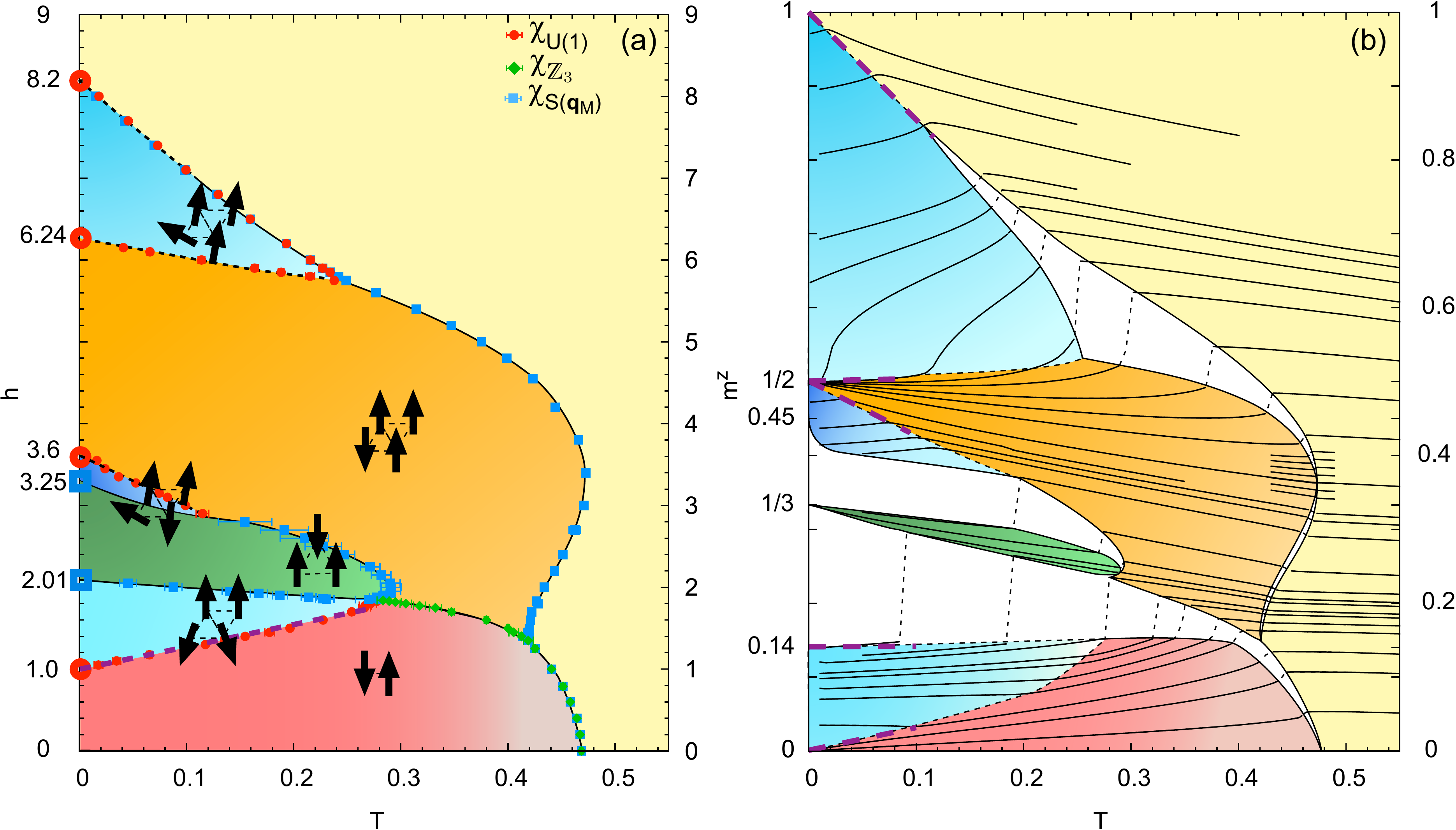}
\caption{ \footnotesize{  (color online) Magnetic phases of a layered triangular-lattice antiferromagnet with  $J_1$=$1$, $J_2$=$0.15$, $J_\perp$=$-0.15$ and \mbox{easy-axis anisotropy $D$=$0.5$}.
(a)  Phase diagram as a function of temperature and magnetic field.  
Open symbols on the h-axis show transitions obtained in mean-field theory.   
Phase boundaries at finite temperature are obtained from Monte Carlo simulation for a cluster of $24$$ \times$$ 24$$ \times 8$ spins, and determined by peaks in the relevant order parameter susceptibilities.  
All phase transitions are first-order, except where shown with a dashed line.   
Thick purple dashed line is obtained through a Landau expansion for the supersolid transition.  
(b)~Phase diagram as a function of temperature and magnetization.   
Solid lines running left-right show cuts at constant magnetic field $h$ taken from simulations.    
The coexistence regions associated with first order phase transitions are coloured white. 
Thick purple dashed lines show phase boundaries  obtained through a  low-T expansion. 
Increasing anisotropy gives rise to an  $m$=1/3 plateau with a different unit-cell from the other phases.}}
\label{fig:D05-panel}
\end{figure*}

%%%%%%%%%%%%%%%%%%%%%%%%%%%%%%

In the same spirit, it is worth re-examining the finite temperature transition between the paramagnet and collinear $m$=1/2 plateau. 
While generically first-order, there exist two points in the phase diagram, $h$$\approx$$1.25$ and $h$$\approx$$3.2$, 
where the jump in magnetisation between the two phases vanishes, indicating the possibility of a continuous behaviour 
[Fig.~\ref{fig:D025-panel}(b)]. 
These two points are a robust feature of the $m$=1/2 plateau for all values of $D$ investigated. 
One possible scenario for this behaviour would be the emergence of a higher symmetry at these special points. 
Sadly, however, this attractive scenario does not survive a closer examination of the data.

%%%%%%%%%%%%%%%%%%%%%%%%%%%

The first point at which $\Delta m^z$=0, at $h$$\approx$$1.25$, appears to be a critical end point, 
where the first order transition between the collinear $m$=1/2 plateau and the paramagnet terminates
on the first order transition into the collinear stripe phase.
This critical end point is distinguished by a unimodal energy histogram. 
The second point at which $\Delta m^z$=0 occurs for $h$$\approx$$3.20$, and can be understood 
in terms of the two different spin wave excitations which connect the collinear $m$=1/2 plateau 
with the 2:1:1 canted at lower field, and with the 3:1 canted state at higher field.
The spin wave-excitation associated with the 2:1:1 canted state {\it lowers} the magnetization 
of the system, while the excitation associated with the 3:1 canted state {\it raises} the 
magnetization of the system.

These spin waves will also determine the sign of the magnetization jump at the 
finite temperature transition from the collinear $m$=1/2 plateau into the paramagnet.
It follows that there will exist a value of magnetic field at which this jump $\Delta m^z$
changes sign, without any dynamically generated symmetry entering into the problem. 
A close examination of energy histograms and the spin structure factor ${\mathcal S}({\bf q})$
at the transition bears out this interpretation.

%%%%%%%%%%%%%%%%%%%%%%%%%%%%%%%%%%%%%%%%%%%%%%%%%%%%%%%%%%%%%%%%%%%%%%%%%%%%%%

\section{Intermediate anisotropy, $D$=0.5}
\label{D=0.5}

%%%%%%%%%%%%%%%%%%%%%%%%%%%%%%%%%%%%%%%%%%%%%%%%%%%%%%%%%%%%%%%%%%%%%%%%%%%%%%

In order to investigate the robustness of the $D$=0.25 results we increase the anisotropy, focusing on a representative value $D$=0.5. As a consequence a new collinear phase emerges at intermediate fields. This is the familiar three-sublattice state with two spins ``up" and one ``down" [Fig.~\ref{fig:configs}(d)], i.e. a collinear one-third magnetisation plateau with ordering vectors at the corners of the Brillouin Zone~[Fig.\ref{fig:BZ-stripes}(d)].  In this case the system sacrifices exchange energy associated with  the second-neighbour interaction  to increase its collinearity.

%%%%%%%%%%%%%%%%%%%%%%%%%%%%%%

This  $m$=1/3 plateau first appears $within$ the 2:1:1 canted phase for $D$$>$0.33, splitting the canted phase in two. Increasing anisotropy rapidly suppresses the ``lower" 2:1:1 canted state so that the $m$=1/3 plateau becomes the third phase under field, above the supersolid. While in the purely nearest-neighbour case this plateau emerges continuously connected to canted phases\cite{miyashita86}, in this situation it is completely unrelated by symmetry to the surrounding phases. Thus this $m$=1/3 can be thought of as  an ``accident" within the natural progression of four-sublattice phases.

%%%%%%%%%%%%%%%%%%%%%%%%%%%%%%

The low-field results at $D$=0.5 have been analysed elsewhere\cite{seabra10} and essentially reproduce the $D$=0.25 case.
At $T$=0 the first-order phase transition from the supersolid into the $m$=1/3 plateau occurs for a mean-field field of
\begin{align}
h=&( D + 8 (J_1 + J_2))/3  \nonumber \\ 
& - 4/3 \sqrt{2(8J_2-D - J_1 ) (D - 2 (J_1 + J_2))}. 
\label{eq:SS1->UUD}
\end{align}
To study this plateau we use the $S^z-S^z$ structure factor associated with the three-sublattice ordering vectors  \mbox{$\{ \textbf{q}_{\sf K} \}$}, cf. Fig.~\ref{fig:BZ-stripes},
 \begin{align}
\mathcal{S}^{zz}(\mathbf{q}_{\sf K} )= \Big \langle\sum_{\{q_K\}} \Big|   \frac{1}{N} \sum_i S_i^z   \textrm{e}^{-i \textbf{q}_{\sf K} \cdotp \textbf{r}_i} \Big| ^2  \Big\rangle.
\label{SqK}
\end{align}

%%%%%%%%%%%%%%%%%%%%%%%%%%%%%%

\begin{figure}[ht]
\includegraphics[width= 7.1087cm]{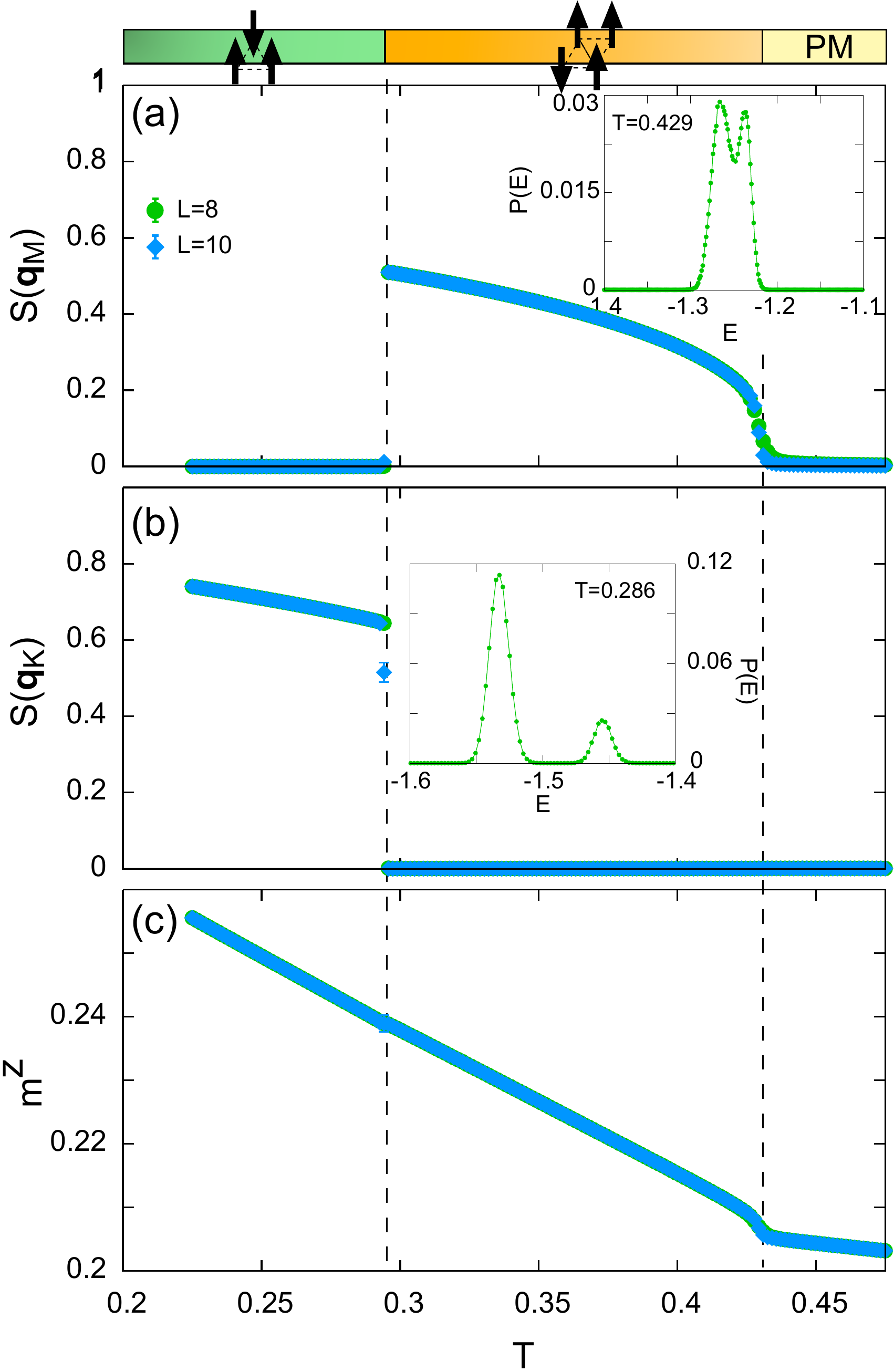}
\caption{\footnotesize{(color online) 
Double phase transition from the paramagnet to $m$=1/2-plateau and then into the $m$=1/3-plateau at \mbox{$D=0.5,h=1.89$}.  
Temperature dependence of the structure factor measured at momenta corresponding to  four-sublattice ordering $\mathcal{S}^{zz}(\mathbf{q}_{\sf M})$ (a), and (b) momenta corresponding to three-sublattice ordering $\mathcal{S}^{zz}(\mathbf{q}_{\sf K})$ (b).  
Both paramagnet-$m$=1/2-plateau and plateau-plateau phase transitions are first-order, as seen in the energy histograms in insets to (a) and (b), however the magnetisation (c) does not exhibits the characteristic jump across the inner transition.}}
\label{fig:D05-UUD-FSS}
\end{figure}

At finite temperature the $m=$$1/3$ plateau is connected through first-order transitions to the $m=$1/2 plateau  [Fig.~\ref{fig:D05-UUD-FSS}],  supersolid and 2:1:1 state phases. These  low-temperature, strongly first-order transitions between  phases with different symmetries are very difficult to simulate, even employing the combined parallel tempering and over-relaxation procedure.

%%%%%%%%%%%%%%%%%%%%%%%%%%%%%%

%%%%%%%%%%%%%%%%%%%%%%%%%%%%%%

For this value of anisotropy the 2:1:1 canted state survives above the $m$=1/3 plateau for a $T$=0 field range of $3.25$$<$$h$$<$$3.60$. While at high field the tendency of the $m=$1/2 plateau to cant through the closing of a spin-gap wins, at lower field the $m$=1/3 plateau is favoured entropically against the canted state. Because these two transitions lie very close to each other at low temperature, it is much more difficult to  find  them with the same accuracy as used in the rest of the phase diagram.

%%%%%%%%%%%%%%%%%%%%%%%%%%%%%%

The nature of the transitions into the supersolid phases is unaffected by the increase of anisotropy.
However the magnetisation jumps [Fig.~\ref{fig:D05-panel}(b)] have become more dramatic. Hence the points with $\delta m^z=0$ connected to the $m=1/2$ plateau, at $h\approx1.4$ and $h\approx3.6$, become more visible.

%%%%%%%%%%%%%%%%%%%%%%%%%%%%%%

The endpoint of the inner plateau-plateau transition is very close to the critical point of the supersolid transition at $h\approx2,T\approx0.3$, in a shape curiously reminiscent of Fig.~\ref{fig:D025-tetraregion}.   The hysteresis associated with these strongly first-order transitions, especially the plateau-plateau one, makes a more in-depth analysis of this region difficult to perform.   Nevertheless, as we shall see below, this feature is not robust for other values of anisotropy.

%%%%%%%%%%%%%%%%%%%%%%%%%%%%%%
 
Another possibly interesting feature is that at an higher  field of $h$$\approx$1.90 the plateau-plateau transition also supports a point for which the magnetisation discontinuity apparently vanishes [Fig.~\ref{fig:D05-UUD-FSS}]. As in the previous case, we find no evidence of the emergence of  higher symmetries or other exotic behaviour at this point. 

%%%%%%%%%%%%%%%%%%%%%%%%%%%%%%%%%%%%%%%%%%%%%%%%%%%%%%%%%%%%%%%%%%%%%%

\label{D=0.675}
\section{Intermediate to strong anisotropy, $D$=0.675}

%%%%%%%%%%%%%%%%%%%%%%%%%%%%%%

\begin{figure*}[htp!]
\includegraphics[width= 17cm]{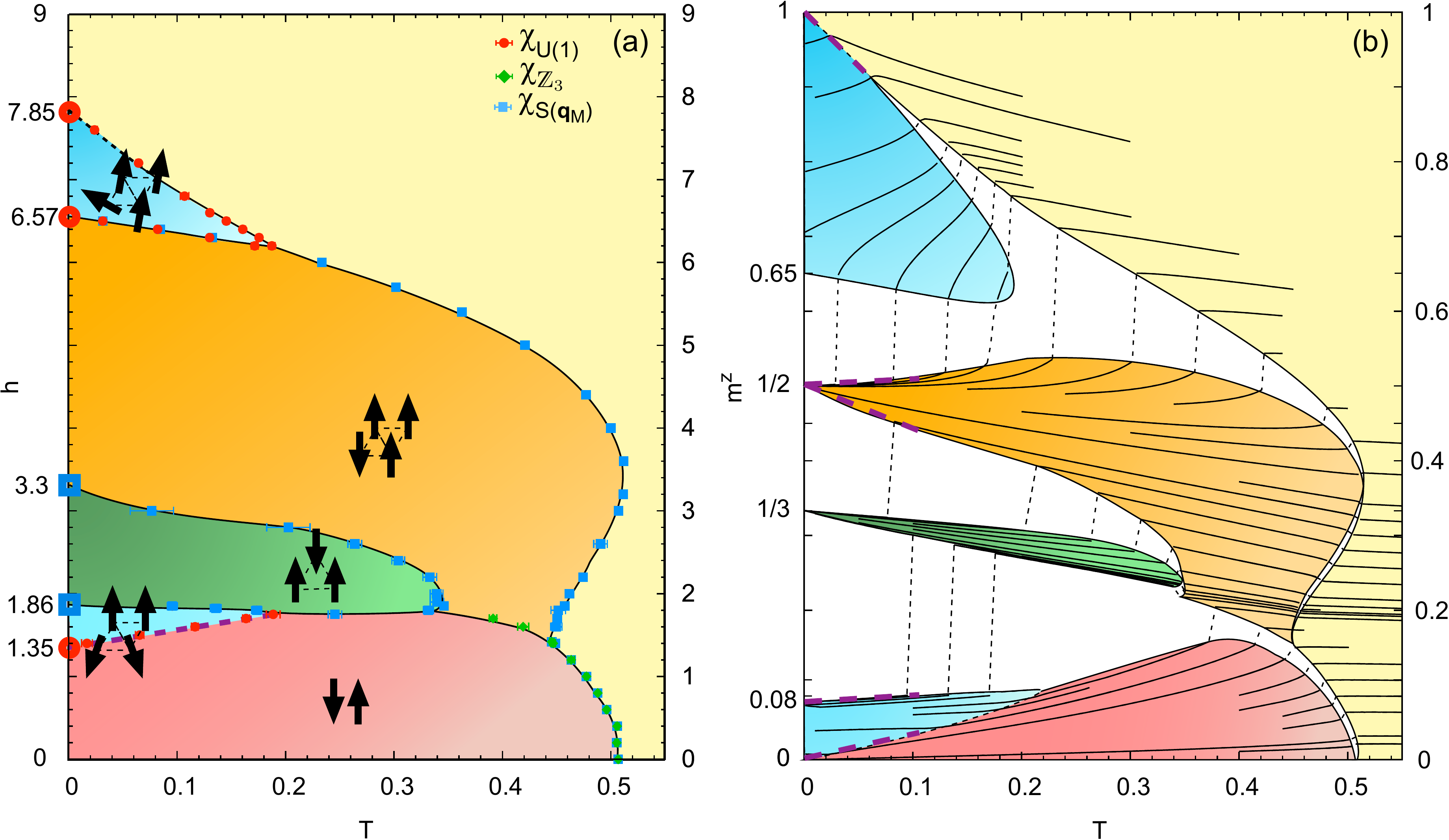}
\caption{ \footnotesize{ (color online) 
Magnetic phases of a layered triangular-lattice antiferromagnet with  $J_1$$=$$1$, $J_2$=$0.15$, $J_\perp$=$-0.15$ and \mbox{easy-axis anisotropy~$D$=$0.675$}.
(a)  Phase diagram as a function of temperature and magnetic field.  
Open symbols on the h-axis show transitions obtained in mean-field theory.   
Phase boundaries at finite temperature are obtained from Monte Carlo simulation for a cluster of $24$$ \times$$ 24$$ \times 8$ spins, and determined by peaks in the relevant order parameter susceptibilities.  
All phase transitions are first-order, except where shown with a dashed line.  
Thick purple dashed line is obtained through a Landau expansion for the supersolid transition. 
(b)~Phase diagram as a function of temperature and magnetization.   
Solid lines running left-right show cuts at constant magnetic field $h$ taken from simulations.    
The coexistence regions associated with first order phase transitions are coloured white.  
Thick purple dashed lines show phase boundaries  obtained through a  low-T expansion.  
Further increasing anisotropy leads to suppression of supersolid phases. }}
\label{fig:D0675-panel}
\end{figure*}

%%%%%%%%%%%%%%%%%%%%%%%%%%%%%%

By further increasing the anisotropy strength we wish to verify the stability of the $D=0.5$ conclusions, especially in the region surrounding the $m$=1/3 plateau. We expect the collinear phases to gain ground and eventually suppress the supersolid states. Therefore we choose to study $D=0.675$ as a representative value. We note that the set of parameters with this value of $D$ gives the bet fit to the magnetic excitations in AgNiO$_2$ \cite{wheeler09}.

%%%%%%%%%%%%%%%%%%%%%%%%%%%%%%

The general shape of the  phase diagram  [Fig.~\ref{fig:D0675-panel}]  is similar to the previous ones. The transition fields have become less dependent on temperature and the phases themselves are more widely separated in the magnetisation-temperature phase diagram, a reflection of the stronger anisotropy. 

%%%%%%%%%%%%%%%%%%%%%%%%%%%%%%

The main difference when compared to lower anisotropy  is the complete suppression of the 2:1:1 canted phase.  This phase ceases to exist   at $T$=0 for $D$$>$$0.625$ resulting in a first-order plateau-plateau transition for  $h=6 (J_1-3 J_2)=3.3$, a value which holds up to the Ising limit \cite{takasaki86}. 

\vspace{5mm}

%%%%%%%%%%%%%%%%%%%%%%%%%%%%%%

The points where both the collinear-supersolid and the plateau-plateau transitions terminate on the~$\mathds{Z}_3$ transition line are now well separated. They occur approximately at the same field $h$$\approx$1.85 but at clearly different temperatures,  $T$$\approx$0.19 and $T$$\approx$0.34 respectively. Hence this implies a direct transition between the stripe phase and the $m$=1/3 plateau as a function of field. We thus interpret the aforementioned closeness of these two points at $D$=0.5 as purely accidental for that specific value anisotropy, and not a robust feature of the model. 

%%%%%%%%%%%%%%%%%%%%%%%%%%%%%%

The previously identified points where the magnetisation jumps disappear at the plateau-stripe critical point and at both plateau-plateau and paramagnet-plateau transitions are still clearly visible.
 We have checked that these features hold for several values of intermediate~t$D$ and thus can be said to be characteristic of the $m$=1/2 plateau phase.

\vspace{5mm}

%%%%%%%%%%%%%%%%%%%%%%%%%%%%%%

A relevant difference is that the 3:1 canted phase has now become separate from the parent \mbox{$m$=$1/2$ plateau} for all temperatures in the magnetisation-temperature phase diagram. The increased anisotropy favours the collinear state, driving the transition first order, instead of the symmetry-allowed second-order transition observed before. At $T$=0 the mean-field  transition field is $h$=6.57, which is lower than the field predicted by the closing of the spin-wave gap  Eq.~(\ref{eq:1/2->SS3}), $h_c(D$=$0.675)$=6.72. This crossover to a first-order transition happens at $D$$\approx$0.525.

%%%%%%%%%%%%%%%%%%%%%%%%%%%%%%

\begin{figure}[ht!]
\includegraphics[width= 7.1087cm]{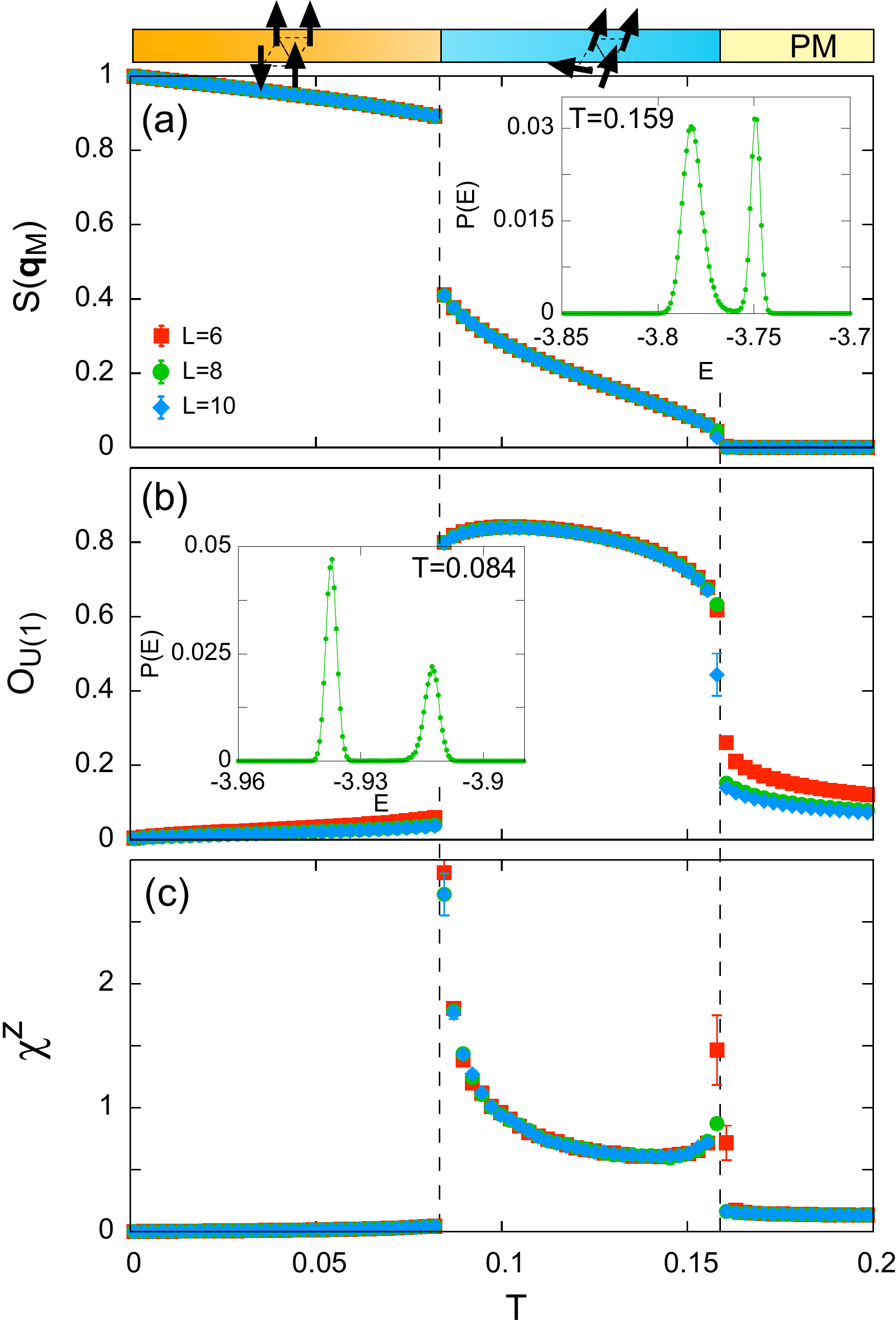}
\caption{\footnotesize{(color online) 
Double first-order transition at \mbox{$D=0.675,h=6.4$} from the paramagnet into 3:1 canted state, and from 3:1 canted state into the $m=$1/2 plateau. 
Temperature dependence of the structure factor $\mathcal{S}^{zz}(\mathbf{q}_{\sf M})$ (a), $U(1)$ order parameter (b) and magnetic susceptibility (c). 
The first-order character of both transitions is clear in the order parameter jumps and double-peaked energy distributions close to the critical temperatures in insets to (a) and (b). 
The lower transition is continuous for smaller~$D$, e.g. $D$=0.25 [Fig.~\ref{fig:D025-h53}]. }}
\label{fig:D0675-h64}
\end{figure}

%%%%%%%%%%%%%%%%%%%%%%%%%%%%%%

The simulation results agree with this interpretation, showing  for all temperatures the hallmarks of a first order transition, jumps in order parameters and magnetisation and double-peaked energy distribution [Fig. \ref{fig:D0675-h64}]. The nature of the paramagnet-canted transition is unchanged, displaying a critical endpoint at $h$$\approx$7.2 and $T$$\approx$0.06 where fluctuations at higher temperatures drive the transition first-order.

\vspace{20 mm}

%%%%%%%%%%%%%%%%%%%%%%%%%%%%%%

\section{Strong anisotropy, $D$=1.5}
\label{D=1.5}

%%%%%%%%%%%%%%%%%%%%%%%%%%%%%%

\begin{figure*}[htp!]
\includegraphics[width= 17cm]{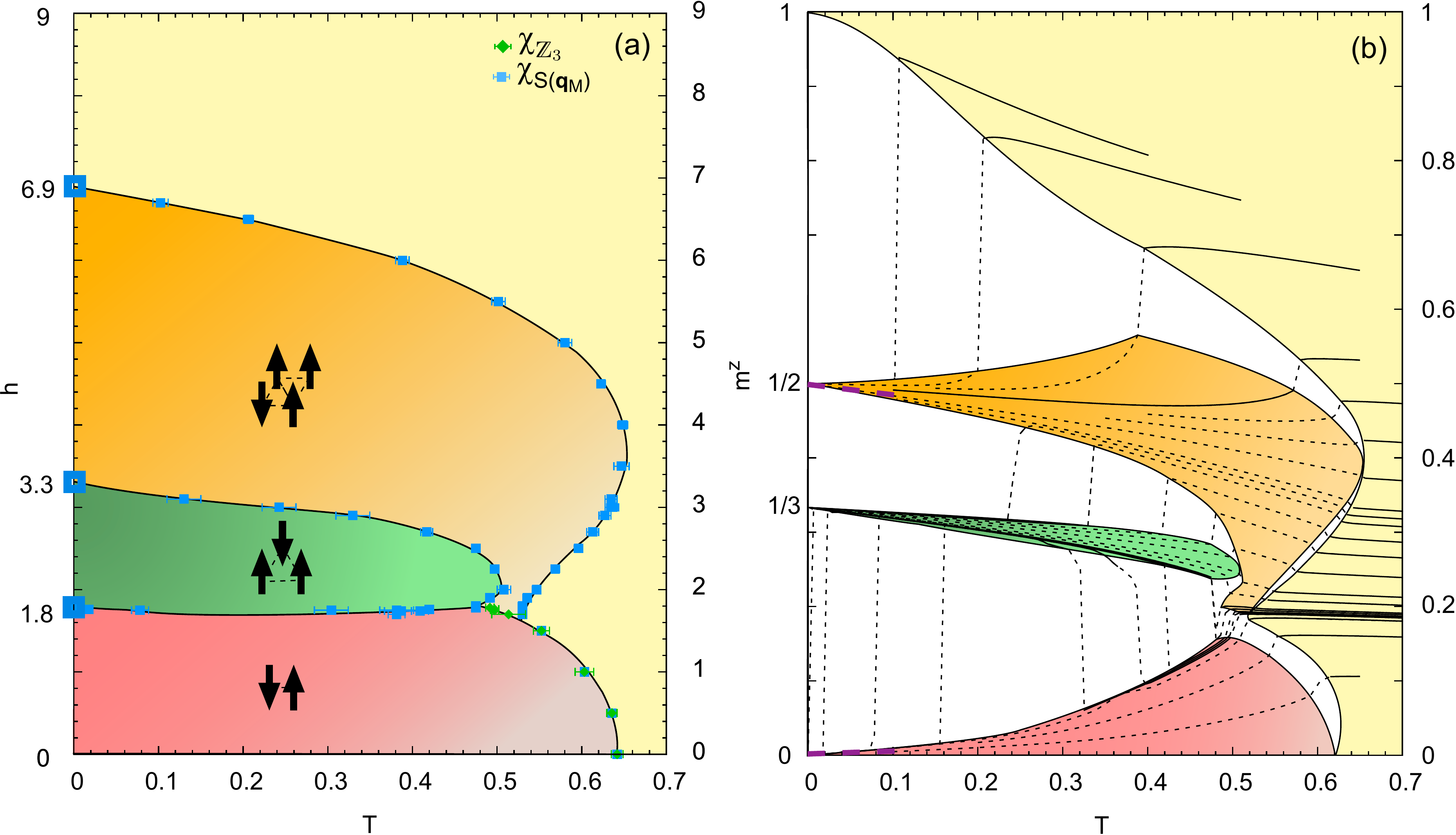}
\caption{\footnotesize{ (color online) (a)  Magnetic phases of a layered triangular-lattice antiferromagnet with  $J_1$=$1$, $J_2$=$0.15$, $J_\perp$=$-0.15$ and \mbox{ strong easy-axis anisotropy $D$=$1.5$}.
(a)  Phase diagram as a function of temperature and magnetic field.  Open symbols on the h-axis show transitions obtained in mean-field theory.   Phase boundaries at finite temperature are obtained from Monte Carlo simulation for a cluster of $24$$ \times$$ 24$$ \times 8$ spins, and determined by peaks in the relevant order parameter susceptibilities.  All phase transitions are first-order.  (b)~Phase diagram as a function of temperature and magnetization.   Solid lines running left-right show cuts at constant magnetic field $h$ taken from simulations.    The coexistence regions associated with first order phase transitions are coloured white. Thick purple dashed lines show magnetisation curves  obtained through a  low-T expansion. At high anisotropy non-collinear phases have disappeared and Ising-like physics is recovered.}}
\label{fig:D15-panel}
\end{figure*}

%%%%%%%%%%%%%%%%%%%%%%%%%%%%%%

Increasing the  anisotropy strength eventually leads to the suppression of all non-collinear phases. At $T$=0, the supersolid state is squeezed out of existence at $D$$>$$0.9$ for $J_2$$=$$0.15$  by the collinear phases which surround it. For anisotropy values of $D$$\gtrsim$1.15 the 3:1 canted phase disappears and there is only a direct plateau-saturation transition, i.e. a magnetisation jump from $m$=1/2 to $m$=1 with increasing field. We thus focus on $D$=1.5 as a representative value.

%%%%%%%%%%%%%%%%%%%%%%%%%%%%%%

The ordered observed phases are those collinear phases which can be accommodated in a three- or a four-site unit cell, i.e. the collinear stripe phase at low field and then the $m$=1/3- and $m$=1/2 magnetisation plateaux. At $T$=0 the transition fields are independent of the anisotropy value and agree with the established results for the Ising model. All phase transitions are first order for all temperatures and fields.

%%%%%%%%%%%%%%%%%%%%%%%%%%%%%%

Nevertheless, even in this strong-anisotropy regime the features above identified previously as robust characteristics of the model are still clearly visible. These include the persistence of the first-order transition between the stripe-collinear and $m$=$1/2$ phases, and the points with  $\Delta m$=0  in both the paramagnet-plateau and plateau-plateau transitions.

%%%%%%%%%%%%%%%%%%%%%%%%%%%%%%

\section{The Ising limit, $D$=$\infty$  }
\label{Ising}

%%%%%%%%%%%%%%%%%%%%%%%%%%%%%%

%%%%%%%%%%%%%%%%%%%%%%%%%%%%%%

It is instructive to compare the strong anisotropy results with the pure Ising limit, \mbox{$D \rightarrow \infty $}, previously investigated at $T$=0  [\linecite{takasaki86}]. 
The most obvious difference is that in the Ising limit the $m$=1/3 plateau is now directly connected to the paramagnetic state. The two values of field where the magnetisation jump at the first-order  paramagnet-m=$1/2$ plateau transition are still present, but one of them now arises where both plateaux and the paramagnet meet.
The inner transition fields all show very little dependence with temperature, indicating the validity of the mean-field picture and that all three phases  possess roughly the same amount of entropy.  As a final comment, we observe that the zero field N\'eel temperature is a monotonically function of anisotropy. Empirically, this can be written as $T_N(D)=T_N(\infty)-\frac{1.81(1)}{2.21(4)+D}$, until saturating at $T_N(\infty)=1.127(3)$ in the Ising limit.

%%%%%%%%%%%%%%%%%%%%%%%%%%%%%%%%%%%%%%%%%%%%%%%%%%%%%%%%%%%%%%%%%%%%

\section {Discussion of results}
\label{results}

%%%%%%%%%%%%%%%%%%%%%%%%%%%%%%%%%%%%%%%%%%%%%%%%%%%%%%%%%%%%%%%%%%%%

\begin{table}[h!]
\caption{Catalogue of phases and associated phase diagrams for different values of $D$.}
\label{table:phases}
\begin{tabular}{|c | c |  c |   c |   c |   c |  c |  c | c | c |   c | c |}
\hline
\hline
%
%%%%% 1st row %%%%%%%%%%%%%%%%%
%
%%%% for every multirow created with content, another one must be added to the next row,
%% with empty content
\multirow{3}{*}{phase} &
\multirow{3}{*}{Fig.} &
\multicolumn{7}{c |}{$D$}       \\\cline{3-9} &  
& 0  & 0.02  &  0.25 & 0.5 & 0.675 & 1.5 &  $\infty$  
\\\cline{3-9} &
& {\scriptsize Fig.\ref{fig:D0-panel}} & {\scriptsize Fig.\ref{fig:D002-panel}} & {\scriptsize Fig.\ref{fig:D025-panel}} & {\scriptsize Fig.\ref{fig:D05-panel}} & {\scriptsize Fig.\ref{fig:D0675-panel}} & {\scriptsize Fig.\ref{fig:D15-panel}} & {\scriptsize Fig.\ref{fig:Ising-panel}} \\
 \hline  
%
%%%%% 2nd row %%%%%%%%%%%%%%%%%
%
collinear stripe  & 
\ref{fig:configs}(a) & 
\checkmark &
\checkmark &
\checkmark &
\checkmark &
\checkmark &
\checkmark &
\checkmark  	  \\   
% 
%%%%% 3rd row %%%%%%%%%%%%%%%%%
%
spin flop                  & 
\ref{fig:configs}(b)  &
\checkmark &
\checkmark &
$\times$ &
$\times$ &
$\times$ &
$\times$ &
$\times$  \\ 
%
%%%%% 4th row %%%%%%%%%%%%%%%%%
%
supersolid               &  
\ref{fig:configs}(c) &
$\times$ &
\checkmark &

\checkmark &
\checkmark &
\checkmark &
$\times$ &$\times$   \\ 
%
%%%%% 5th row %%%%%%%%%%%%%%%%%
%
$m$=$1/3$ plateau &  
\ref{fig:configs}(d) &
$\times$  &
$\times$ &
$\times$ &
\checkmark &
\checkmark &
\checkmark &
\multicolumn{1}{c|}{\checkmark }   \\ 
%
%%%%% 6th row %%%%%%%%%%%%%%%%%
%
2:1:1 canted           &  
\ref{fig:configs}(e) & 
$\times$ &
\checkmark &
\checkmark &
\checkmark &
$\times$ &
$\times$ &
\multicolumn{1}{c|}{$\times$ } \\ 
%
%%%%% 7th row %%%%%%%%%%%%%%%%%
%
$m$=$1/2$ plateau    &  
\ref{fig:configs}(f)   &
\checkmark &
\checkmark &
\checkmark &
\checkmark &
\checkmark &
\checkmark &
\multicolumn{1}{c|}{\checkmark }   \\ 
%
%%%%% 8th row %%%%%%%%%%%%%%%%%
%
3:1 canted              &  
\ref{fig:configs}(g) & \checkmark &
\checkmark &
\checkmark &
\checkmark &
\checkmark &
$\times$ &
\multicolumn{1}{c|}{$\times$ }   \\  [1ex]       % [1ex] adds vertical space 
\hline \hline
\end{tabular}
\end{table}

%%%%%%%%%%%%%%%%%%%%%%%%%%%%%%%%%%%%%%%%%%%%%%%%%%%%%%%%%%%%%%%%%%%%

In this paper, we have explored the magnetic-field properties of a frustrated easy-axis Heisenberg model, 
originally introduced to explain the magnetic properties of the hexagonal delafossite 2H-AgNiO$_2$~[\onlinecite{wheeler09}].
We have used extensive Monte Carlo simulations, combined with Landau theory and spin-wave 
analysis, to obtain a set of magnetic phase diagrams for a set of values of single-ion anisotropy,
%$D$=0 [Fig.~\ref{fig:D0-panel}],  
%$D$=0.02 [Fig.~\ref{fig:D002-panel}], 
%$D$=0.25 [Fig.~\ref{fig:D025-panel}], 
%$D$=0.5 [Fig.~\ref{fig:D05-panel}], 
%$D$=0.675 [Fig.~\ref{fig:D0675-panel}], 
%$D$=1.5 [Fig.~\ref{fig:D15-panel}] and 
%$D$=$\infty$ [Fig.~\ref{fig:Ising-panel}], 
ranging from the Heisenberg [$D$=0] to the Ising [$D$=$\infty$] limits.  
The model exhibits a spectacularly rich set of phase diagrams, with a wide variety of competing collinear 
and non-collinear states, some of which are magnetic supersolids in the sense of Matsuda and Tsuneto 
or Liu and Fisher.
A catalogue of these phases, together with the values of $D$ for which they occur, is given in Table~\ref{table:phases}.

%%%%%%%%%%%%%%%%%%%%%%%%%%%%%%

The key to understanding the properties of the model for small values of anisotropy lies in 
the Heisenberg limit, $D \equiv 0$ [Fig.~\ref{fig:D0-panel}].
Here the system breaks up into two, decoupled, sublattices [Fig~\ref{fig:twosublattice}], 
and the ordered states seen at finite temperature are selected by considerations of entropy, 
rather than energy.
However the difference in entropy between these phases is very small, and so any finite magnetic 
anisotropy can stabilize new phases at low temperature, as observed for $D=0.02$ [Fig.~\ref{fig:D002-panel}].
These new phases include the collinear stripe state observed in AgNiO$_2$~[\onlinecite{wawrzynska07}], 
and the novel magnetic supersolid [Fig~\ref{fig:configs}(c)] introduced in [\onlinecite{seabra10}].

%%%%%%%%%%%%%%%%%%%%%%%%%%%%%%

That such a supersolid should be {\it stabilized} by easy-axis anisotropy might at first seem surprising, since easy-axis 
anisotropy naturally suppresses the in-plane magnetization which lends a magnetic supersolid its ``superfluid'' character.
In this case, however, the supersolid is driven by the balance of competing exchange interactions 
and anisotropy, and not by any delicate order-from-disorder effect.
This energetic origin of the supersolid makes it very robust --- it is present at low temperatures for a wide range 
of values of anisotropy $0<D \lesssim 0.9$, as illustrated in the phase diagrams for 
$D$=0.02 [Fig.~\ref{fig:D002-panel}], $D$=0.25 [Fig.~\ref{fig:D025-panel}], 
$D$=0.5 [Fig.~\ref{fig:D05-panel}] and $D$=0.675 [Fig.~\ref{fig:D0675-panel}].  

%%%%%%%%%%%%%%%%%%%%%%%%%%%%%%

The transition from the collinear ``stripe'' state [Fig.~\ref{fig:configs}(a)] into the magnetic supersolid, 
as a function of magnetic field, can be viewed as the condensation of spin wave excitations with 
{\it finite} momentum.
This sets the new state apart from other, known examples of magnetic supersolids, 
where the mode which condenses has zero momentum.  
As a consequence the supersolid does not simply interpolate between collinear stripe (solid)
and spin-flop (superfluid) states, or terminate in a tetracritical point, as envisaged by Liu and Fisher~\cite{liu73}.

%Nor is the supersolid associated with a highly degenerate region at $T$=0, as in models with \emph{XXZ} 
%and \emph{easy-plane} anisotropy supporting a biconical phase~\cite{holtschneider07, holtschneider08}.  
%%
%And in these humble, energetic, origins lies the key to the robustness of the new supersolid, which persists for 
%an extremely wide range of parameters ---cf. Fig.~\ref{fig:mft-field}.

%%%%%%%%%%%%%%%%%%%%%%%%%%%%%%

Instead, at higher temperatures, the supersolid is ``squeezed out'' of the magnetic phase diagram by 
competing phases with higher entropy --- the collinear ``stripe'' phase from which it descends, 
%[$D$=0.02, Fig.~\ref{fig:D002-panel};
%$D$=0.25,  Fig.~\ref{fig:D025-panel};
%$D$=0.5 Fig.~\ref{fig:D05-panel} and 
%$D$=0.675 Fig.~\ref{fig:D0675-panel}]
its canted ``spin-flop'' analogue [$D$=0.02; Fig.~\ref{fig:D002-panel}] 
and, at higher values of $D$, 
collinear \mbox{$m=1/3$} [$D$=0.25; Fig.~\ref{fig:D025-panel} and 
$D$=0.5; Fig.~\ref{fig:D05-panel}] 
and \mbox{$m=1/2$} magnetization plateaux [$D$=0.675; Fig.~\ref{fig:D0675-panel}].   
We anticipate that increasing the second-neighbour interaction $J_2$ (fixed in these simulations at 
$J_2 = 0.15J_1$),  will make the magnetic supersolid more robust at higher temperatures.

%%%%%%%%%%%%%%%%%%%%%%%%%%%%%%

A similar story is repeated with the two other magnetic supersolid phases found at higher values 
of magnetic fields : an unusual 2:1:1 canted state [Fig.~\ref{fig:configs}(e)], and a 3:1 canted state [Fig.~\ref{fig:configs}(g)].
The 2:1:1 canted state interpolates between the magnetic supersolid and a collinear 
$m$=1/2 plateau.
Like the supersolid, it owes its stability to a balance of competing interactions, 
and is displaced at high temperatures by the collinear $m$=1/2 plateau, which has superior entropy.
It not only exhibits a substantial staggered magnetization in the $S^x$--$S^y$ plane, but also a finite
(if small) uniform magnetization $m_\perp$, perpendicular to the magnetic field.
In turn, the 3:1 canted phase interpolates between the collinear $m$=1/2 plateau 
and saturation.
Like the 2:1:1 canted phase it exhibits a small magnetization $m_\perp$ in the 
$S^x$--$S^y$ plane.
It is the only supersolid phase present for $D=0$ (where $m_\perp \equiv 0$), 
[Fig.~\ref{fig:D0-panel}], and the only supersolid phase favoured by considerations of entropy alone.  
For this reason it is also the only supersolid connected to the paramagnetic phase at high 
temperatures.

%%%%%%%%%%%%%%%%%%%%%%%%%%%%%

The phase transitions which link these different phases also evolve as a function of $D$.
Each of the canted phases is connected to neighbouring collinear phases by a soft
spin-wave mode within the collinear state, and so these transitions can be continuous.
However phase transitions at high temperatures are generically first order, 
and become more strongly so as anisotropy is increased, as observed by e.g. 
comparing the phase diagrams for $D=0.5$ [Fig.~\ref{fig:D05-panel}], 
$D=0.675$ [Fig.~\ref{fig:D0675-panel}] and $D=1.5$ [Fig.~\ref{fig:D15-panel}].
This trend is seem most clearly in phase diagrams plotted as a function
of temperature and magnetization, which begin to exhibit large regions of phase coexistence 
as $D$ is increased.

%%%%%%%%%%%%%%%%%%%%%%%%%%%%%

The anaysis in this paper clarifies the origin of the novel magnetic supersolid introduced 
in~[\onlinecite{seabra10}], explores its unusual properties at finite temperature, and confirms 
its robustness against other competing phases for a wide range of parameters.
However a number of interesting questions remain.
One of these is, how would the system respond for magnetic field {\it not} aligned with the easy axis~?
An obvious limiting case is a field perpendicular to the easy axis.
In this case, the two sublattices of the collinear stripe phase can respond to field simply by canting, 
and (zero temperature) mean-field calculation indicate that a canted stripe phase interpolates to 
saturation for all finite $D$.
The second obvious limit is that of a field at a small angle to the easy axis.
In this case, the states described in this paper will generally survive, but the symmetries they break
will be modified by the presence of a component of magnetic field in the $S^x$-$S^y$ plane.  
This in turn will lead to a modification of the phase transitions between them.
The limit of small angle is also of relevance to experiments which measure magnetization through 
magnetic torque, discussed in the context of AgNiO$_2$ above.
Away from either of these limits, for intermediate angle, the story is not so simple --- the balance of energy 
and entropy can favour many different states.
This remains an interesting topic for future study.

%%%%%%%%%%%%%%%%%%%%%%%%%%%%

Another interesting open question is the study of the quantum mechanical effects in this spin model.
The succession of continuous phase transitions observed in our model for moderate values of anisotropy,
raises the intriguing possibility that a quantum magnet like AgNiO$_2$ could exhibit 
a series of quantum phase transitions as a function of magnetic field.  
Concentrating on the supersolid phases, two-dimensional quantum models with broadly similar 
Hamiltonians support supersolids both at T=0 [\linecite{wessel05,heidarian05,melko05}] and finite 
temperature\cite{boninsegni05,laflorencie07}. 
However, recent calculations suggest that the supersolid phase in the original three-dimensional 
model considered by Liu and Fisher may not survive quantum fluctuations \cite{ueda10}. 
Each case therefore needs to be considered on its own merits.
One simple, phenomenological, way to incorporate quantum fluctuations is through the addition an effective biquadratic 
interaction term, mimicking the selection of collinear states by quantum fluctuations\cite{larson09}.  
We have checked within classical Monte Carlo simulations that such a term does not change the nature of the 
transition into the supersolid phase.
On the strength of this, and on general grounds, we anticipate that the supersolid introduced in
[\onlinecite{seabra10}] will prove robust against quantum effects.
However this remains to be tested.

%%%%%%%%%%%%%%%%%%%%%%%%%%%%%

\section{Application to experiment}
\label{experiment}

%%%%%%%%%%%%%%%%%%%%%%%%%%%%%%%%%%%%%%%%%%%%%%%%%%%%%%%%%%%%%%%%%%%%

The advent of high field facilities offering static fields of up to 45T, and pulsed fields of up 
to 600T, has made it 
possible to explore the high-field properties of a wide range of magnets for the first time.
Frustrated magnets are foremost among these, with model systems like TlCuCl$_3$ providing a perfect 
opportunity to study how the Bose-Einstein condensation of magnons gives rise to a magnetic superfluid~[\linecite{giamarchi08}].
Generically, these systems exhibit some degree of magnetic anisotropy, either at the level of a single ion, 
or in their exchange interactions.  

%%%%%%%%%%%%%%%%%%%%%%%%%%%%%%%%%%%%%%%%%%%%%%%%%%%%%%%%%%%%%%%%%%%%

Triangular lattice antiferromagnets are no exception.
Examples with weak easy-axis anisotropy (relative to exchange interactions) 
include the quasi-two dimensional halides VBr$_2$ and VCl$_2$\cite{hirakawa83,kadowaki87}, 
which contain antiferromagnetically coupled spin-3/2 V$^{2+}$ ions on a triangular lattice.
The insulating oxide Rb$_4$Mn(MoO$_4$)$_3$~\cite{ishii09} and the multiferroic material RbFe(MoO$_4$)$_2$~\cite{svistov03, svistov06, smirnov07}, 
are well-described by a spin-5/2 nearest-neighbour Heisenberg antiferromagnet on a triangular lattice with moderate easy-axis anisotropy
(Rb$_4$Mn(MoO$_4$)$_3$), or easy-plane anisotropy (RbFe(MoO$_4$)$_2$). 
The multiferroic material KFe(MoO$_4$)$_2$ has similar underlying chemistry to RbFe(MoO$_4$)$_2$, but a distortion 
of the triangular lattice leads to a somewhat more complex phenomenology~\cite{smirnov09}.
Much stronger single-ion anisotropy, of order of the exchange interactions, is found in the hexagonal Ni halides
CsNiCl$_3$, CsNiBr$_3$\cite{sano89} and in the Cr oxide LiCrO$_2$\cite{kadowaki95}, where triangularly 
coordinated $S$=3/2 spins order on the magnetic easy-axis.  The insulating oxide 

%%%%%%%%%%%%%%%%%%%%%%%%%%%%%%%%%%%%%%%%%%%%%%%%%%%%%%%%%%%%%%%%%%%%

Further examples are found in the delafossite family, including $S$=5/2 CuFeO$_2$~\cite{ye07}, 
$S$=3/2 PdCrO$_2$\cite{takatsu09} and $S$=1 AgNiO$_2$\cite{wawrzynska07,wheeler09}
with easy-axis anisotropy, and $S$=3/2 CuCrO$_2$\cite{poienar10} with easy-plane anisotropy.
Notably, all of these delafossite materials require a frustrated model such as Eq.~(\ref{eq:H}), with second (or further) 
neighbour interactions, to accurately describe their magnetic properties.
Here we have concentrate on AgNiO$_2$, where powder neutron scattering studies provide clear 
evidence for a collinear, stripe ground state in zero magnetic field~\cite{wawrzynska07}, 
as would be expected for the frustrated, easy--axis model considered in this paper with moderate $J_2$ and $D$.

%%%%%%%%%%%%%%%%%%%%%%%%%%%%%%%%%%%%%%%%%%%%%%%%%%%%%%%%%%%%%%%%%%%%

At present, single crystals of AgNiO$_2$ are too small for inelastic neutron scattering experiments 
to be performed on them.
However angle--integrated spin--wave spectra can be measured in powder samples.
These show clear evidence of a spin-gap, and spectra are well-described by the present model 
[Eq.~(\ref{eq:H})], with parameters $J_1$=1.32meV, $J_2$=0.20meV, 
$D$=1.78meV, $J_\perp$=-0.14 meV~\cite{wheeler09}.
For spin $S$=1, the effective value of $D$ seen in quantum spin wave spectra is renormalized 
by a factor two relative to its classical value~\cite{wheeler09}, and so comparison should be 
made with classical Monte Carlo simulation results for 
$D$=0.675$\times J_1$.  
Thus, under the (strong) assumption that this simple spin model provides an adequate description of 
AgNiO$_2$ in high magnetic field, the field-temperature phase diagram for AgNiO$_2$ should be of the 
form shown in 
Fig.~\ref{fig:D0675-panel} [$D$=0.675].  
Given the simplifications inherent in using {\it any} spin model to describe a metal, the critical fields 
and transitions temperatures predicted here should be approached with some caution.
are unlikely to be quantitatively correct.
However our Monte Carlo simulation results should provide a reasonable first guide to the different
phases occuring in AgNiO$_2$ in magnetic field, and the nature of the phase transitions between them.
%

%%%%%%%%%%%%%%%%%%%%%%%%%%%%%%%%%%%%%%%%%%%%%%%%%%%%%%%%%%%%%%%%%%%%

The actual phase transitions which occur in AgNiO$_2$ for fields of up to 40T, have been studied 
through meaurements of magnetic torque
$
\vec{\tau} = {\bf m} \times {\bf h}
$, 
heat capacity and electrical transport~\cite{coldea09}.
These experiments indicate that a continuous (or very weakly first order) phase transition out of the collinear 
stripe state occurs at about 12.5T (at a base temperature of 1.5K).
This phase transition is accompanied by a weak anomaly in the specific heat.
And, crucially, it occurs at a critical field which increases steadily with temperature, indicating that the 
collinear stripe phase has a higher entropy than its successor.
These are {\it exactly} the characteristics of the phase transition from the collinear stripe phase 
into the supersolid for the frustrated easy-axis model Eq.~(\ref{eq:H}).

%%%%%%%%%%%%%%%%%%%%%%%%%%%%%

In Ref.~[\onlinecite{seabra10}] we therefore proposed that a novel magnetic supersolid is realised 
in AgNiO$_2$ for magnetic fields greater than 12.5$T$.
We further suggested that torque was a good tool for distinguishing phase transitions in easy-axis magnets,
exhibiting changes in sign as well as in magnitude,
and made explicit predictions for the torque signature of the proposed magnetic supersolid~\cite{seabra10}.
It is also interesting to note that torque clearly distinguishes those phases, such as the 2:1:1 canted phase 
and the 3:1 canted phase above, which have a finite magnetization $m_\perp$ in the $S^x$--$S^y$ plane.

%%%%%%%%%%%%%%%%%%%%%%%%%%%%%

The results of this paper confirm that the scenario presented in [\onlinecite{seabra10}] is robust across a wide range 
of parameter space, and does not require any special assumptions or fine-tuning of the model.
In this context it would be interesting to look for evidence of the spin gap closing at the putative supersolid
transition in AgNiO$_2$, either from powder neutron scattering, or NMR relaxation rates.
However it is important also to remember this model is a gross simplification of the physics of AgNiO$_2$, 
which contains itinerant electrons as well as local moments.
The development of a more realistic model for AgNiO$_2$, which takes these itinerant electrons into account, 
remains an important avenue for future study.

%%%%%%%%%%%%%%%%%%%%%%%%%%%%%%%%%%%%%%%%%%%%%%%%%%%%%%%%%%%%%%%%%%%%

\section{Conclusions}
\label{conclusions}

In this paper we have studied a realistic, three-dimensional spin model motivated by the hexagonal delafossite 2H-AgNiO$_2$.
We have obtained the full magnetic phase diagram of this model as a function of temperature and magnetic field
for values of easy-axis anisotropy $D$ ranging from the Heisenberg ($D$=0) to the Ising ($D$=$\infty$) limits.
We uncovered a rich variety of different magnetic phases, including several phases which are magnetic supersolids 
(in the sense of Matsuda and Tstuneto or Liu and Fisher), one of which may already have been observed in AgNiO$_2$\cite{coldea09}.  
We explored how this particular supersolid, first introduced in~[\onlinecite{seabra10}],  arises through the closing of 
a gap in the spin-wave spectrum, and how it competes with neighbouring phases as the easy-axis anisotropy is increased.  
This novel phase was shown to have qualitatively different finite-temperature properties from any 
previously studied magnetic supersolid, and to be remarkably robust against changes in parameters.
These results suggest that magnetic supersolids in frustrated systems can have a richer phonomenolgy, 
and be far more robust, than previously supposed.

%%%%%%%%%%%%%%%%%%%%%%%%%%%%%%%%%%%%%%%%%%%%%%%%%%%%
%%%%%%%%%%%%%

\section*{Acknowledgments}

The authors thank  Tony Carrington, Andrey Chubukov, Amalia and Radu Coldea, Andreas L\"{a}uchli, Yukitoshi Motome, 
Wolfgang Selke and Mike Zhitomirsky for helpful comments on this work.
Numerical simulations made use of the Advanced Computing Research Centre, University of Bristol.   
This work was supported by FCT Grant No.~SFRH/BD/27862/2006 and EPSRC Grants No.~EP/C539974/1
and EP/G031460/1.

%%%%%%%%%%%%%%%%%%%%%%%%%%%%%%%%%%%%%%%%%%%%%%%%%%%%%%%%%%%%%%%%

\appendix

%%%%%%%%%%%%%%%%%%%%%%%%%%%%%%

\section{Calculation of spin stiffness}
\label{spinstiffness}

Spin stiffness is defined as a generalised elasticity coefficient representing the free-energy cost of applying a twist 
to the boundary conditions with gradient $\delta\phi$ along a given direction~$\mathbf{ \hat{e} }$. 
Since there is a direct mapping between spin stiffness and superfluid density \cite{ohta79}, a phase with 
non-vanishing $\rho_S$ is said to have superfluid character. 
In finite magnetic field it is sufficient to consider twists around $S^z$, i.e in the $S^x$-$S^y$ plane. 
For a single pair of spins we write
\begin{align}
\mathbf{S}_i\cdot \mathbf{S}_j'=\cos \big(\phi_i-\phi_j+\delta\phi. \textbf{\^{e}}.(\textbf{r}_i-\textbf{r}_j  ) \big),
\label{eq:rhos-2}
\end{align}
where each spin is expressed in polar coordinates
\begin{align}
\mathbf{S}_i=(\cos\phi_i\sin\theta_i, \sin\phi_i\sin\theta_i, \cos\theta_i).
\label{eq:rhos-2.1}
\end{align}
Spin stiffness is then given by the second derivative of the free energy with respect to the twisting angle $\delta\phi$ 
\begin{align}
\rho_s[\textbf{\^{e}}] =\frac{\partial^2\mathcal{F}}{\partial(\delta\phi)^2}\Big|_{\delta\phi=0} \hspace{-5pt} = \Big\langle \pd{^2\mathcal{H}}{(\delta\phi)^2} \Big\rangle \Big|_{\delta\phi=0}\hspace{-5pt}-\frac{1}{T} \Big\langle  \Big(\pd{\mathcal{H}}{(\delta\phi)}\Big) ^2 \Big\rangle \Big|_{\delta\phi=0}.
\label{eq:rhos-1}
\end{align}
Applying Eq. (\ref{eq:rhos-1}) to both first- and second-neighbour interactions and normalising per unit area
(i.e. per spin), spin stiffness reads \cite{ohta79,teitel83}
\begin{align}
&\rho_s(\textbf{\^{e}})=  -\frac{2  }{\sqrt{3}N}  \Bigg\langle J_1\hspace{-5pt}\sum_{{\langle i,j \rangle}_1}\hspace{-3pt}  (\textbf{\^{e}} .\mathbf{r}_{ij})^2 . \mathbf{S}^\perp_i.\mathbf{S}^\perp_j 
\hspace{-2pt}+\hspace{-2pt}
 J_2\hspace{-5pt}\sum_{{\langle i,j \rangle}_2}\hspace{-3pt} (\textbf{\^{e}} .\mathbf{r}_{ij})^2 .\mathbf{S}^\perp_i.\mathbf{S}^\perp_j    \hspace{-2pt} \Bigg\rangle \nonumber \\
&
- \frac{2}{\sqrt{3}NT} \Bigg\langle  \hspace{-3pt} \Big\{  \hspace{-1pt}J_1\hspace{-5pt}\sum_{{\langle i,j \rangle}_1}\hspace{-3pt}   (\textbf{\^{e}} .\mathbf{r}_{ij}) . \mathbf{S}^\perp_i 
\hspace{-3pt} \times \hspace{-2pt} \mathbf{S}^\perp_j
\hspace{-2pt}+\hspace{-2pt}
  J_2\hspace{-5pt}\sum_{{\langle i,j \rangle}_2} \hspace{-3pt} (\textbf{\^{e}} .\mathbf{r}_{ij}) . \mathbf{S}^\perp_i              \hspace{-3pt}    \times \hspace{-2pt}
      \mathbf{S}^\perp_j \Big\}^2     \Bigg\rangle ,
\label{eq:rhos-3}
\end{align}
where $\mathbf{r}_{ij}$$=$$\textbf{r}_i$$-$$\textbf{r}_j$. %
Since the parallel tempering method restores the full lattice symmetries, we average $\rho_S$ over the three symmetric directions in the lattice \mbox{$\textbf{\^e}$$=$$(\textbf{\^e}_x,\textbf{\^e}_y)\hspace{-3pt}=\{(1,0),(1/2,\sqrt{3}/2),(-1/2,\sqrt{3}/2)$}\}.

%%%%%%%%%%%%%%%%%%%%%%%%%%%%

\section{Low temperature expansion and Landau theory for supersolid transition }
\label{appendix}

The low-temperature excitations of a classical spin model of the form Eq.~(\ref{eq:H}) are  frozen spin waves, in contrast with the dynamic spin waves found in quantum magnets. These excitations have a dispersion which can be calculated through an expansion of  each spin fluctuations  around its preferred $T$=0 configuration.
In the sufficiently generic case of two spins in a common plane canted by an angle $2\theta$,
the Heisenberg interaction is approximated by  
\begin{align}
\label{eq:CSW-spin-spin}
\mathbf{S}_i \cdotp \mathbf{S}_j \approx & \cos{2\theta}-\frac{1}{2}\big(x_i^2 + x_j^2 + y_i^2 + y_j^2\big)\cos{2\theta} \nonumber\\
& +  x_ix_j\cos{2\theta} +y_iy_j,
\end{align}
where $x_i$ and $y_i$ denote small fluctuations about the ordered state  in the local $S^x_i$ and $S^y_i$ directions respectively. The rotation into a common frame is  arbitrarily chosen to be performed around the $S^y$ direction.
%%%%%%%%%%%%%%%%%%%%%%%%%%%%%%

The resulting low-temperature Hamiltonian is solved by Fourier transform, with two spin-wave  modes 
$\epsilon_\alpha(\mathbf{k})$ per $n$ sublattice, i.e.
\begin{eqnarray}
\label{eq:Csw-H}
\mathcal{H}=  \mathcal{H}_{0} + \frac{1}{2}\sum^{2n}_\alpha  \sum^{N/n}_{\mathbf{k} \in M_{BZ} } \epsilon_\alpha(\mathbf{k}) x_{\alpha_{\mathbf k}} x_{\alpha_{ \mathbf{-k}}}.
\end{eqnarray}

%%%%%%%%%%%%%%%%%%%%%%%%%%%%%%

We now illustrate this procedure for  the supersolid state with four different sublattices : while two are collinear with $S^z$ in the positive direction the other two are canted by an angle of $\theta$ from $S^z$ in opposite directions. After writing down the interactions between sublattices in matrix form, the spin-wave dispersion modes can be obtained by diagonalisation of
\begin{align}
\label{eq:Mx}
M^x_{\mathbf{k}} \hspace{-2pt} = \hspace{-2pt} 2\left[\hspace{-4pt} \hspace{-2pt} \begin{array}{cccc}
{V}_\mathbf{k}   & X_\mathbf{k}  & Y_\mathbf{k}.c_\theta & Z_\mathbf{k}.c_\theta  \\
X_\mathbf{k}   & {V}_\mathbf{k}  & Z_\mathbf{k}.c_\theta  & Y_\mathbf{k}.c_\theta  \\
Y_\mathbf{k}.c_\theta  & Z_\mathbf{k}.c_\theta & {W}_\mathbf{k} + D.c_{2\theta}  & X_\mathbf{k}.c_{2\theta} \\
Z_\mathbf{k}.c_\theta  & Y_\mathbf{k}.c_\theta & X_\mathbf{k}.c_{2\theta} & {W}_\mathbf{k}+D.c_{2\theta}   \\ 
 \end{array} \hspace{-4pt}\right] ,
 \end{align} 
\begin{align}
\label{eq:My}
M^y_{\mathbf{k}} = 2\left[\begin{array}{cccc}
{V}_\mathbf{k}  & X_\mathbf{k}  & Y_\mathbf{k} & Z_\mathbf{k} \\
X_\mathbf{k}  & {V}_\mathbf{k} & Z_\mathbf{k} & Y_\mathbf{k} \\
Y_\mathbf{k}  & Z_\mathbf{k} & {W}_\mathbf{k}+D.{c^2_\theta} & X_\mathbf{k} \\
Z_\mathbf{k}  & Y_\mathbf{k} &  X_\mathbf{k} & {W}_\mathbf{k} +D.{c^2_\theta} \\ 
 \end{array} \right], 
 \end{align} 
where $c_\theta$$=$$\cos{\theta}$ and the other coefficients are given by
\begin{align}
\label{eq:factors}
&{V}_\mathbf{k}= (J_1+J_2)(2c_\theta-1) + J_{\perp} (\gamma_z(\mathbf{k})-1)  +D + \frac{h}{2} ,\nonumber \\
&{W}_\mathbf{k}= (J_1+J_2)(2c_\theta-c_{2\theta}) + J_{\perp} (\gamma_z(\mathbf{k})-1)   - \frac{h}{2}c_{\theta},\nonumber \\
&X_\mathbf{k}= J_1\gamma_1^{AB}(\mathbf{k})+J_2\gamma_2^{AB}(\mathbf{k}), \nonumber \\
&Y_\mathbf{k}=- J_1\gamma_1^{+}(\mathbf{k}) -J_2\gamma_2^{-}(\mathbf{k}), \nonumber \\
&Z_\mathbf{k}=- J_1\gamma_1^{-}(\mathbf{k}) -J_2\gamma_2^{+}(\mathbf{k}).
 \end{align} 

%%%%%%%%%%%%%%%%%%%%%%%%%%%%%%

The lattice structure factors are $\gamma_1^{AB}(\mathbf{k})$=$\cos{k_x}$, $\gamma_2^{AB}(\mathbf{k})$=$\cos{\sqrt{3}k_y}$, $\gamma_z(\mathbf{k})$=$\cos k_z$, \mbox{$\gamma_1^{\pm}(\mathbf{k})$=$\cos((k_x\pm\sqrt{3}k_y)/2)$} and  $\gamma_2^{\pm}(\mathbf{k})$=$\cos((3k_x\pm\sqrt{3}k_y)/2)$. The four-sublattice structure results in 8 dispersing modes $\epsilon(\mathbf{k})$. 
The thermodynamics at low-T  are accessed via the free energy,
\begin{align}
\label{eq:free-energy}
\frac{\mathcal{F}}{N}=\frac{E_0}{N} -T\ln T \hspace{-2pt} + \hspace{-2pt} \frac{1}{2} \frac{T}{N} \sum^{2n}_\alpha \hspace{-4pt} \sum^{N/n}_{\mathbf{k} \in M_{BZ} } \hspace{-8pt} \ln \epsilon_\alpha(\mathbf{k}) \hspace{-2pt} + \hspace{-2pt} O(T^2).
\end{align}

%%%%%%%%%%%%%%%%%%%%%%%%%%%%%%

The free energy of the supersolid phase near its phase transition can be obtained as a Landau expansion in powers of the order parameter measuring the $U(1)$ broken symmetry 
\begin{align}
\label{eq:GL-1}
\mathcal{F}\approx \mathcal{F}_0 \frac{a}{2}|O_{U(1)}|^2 + \frac{b}{4}|O_{U(1)}|^4 + ...  \hspace{0.5cm}.
\end{align}
 The $U(1)$ order parameter [Eq.~(\ref{eq:U1-op})] is by construction proportional to the canting angle $\theta$, as represented in the cartoon Fig.~\ref{fig:configs}(c), and therefore near the transition 
  \begin{align}
O_{U(1)}=2\sin\theta\sim\theta\label{eq:GL-a}.
\end{align}
The  coefficient $a$ collects the quadratic contributions from the spin wave modes
 \begin{align}
\label{eq:GL-h(T)}
a&=h_{\sf SSD}(T)-h, \\
h_{\sf SSD}(T)&=2D+2 T \frac{\partial^2}{\partial\theta^2} \Big\{  \frac{1}{2N} \sum_{\alpha,\mathbf{k} } \ln \epsilon_\alpha(\mathbf{k}) \Big\}  \Big|_{\theta=0},
\end{align}
where the critical field is calculated near $T$=0. The prefactor for the temperature term is obtained by numerically integrating the derivatives of the eight spin-wave branches over the four-sublattice magnetic Brillouin Zone, corresponding to the finite lattice used in simulation. The resulting critical fields are plotted as a yellow line on the  (a) panel of \mbox{Figs. \ref{fig:D002-panel},\ref{fig:D025-panel}, \ref{fig:D05-panel} and \ref{fig:D0675-panel}}, yielding a very good agreement with simulation, even at relatively high temperatures.

%%%%%%%%%%%%%%%%%%%%%%%%%%%%%%%%%%%%%%%%%%%%%%%%%%

\end{document}